\documentclass{article}
\usepackage{amsmath,amsfonts,amsthm,amssymb,amscd,latexsym, bm,bbm}
\usepackage{hyperref}
\usepackage[usenames,dvipsnames]{color}
\hypersetup{colorlinks,
citecolor=Blue,
linkcolor=Blue,
urlcolor=Blue}

\usepackage[T2A]{fontenc}
\usepackage{graphicx}
\usepackage{mathrsfs}

{\theoremstyle {definition} \newtheorem {defi} {Definition} [section] }
{\theoremstyle {plain}  \newtheorem {theo} [defi] {Theorem}}
{\theoremstyle {plain}  \newtheorem {cor} [defi]{Corollary}}
{\theoremstyle {plain}  \newtheorem {lem} [defi]{Lemma}}
{\theoremstyle {plain}  \newtheorem {prop} [defi]{Proposition}}
{\theoremstyle {plain} \newtheorem {rem}[defi] {Remark}}
{\theoremstyle {remark} }

\numberwithin{equation}{section}
\newtheorem*{defi*}{Definition}
\newtheorem*{problem*}{Problem}

\newtheorem*{rem*}{Remark}
\newtheorem*{note*}{Note}

\makeatletter \@addtoreset{equation}{section} \makeatother

\newcommand{\mC}{\mathbb{C}}
\newcommand{\mR}{\mathbb{R}}
\newcommand{\mT}{\mathbb{T}}
\newcommand{\mZ}{\mathbb{Z}}
\newcommand{\mN}{\mathbb{N}}

\newcommand{\mI}{\mathbb{I}}

\newcommand{\BB}{{\cal B}}
\newcommand{\CC}{{\cal C}}
\newcommand{\DD}{{\cal D}}
\newcommand{\EE}{{\cal E}}

\newcommand{\GG}{{\cal G}}

\newcommand{\LL}{{\cal L}}
\newcommand{\MM}{{\cal M}}
\newcommand{\PP}{{\cal P}}

\newcommand{\UU}{{\cal U}}
\newcommand{\YY}{{\cal Y}}
\newcommand{\RR}{{\cal R}}

\newcommand{\JJ}{{\cal J}}
\newcommand{\TT}{{\cal T}}

\newcommand{\XX}{{\cal X}}
\newcommand{\fB}{\mathfrak{P}}
\newcommand{\msP}{\mathscr{P}}

\newcommand{\CCC}{{\mC^{|\CC|}}}

\newcommand{\eps}{\varepsilon}
\newcommand{\ph}{\varphi}

\newcommand{\Imm}{\operatorname{Im}}
\newcommand{\Ree}{\operatorname{Re}}

\newcommand{\diag}{\operatorname{diag}}

\newcommand{\Supp}{\operatorname{Supp}}

\newcommand{\ka}{\kappa}
\newcommand{\om}{\omega}
\newcommand{\Om}{\Omega}
\newcommand{\tht}{\theta}
\newcommand{\al}{\alpha}
\newcommand{\ga}{\gamma}
\newcommand{\la}{\lambda}
\newcommand{\La}{\Lambda}
\newcommand{\del}{\delta}
\newcommand{\Del}{\Delta}

\newcommand{\ov}{\overline}
\newcommand{\wid}{\widetilde}

\newcommand{\ssk}{\smallskip}
\newcommand{\msk}{\medskip}

\newcommand{\chp}{\partial}
\newcommand{\MO}{\mbox{\bf E}\,}
\newcommand{\MOOM}{\mbox{\bf E}_{\Omega_R}\,}

\newcommand{\PR}{\mbox{\bf P}\,}
\newcommand{\PRQ}{\mbox{\bf Q}\,}
\newcommand{\ds}{\displaystyle{}}
\newcommand{\ra}{\rightarrow}
\newcommand{\ran}{\rangle}
\newcommand{\lan}{\langle}
\newcommand{\volna}{\thicksim}
\newcommand{\sm}{\setminus}
\newcommand{\raw}{\rightharpoonup}
\newcommand{\ass}{\quad\mbox{as}\quad}
\newcommand{\assN}{\quad\mbox{as}\quad\eps\ra 0\quad\mbox{uniformly in }\CC}

\newcommand{\bee}{\begin{equation}}
\newcommand{\eee}{\end{equation}}
\newcommand{\btt}{\begin{theo}}
\newcommand{\ett}{\end{theo}}
\newcommand{\bl}{\begin{lem}}
\newcommand{\el}{\end{lem}}
\newcommand{\bpp}{\begin{prop}}
\newcommand{\epp}{\end{prop}}
\newcommand{\bcc}{\begin{cor}}
\newcommand{\ecc}{\end{cor}}
\newcommand{\bdd}{\begin{def}}
\newcommand{\edd}{\end{def}}
\newcommand{\brr}{\begin{rem}}
\newcommand{\err}{\end{rem}}

\def\dbar{{\mathchar'26\mkern-12mu d}}

\begin{document}
\large
\title
{Nonequilibrium statistical mechanics of weakly stochastically perturbed system of oscillators}
\author{A. Dymov\footnote{Universit\'e de Cergy-Pontoise, CNRS, Mathematics Department, 
F-95000 Cergy-Pontoise.} \footnote{Steklov Mathematical Institute, Moscow, Russia; 
e-mail: adymov88@gmail.com.}}
\date{}
\maketitle

\begin{abstract}
We consider a finite region of a $d$-dimensional lattice, $d\in\mN$, of weakly coupled harmonic oscillators. The coupling is provided by a nearest-neighbour potential (harmonic or not) of size $\eps$. Each oscillator weakly interacts by force of order  $\eps$ with its  
own stochastic Langevin thermostat of arbitrary positive temperature.  We investigate limiting as $\eps\ra 0$ behaviour of solutions of the system and of the local energy of oscillators on long-time intervals of order $\eps^{-1}$ and in a stationary regime. We show that it is governed by an effective equation which is a dissipative SDE with nondegenerate diffusion. 
Next we assume that the interaction potential is of size $\eps\la$, where $\la$ is another small parameter, independent from $\eps$. Solutions corresponding to this scaling describe small low temperature oscillations.  We prove that in a stationary regime, under the limit $\eps\ra 0$, the main order in $\la$ of the averaged  Hamiltonian energy flow 
is 
proportional to the gradient of temperature. We show that  the coefficient of proportionality, which we call the conductivity, admits  a representation through stationary space-time correlations of the energy flow. 
Most of the results and convergences we obtain are uniform with respect to the number of oscillators in the system. 
\end{abstract}

\tableofcontents

\section{Introduction}
\subsection{Motivation and set up}
Investigation of the energy transport in crystals is one of the central problems in nonequilibrium statistical mechanics. In particular, a derivation of the Fourier law and the Green-Kubo formula from the microscopic dynamics of particles is of great interest (see \cite{Pei,Leb}). In the classical setting one investigates the energy transport in a Hamiltonian system, interacting with thermal baths of different temperatures. This interaction is weak in geometrical sense: the thermal baths are coupled with the Hamiltonian system only through a boundary of the latter.  
Unfortunately, for the moment of writing this problem seems to be out of reach, due to the weakness of the interaction. In this setting even the existence of a stationary state in the system is not obvious (see \cite{Eck,EckH,RBT,Car,HM,CEPo, CE})
\footnote{See also \cite{Tr} and \cite{Dym12} for a similar problem in deterministic setting.}. That is why usually one modifies the system in order to get additional ergodic properties. Standard ways to achieve that are {\it i)} to consider a weak perturbation of the hyperbolic system of independent particles (\cite{DL},\cite{R});  {\it ii)} to perturb each mode of the Hamiltonian system by sufficiently strong noise  (\cite{BLL, BO, BBO, BLLO, LO, BeKLeLu, BO1}).  
Following the second way, usually one perturbs the Hamiltonian dynamics by very special noise, 
preserving the energy of the system (\cite{BO,BBO,LO,BeKLeLu,BO1}),  
or one couples each particle of the Hamiltonian system with its own stochastic Langevin-type thermostat (\cite{BLL,BLLO,BeKLeLu}). In the latter case the stochastic perturbation has rather natural structure, however the energy of the system is not conserved. 

Let us discuss some of results mentioned above in more details. 
In \cite{BO} the authors study a chain of harmonic oscillators, perturbed by energy preserving stochastic exchange of momentum between neighbouring oscillators. Coupling the first and the last oscillator to thermal baths of different temperatures $\TT_1$ and $\TT_2$, they prove the Fourier law, i.e. that  the stationary averaged  flow of energy is approximately proportional to $(\TT_1-\TT_2)/N$ for a large number of particles  $N$ and a small difference $\TT_1-\TT_2$.
In \cite{BBO} the authors consider a system of oscillators, where each oscillator is perturbed by a stochastic dynamics, conserving momentum and energy. They study existence of the Green-Kubo conductivity.
In \cite{BLL} the authors consider a system of harmonic oscillators, where each oscillator is coupled with a stochastic Langevin thermostat. Fixing temperatures of the "exterior" left and right thermostats they find and study the self-consistent temperature profile for the "interior" thermostats, i.e. a temperature profile such that in a stationary regime there is no average flow of energy between  the "interior" thermostats and the system of oscillators.
Then they prove that with respect to this temperature profile the Fourier law and the Green-Kubo formula hold.

In all the works listed above the stochastic perturbation is of order one. It is natural to study the case when it goes to zero. Such situation was investigated in \cite{BOS, BeHu, BeHuLeLO}.
In \cite{BOS} the authors study the FPU-chain, where the nonlinearity is replaced by an energy preserving stochastic exchange of momentum between neighbouring nodes. They investigate the energy transport under the limit when the strength of this exchange tends to zero.
In \cite{BeHu} the authors consider a chain of harmonic oscillators, where each oscillator is weakly perturbed by energy preserving noise of size $\eps$ and anharmonic potential of size $\nu$, $\nu\leq\eps$. They investigate limiting as $\eps\ra 0$ behaviour of an upper bound for the Green-Kubo conductivity.
In  \cite{BeHuLeLO} the authors  study an infinite chain of cells weakly coupled by a potential of size $\nu$, where each cell is weakly perturbed by  energy preserving  noise of size $\eps$. Expanding the Green-Kubo conductivity in a formal series $\ka(\nu,\eps)=\sum\limits_{n\geq 2}\nu^n\ka_n(\eps)$, they find the term $\ka_2(\eps)$ and  investigate  its limiting behaviour as $\eps\ra 0$. It is argued that  under the formal limit $\nu\ra 0$  the Fourier law holds with the conductivity $\ka_2(\eps)$.

In the present work 
we consider a system of oscillators, situated in nodes of a bounded set $\CC\subset\mZ^d$, $d\in\mN$, and interacting via a nearest-neighbour type potential. The system is given by the Hamiltonian 
\begin{equation}
\label{oH}
H^\nu(p,q)=\sum\limits_{j \in\CC} \Big(\frac{p_j^2}{2} + U_j(q_j)\Big) + \frac{\nu}{2}\sum\limits_{j,k \in\CC: |j-k|=1} V(q_j,q_{k}) , 
\end{equation}
where $(p,q)=(p_j,q_j)_{j\in\CC}\in\mR^{2|\CC|}$, $\nu>0$ and for $j\in\mZ^d$ by $|j|$ we denote the $l_1$-norm. The interaction potential $V$ satisfies $V(x,y)\equiv V(y,x)$, is $C^2$-smooth and has at most a polynomial growth at infinity (for precise assumptions see Section \ref{osec:ass}). The pinning potentials $U_j$ will be specified later. Following the second strategy above, we couple each oscillator with its own stochastic Langevin thermostat of arbitrary temperature $\TT_j$ by coupling of size $\eps$, where the temperatures satisfy
\bee\label{otemp}
0<\TT_{j}< C<\infty,
\eee
with a constant $C$ independent from $j$. 
The resulting system obeys the equation
\bee\label{oini_ee}
\frac{d}{dt}q_j=p_j,\quad 
\frac{d}{dt}p_j=-\chp_{q_j} U_j(q_j) - \nu\sum\limits_{k:|k-j|=1}\chp_{q_j}V(q_j,q_k) - \eps p_j + \sqrt{2\eps\TT_j}\frac{d}{dt}\beta_j, \quad j\in\CC,
\eee
where $\beta = (\beta_j)_{j \in\CC}\in \mR^{|\CC|}$  is a standard $|\CC|$-dimensional Brownian motion.  
Our goal is to investigate limiting as $\eps\ra 0$ behaviour of solutions and of energy transport for eq. (\ref{oini_ee}) on long time intervals and in a stationary regime. So, as in the classical setting given above, we are interested in the case of weak interaction between the Hamiltonian system and the thermal baths, 
however the weakness is understood in a different, non-geometrical sense.
Unfortunately, we will have to assume that the coupling constant $\nu$ goes to zero simultaneously with $\eps$, with some precise scaling which will be discussed later.

Our system admits a clear physical interpretation, so not only it can be considered as a toy model for a purely Hamiltonian situation, but it is of independent interest. Indeed, since the stochastic perturbation is provided by the weak coupling with the Langevin thermostats, one can think about the transport of energy through a crystal plugged in some medium of a given temperature distribution and weakly interacting with it. However, since the energy of the system is not conserved, in our case non-Hamiltonian effects seem to be stronger than in the systems above, perturbed by vanishing energy conserving noises. In order to have some kind of energy preservation, one could consider the self-consistent temperature profile, if the latter exists. 

Since the systems of statistical physics are of very large dimension, it is crucial to control the dependence of the systems on their size. Let us emphasize that most of results and convergences we obtain are uniform with respect to the choice of the set $\CC$ (in particular, with respect to its size), so they satisfy this physical requirement.  

We started our study  in \cite{Dym14}. There we considered a system of rotators, where the uncoupled rotators were assumed nonlinear, while the coupling constant $\nu$ was chosen as $\nu=\eps^a,\,a\geq 1/2$. In the case $1/2\leq a <1 $ the rotators were required to have alternated spins (in such a way we excluded from the system too strong resonances). Using the averaging method of Khas'minski-Freidlin-Wentzell type in the form developed in \cite{KuPi}, we found an autonomous (stochastic)  equation for the local energy which described  its limiting  (as $\eps\ra 0$) behaviour on long time intervals of order $\eps^{-1}$ and in a stationary regime, uniformly with respect to the choice of the set $\CC$. However, it turned out that this equation  did not feel the Hamiltonian interaction of rotators, so that  
in the considered system  there was no Hamiltonian energy flow under the limit $\eps\ra 0$, neither on time intervals of order $\eps^{-1}$ nor in the stationary regime. The reason is that under natural assumptions resonances of the system of uncoupled nonlinear rotators have  Lebesgue measure zero. A weak coupling does not change the situation significantly, so that the noise pushes out rapidly the perturbed dynamics to nonresonant area, where the Hamiltonian terms average out. 
In \cite{FW06}
\footnote{See also Chapter 9.3 in \cite{FW}.} 
similar results were obtained for a system similar to (\ref{oini_ee}), where the uncoupled oscillators were nonlinear and $\nu=\eps$ (however, no information was provided about dependence of the results on the choice of the set $\CC$ and behaviour of the system in a stationary regime).

In order to observe the Hamiltonian energy flow, in the present paper we study the situation when the uncoupled oscillators are linear and have the same frequencies, so that
\bee\label{oUj'}
U_j(q_j)=\frac{q_j^2}{2} \quad\mbox{for all}\quad  j\in\CC.
\eee
In this case eq. (\ref{oini_ee}) is strongly resonant, see below. We choose the coupling constant $\nu$ as
\bee\label{onueps}
\nu=\la\eps, \quad\mbox{where}\quad 0<\la\leq 1,
\eee
since we are not able to treat the case when $\eps=o(\nu)$, while the choice (\ref{onueps}) already provides nontrivial results. In the first part of the paper, devoted to the study of the limiting as $\eps\ra 0$ behaviour of solutions of eq. (\ref{oini_ee})-(\ref{onueps}), we assume that $\la$ is a fixed number satisfying $0<\la\leq 1$. In the second part, treating the limiting (as $\eps\ra 0$) behaviour of the energy flow, we assume that $\la$ is a small parameter, independent from $\eps$ (but usually also fixed). 
Eq. (\ref{oini_ee})-(\ref{onueps}) with $\la\ll 1$ describes small amplitude low temperature oscillations, 
see Appendix \ref{oapp:lt} for explanation.

Since $\la$ will be mostly fixed while $\eps$ will go to zero, natural time scale for eq. (\ref{oini_ee})-(\ref{onueps}) is $t\volna \eps^{-1}$. It is convenient to introduce a slow time $\tau=\eps t$, so that $\tau\volna 1$. Then eq. (\ref{oini_ee})-(\ref{onueps}) takes the form
\bee\label{oini_er}
\dot q_j=\eps^{-1} p_j,\quad 
\dot p_j=-\eps^{-1}  q_j - \la\sum\limits_{k:|k-j|=1}\chp_{q_j}V(q_j,q_k) - p_j + \sqrt{2\TT_j}\dot\beta_j, \quad j\in\CC,
\eee  
where the dot denotes the derivative with respect to $\tau$.
It is well known that eq. (\ref{oini_er})  has a unique solution, defined for all $\tau\geq 0$ (see \cite{Khb}),  and is mixing \nolinebreak
\footnote{I.e. it has a unique stationary measure and its solutions weakly converge as $\tau\ra\infty$ to this measure in distribution.} (see Appendix \nolinebreak\ref{oapp:mixing}).

\subsection{Main results: the limiting dynamics}
Our first goal is to investigate the limiting as $\eps\ra 0$ dynamics of eq. (\ref{oini_er}) for fixed $\la$ satisfying $0<\la\leq 1$.
Introduce the action-angle variables for the system of uncoupled oscillators, given by the Hamiltonian $H^0$, 
\bee\label{oacang}
I=(I_j)_{j\in\CC}\in\mR^{|\CC|},\; I_j:=\frac{p_j^2+q_j^2}{2} \quad\mbox{and}\quad\ph=(\ph_j)_{j\in\CC}\in\mT^{|\CC|},\; \ph_j:=\arg (p_j+i q_j), 
\eee
where $i$ denotes the imaginary unit and we put $\arg 0:=0$. 
Then eq. (\ref{oini_er})  takes the form 
\bee
\label{ointroaa}
\dot I_j =\ldots, \quad
\dot\ph_j =\eps^{-1}+ \ldots, \quad j\in \CC,
\eee
where the dots stand for terms of order one (as $\eps\ra 0$). 
Eq. (\ref{ointroaa}) describes a strongly resonant fast-slow system: the actions change slowly while the dynamics of angles can be decomposed to a fast rotation along the diagonal $\bm 1=(1,\ldots,1)$ of the torus $\mT^{|\CC|}$, and a slow drift in $|\CC|-1$  ortogonal directions $\xi^m$, $\xi^m\cdot \bm 1=0$, $1\leq m\leq |\CC|-1.$ A usual way of studying of the limiting as $\eps\ra 0$ dynamics of such systems is the method of resonant averaging (see \cite{AKN}). 
The latter suggests to pass to the variables $(I,\psi,\eta)\in\mR^{|\CC|}\times\wid\mT^{|\CC|}$, where $(\psi,\eta)=(\psi_1\ldots,\psi_{|\CC|-1},\eta)\in\wid\mT^{|\CC|}$ are coordinates of the vector of angles $\ph$ with respect to the basis $(\xi^1,\ldots,\xi^{|\CC|-1}, \bm 1)$, and $\wid\mT^{|\CC|}$ denotes the torus $\mT^{|\CC|}$ expanded in a suitable way.
 Then eq. (\ref{ointroaa}) takes the form 
\bee\label{ointroaa1}
\dot I_j=\ldots,\quad \dot\psi_m=\ldots,\quad \dot\eta=\eps^{-1}+\ldots,\quad\mbox{where}\quad 1\leq m\leq |\CC|-1.
\eee  
Eq. (\ref{ointroaa1}) is a fast-slow system with the unique fast variable $\eta$, so the limiting as $\eps\ra 0$ behaviour of the slow variables $(I,\psi)$ is expected to be given by the averaging of $(I,\psi)$-equations with respect to $\eta$, which further on we call the {\it resonant averaging}. However, in order to establish this we can not use the usual scheme which suppose to work in the action-angle variables, since writing the $\ph$-equations in more details we find $\dot\ph_j =\eps^{-1}+ I_j^{-1}\cdot\ldots$, $j\in\CC$, so that they have singularities at the locus   
$\{I=(I_j)_{j\in\CC}:\, I_j=0 \mbox{ for some } j\in\CC \}$. Moreover, the resonant averaging of the $I$-equations has weak singularities there. 
To overcome this difficulty we use a method developed in \cite{Kuk10,Kuk13} for a non-resonant case and in \cite{KM} for a resonant one. It suggests to find and study an effective equation, written in the regular $(p,q)$-variables, which governs the limiting as $\eps\ra 0$ dynamics of the $(I,\psi)$-components  of solutions of eq. (\ref{oini_er}). To this end, in a general situation one should use that the $(I,\psi)$-components of solutions of the effective equation have to satisfy the resonant averaging of the $(I,\psi)$-equations from (\ref{ointroaa1}). However,
following \cite{KM}, we note that in our case, where the system of uncoupled oscillators is linear, there exists a simple way to find the effective equation without using the $(I,\psi)$-coordinates, so that the singularities does not appear at all. It is based on a passage to a fast-rotating coordinate system, where eq. (\ref{oini_er}) has uniformly in $\eps$ bounded coefficients, see Section \ref{osec:eff}. The effective equation has the form 
\bee
\label{ointroeff}
\dot q_j=\la \chp_{p_j} H^{res} - \frac{q_j}{2} +  \sqrt{\TT_j}\dot \beta^1_j, \quad
\dot p_j=-\la \chp_{q_j} H^{res} - \frac{p_j}{2} + \sqrt{\TT_j} \dot \beta^2_j, \quad j\in\CC,
\eee
where $(\beta^k_j)_{k=1,2,\,j\in\CC}$ are standard independent Brownian motions, while
$H^{res}(p,q)$ denotes the resonant averaging of the  interaction potential $\frac12\sum\limits_{|k-j|=1}V(q_j,q_k)$ (see (\ref{hress})).
\footnote{Despite that the interaction potential does not depend on the moments $p_k$, its resonant averaging may depend on them.}

Eq. (\ref{ointroeff}) has a unique solution, this solution is defined globally,  and (\ref{ointroeff}) is  mixing (see \cite{Khb, Ver87,Ver97}).  
Let us fix $T\geq 0$ and some "not very bad" random initial conditions $(p_0,q_0)$.
\footnote{By "not very bad" we mean that the components $p^2_{0j},q^2_{0j}$ have finite exponential moments which are bounded uniformly with respect to $j\in\CC$ and with respect to the choice of the set $\CC\subset\mZ^d$. See ass. {\it HI} in Section \ref{osec:ass}.}
We prove
\btt
\label{otheo:introav}
Let $(p^\eps,q^\eps)(\tau)$ and $(p,q)(\tau)$ be unique solutions of equation (\ref{oini_er}) and the effective equation (\ref{ointroeff}), satisfying $\DD(p^\eps,q^\eps)(0)=\DD(p,q)(0)=\DD(p_0,q_0)$. Then
\bee\label{6468763875964}
\DD\big(I(p^\eps,q^\eps)(\cdot)\big) \raw \DD\big(I(p,q)(\cdot)\big) \ass \eps\ra 0\quad\mbox{on } C([0,T],\mR^{|\CC|}) \mbox{ uniformly in $\CC$}.
\eee
\ett
Theorem \ref{otheo:introav} is a corollary of Theorem \ref{otheo:a}. This is explained in Section \ref{sec:avth}. 
Here $\DD(\cdot)$ denotes the distribution of a random variable. 
By saying that a convergence is uniform in $\CC$, we mean that it holds  uniformly with respect to the choice of the finite set $\CC\subset\mZ^d$. In particular, with respect to the number of nodes $|\CC|$. 
Uniformity in $\CC$ of the weak convergences of measures through all the text is understood in the sense of {\it finite-dimensional projections}. It means that for any finite set of nodes $\La\subset\mZ^d$, projections of considered measures to $\La$ converge with a rate, independent from the choice of the set  $\CC\supseteq\La$. 
For instance, for convergence (\ref{6468763875964}) this means
$$
\MO f\big(I(p^\eps,q^\eps)(\cdot)\big)\ra \MO f\big(I(p,q)(\cdot)\big) \ass\eps\ra 0 \quad\mbox{uniformly in }\CC,  
$$
for any  continuous bounded functional  $f:C([0,T],\mR^{|\CC|})\mapsto \mR$, satisfying
$\Supp f\subseteq\La$, where $\Supp f$ is defined in Section \ref{osec:ass}, Agreements.6.
Here and further on $\MO$ denotes the expectation.

\begin{rem}\label{orem:jd}
In a similar, but technically more complicated way, it can be shown that the limiting (as $\eps\ra 0$) behaviour  of joint actions $I$ and slow components on angles $\psi$ is also governed by the effective equation,  see \cite{KM}. Since it is not needed for study of the limiting transport of energy, we skip it.
\end{rem}
Let $\mu^\eps$ be the unique stationary measure of eq. (\ref{oini_er}). 
Each limiting point (as $\eps\ra 0$) of the family of measures $\{\mu^\eps,\,0<\eps\leq 1\}$ is an invariant measure of  eq. (\ref{oini_ee})-(\ref{onueps}) with $\eps=0$. This equation describes a system of uncoupled harmonic oscillators, so it has plenty of invariant measures. The next theorem ensures that, in fact, only one of them is the limiting point and distinguishes it.
\btt
\label{otheo:smintro} Let $\mu^\eps$ and $\mu$ be the unique stationary measures of equation (\ref{oini_er}) and the effective equation (\ref{ointroeff}) correspondingly. Then $\mu^\eps\raw \mu$ as $\eps\ra 0$. 
Under some additional assumption this convergence holds uniformly in $\CC$.
\ett
See for details Theorem \ref{otheo:sm}.
 Roughly speaking, the additional assumption above demands the effective equation for an infinite system of oscillators to have a unique stationary measure, in some class of measures. In particular, it holds if $\la$ is sufficiently small
and second partial derivatives of the interaction potential $V$ are bounded.

To establish the uniformity in $\CC$ of convergences from Theorems \ref{otheo:introav} and \ref{otheo:smintro}
we employ a method developed in \cite{Dym14}. Its main tool is Lemma \ref{olem:est}, where using $l_2$-norms with exponential decaying weight, we obtain uniform in $\CC$, $\eps$ and $\tau$ estimates for solutions of equation \nolinebreak (\ref{oini_er}).

In Appendix \ref{oapp:mixing} we prove that rate of mixing of equation (\ref{oini_er}) is independent from $\eps$. Jointly with Theorems \ref{otheo:introav} and \ref{otheo:smintro} this implies the following result.
\btt\label{otheo:Iunifintro}
(i) The convergence $\DD\big(I(p^\eps,q^\eps)(\tau)\big)\raw \DD\big(I(p,q)(\tau)\big)$ as $\eps\ra 0$ provided by Theorem \nolinebreak \ref{otheo:introav} holds uniformly in $\tau\geq 0$. \\
(ii) We have
$$
\lim\limits_{\eps\ra 0} \lim\limits_{\tau\ra\infty} \DD\big(I(p^\eps,q^\eps)(\tau)\big)=\lim\limits_{\tau\ra\infty}\lim\limits_{\eps\ra 0} \DD\big(I(p^\eps,q^\eps)(\tau)\big)=
\lim\limits_{\substack{\eps\ra 0 \\ \tau\ra\infty}} \DD\big(I(p^\eps,q^\eps)(\tau)\big)= \Pi_{I*} \mu,
$$
where $\mu$ is the unique stationary measure of the effective equation (\ref{ointroeff}) and $\Pi_I$ denotes the projection to the space of actions.
\ett
For details see Theorem \ref{otheo:Iunif}.
Define the local energy of a $j$-th oscillator as
\bee
\label{oH_j}
\EE^{\nu}_j(p,q):=\frac{p_j^2+q_j^2}{2} + \frac{\nu}{2}\sum\limits_{k \in\CC: |j-k|=1} V(q_j,q_k), \quad \mbox{so that} \quad H^{\nu}=\sum\limits_{j\in\CC}\EE^{\nu}_j. 
\eee
Let $(p^\eps,q^\eps)(\tau)$ be a solution of eq. (\ref{oini_er}). 
Since $\EE^{\nu}_j=I_j+O(\nu)$, we obtain the following result whose full version is given in Proposition \nolinebreak\ref{olem:introen}.
\bpp
The limiting (as $\eps\ra 0$) behaviour of the vector of local energy \\
$(\EE_j^{\nu}(p^\eps,q^\eps))_{j\in\CC}$ coincides with that of the vector of actions $I(p^\eps,q^\eps)$, 
while the latter is described by Theorems \nolinebreak \ref{otheo:introav}-\ref{otheo:Iunifintro}. 
\epp

\subsection{Main results: the energy transport} 
Our next goal is to study the limiting (as $\eps\ra 0$) dynamics of a stationary Hamiltonian energy flow in system (\ref{oini_er}). We will assume that $\la$ is sufficiently small.
\footnote{Here and further on the rate of smallness of $\la$ is independent neither from $\eps$ nor from  the choice of the set $\CC$.}

The local energy $\EE_j^{\nu}$ changes due to the Hamiltonian coupling of the $j$-th oscillator with neighbouring oscillators and with the $j$-th  thermostat. Applying the Ito formula, we get
\bee
\label{oH'intro}
\dot\EE^{\nu}_j=\eps^{-1}\{\EE_j^{\nu}, H^{\nu}\}+\mbox{ (thermost. term)}_j= \frac{\la}{2}\sum\limits_{k:|k-j|=1} \JJ_{kj}+ \mbox{ (thermost. term)}_j,
\eee
where $\{\cdot,\cdot\}$ denotes the Poisson bracket, (thermost. term)$_j: = -p_j^2 + \TT_j + \sqrt{2\TT_j} \dot\beta_j$,
and
\bee\label{oef}
\JJ_{kj}(p,q):=\Big\{\frac{p_j^2+q_j^2}{2}-\frac{p_k^2+q_k^2}{2}, V(q_j,q_k) \Big\}= p_k\chp_{q_k} V(q_k,q_j)-p_j\chp_{q_j}V(q_k,q_j)
\eee
is the Hamiltonian energy flow from the $k$-th oscillator to the $j$-th one, normalized with respect to $\la$ (see \cite{Leb}).
Or, for short, the energy flow. We prove
\btt 
\label{otheo:introF}
Let $\mu^{\eps,\la}$ be the unique stationary measure of eq. (\ref{oini_er}). Assume $\la$ to be sufficiently small.
Then, under suitable assumptions for the interaction potential $V$, there exists a $C^1$-smooth strictly positive function $\kappa:\mR^2_+\mapsto \mR_+$ satisfying $\kappa(x,y)\equiv\kappa(y,x)$,  such that for every $j,k\in\CC$, $|j-k|=1$, we have
\bee
\label{ointroF} 
\lan\mu^{\eps,\la},\JJ_{kj}\ran \ra \la\kappa(\TT_k,\TT_j)(\TT_{k}-\TT_{j}) + o(\la)\quad \mbox{as}\quad \eps\ra 0 
\quad\mbox{uniformly in $\CC$,}
\eee
where $o(\la)/\la\ra 0$ as $\la\ra 0$ uniformly in $\CC$.
\ett
For details see Theorem \ref{otheo:F}. We emphasize that the function $\ka(\TT_k,\TT_j)$, which we call the {\it conductivity}, depends only on the temperatures $\TT_k,\TT_j$, and not on the choice of the set $\CC$.
Convergence (\ref{ointroF}) shows that in the stationary regime the averaged flow of energy  is approximately proportional (uniformly in $\CC$) to the local temperature gradient, if the noise and the interaction between oscillators are sufficiently weak. The coefficient of proportionality $\kappa$  is strictly positive and depends on the temperatures in a sufficiently smooth way. 
To prove Theorem \ref{otheo:introF} we note that Theorem \ref{otheo:smintro} implies 
$\lan\mu^{\eps,\la},\JJ_{kj}\ran \ra \lan \mu^\la, \JJ_{kj}\ran$ as $\eps\ra 0$ uniformly in $\CC$, where $\mu^\la$ is the unique stationary measure of the effective equation (\ref{ointroeff}). Then we analyse the development of the measure $\mu^\la$ in $\la$. 
\brr For the reader interested in the limiting (as $\eps\ra 0$) behaviour of the stationary thermostatic energy flow, we note that it can be easily obtained  from eq. (\ref{oH'intro}). Indeed, let us average the both sides of (\ref{oH'intro}) with respect to the stationary measure $\mu^{\eps,\la}$. The left-hand side vanishes, while the limiting behaviour of the first term in the right-hand side is given by  Theorem \ref{otheo:introF}. 
\err

Now for simplicity we fix the dimension of the lattice $d=1$ and put $\CC=\{0,1,\ldots,N\}.$ 
Let us study the limit $N\ra\infty$. For $x\in[0,1]$ denote $j_x:=[xN]$ and $\JJ(x):=\JJ_{j_x+1j_x}$. If we put the oscillators to the interval $[0,1]$, a $j$-th oscillator to the point $j/N$, then the function $\JJ(x)$ will describe the energy flow between the nearest to the point $x$ oscillators, so it can be seen as the energy flow through the point $x$.  Choose a temperature profile $\TT_j:=\TT(j/N)$, where $\TT(x)$ is a $C^1$-smooth positive function, defined for $x\in [0,1]$. Then Theorem \ref{otheo:introF} immediately implies
\bee\label{ohydro}
\lim\limits_{N\ra\infty}\lim\limits_{\la\ra 0}\lim\limits_{\eps\ra 0} \frac{N}{\la}\lan\mu^{\eps,\la},\JJ(x)\ran=\hat\kappa(\TT(x))\frac{d}{dx}\TT(x),\quad\mbox{where}\quad\hat\kappa(y):=\kappa(y,y).
\eee
If the temperature profile is linear, i.e. $\TT(x)=\TT_1 x+ \TT_0(1-x)$, where $\TT_0,\TT_1>0$, then (\ref{ohydro}) implies that
\bee\label{ohydro1}
\lim\limits_{N\ra\infty}\lim\limits_{\la\ra 0}\lim\limits_{\eps\ra 0}\frac{N}{\la\delta\TT}\lan\mu^{\eps,\la},\JJ(x)\ran=\hat\kappa(\TT(x)),
\eee
where $\delta\TT:=\TT_1-\TT_0$. This resembles the Fourier law, see \cite{Leb}.

Next we study stationary space-time correlations of the energy flow. In particular, we get the following result. 
\btt\label{otheo:GKintro}
For any $\hat\TT>0$ and $N\in\mN$ we have 
\bee\label{o8345}
\hat\kappa(\hat\TT) =\lim\limits_{\la\ra 0}\lim\limits_{\eps\ra 0} \frac{1}{\hat\TT^2 N}
\int\limits_{0}^\infty \big\lan\mu^{\eps,\la}, \sum\limits_{j=0}^{N-1} \JJ_{jj+1} \fB_\tau^{\eps,\la}\big(\sum\limits_{j=0}^{N-1} \JJ_{jj+1} \big) \big\ran\,d\tau,
\eee
where $\mathfrak{P}^{\eps,\la}_\tau$ denotes the Markov semigroup associated with  equation (\ref{oini_er}) with the temperatures satisfying  $\TT_j=\hat\TT$ for every $j\in\CC$, while  $\mu^{\eps,\la}$ denotes  its unique stationary measure (which is the Gibbs measure at the temperature $\hat\TT$).
\ett
See for details Section \ref{osec:GK}. Relation (\ref{o8345}) resembles the Green-Kubo formula, see \cite{Sp},\cite{Leb} and \cite{LepLiPo}. 
\footnote{In formula (\ref{o8345}) we could also send $N$ to infinity, however it does not make sense:  limit (\ref{o8345}) is the same for all $N$.} 
To prove Theorem \ref{otheo:GKintro} we employ the averaging technique developed above, an explicit formula for the conductivity $\kappa$, given in Theorem \ref{otheo:F} (which is a full version of Theorem \nolinebreak\ref{otheo:introF}) and the uniformity in $\eps$ of the rate of mixing for eq. (\ref{oini_er}).

Despite that (\ref{ohydro1}) and (\ref{o8345}) resemble the Fourier law and the Green-Kubo formula, of course they are not.

\subsection{Assumptions and agreements}
\label{osec:ass}

{\bf Assumptions.} Throughout the text we assume that the following hypotheses hold.

{\bf HV.} 
{\it The function $V:\mR^2\ra \mR$ is $C^r$-smooth, $r\geq 2,$ and $V(x,y)= V(y,x)$ for any $x,y\in\mR$. There exists a constant $C>0$ such that}
\begin{align}\label{oHG1}
|V(x,y)|\leq &C(1+ x^2+y^2),  \\
\label{oHG2}
|\chp_{z_1}V(x,y)|\leq C(1+|x|+|y|),&\; |\chp^2_{z_1 z_2}V(x,y)|\leq C(1+x^2+y^2),
\end{align}
{\it where $z_1,z_2\in\{x,y\}$. Moreover, higher orders partial derivatives of $V$  have at most a polynomial growth at infinity.} 

{\bf HI.} {\it There exist constants $\al_0>0$ and $C>0$, independent from the choice of the set $\CC$, such that the initial conditions $(p_0,q_0)$ satisfy}
$$\MO e^{\al_0 (p_{0j}^2+q_{0j}^2)}< C\quad \mbox{for every}\quad j \in\CC.$$
\begin{rem}
\label{orem:HG1}
We use relations (\ref{oHG1})-(\ref{oHG2}) only to obtain uniform in $\CC$, $\eps$ and $\tau$  a-priori estimates (\ref{oestimates}) for solutions and stationary measures of eq. (\ref{oini_er}), and to prove that eq. (\ref{oini_er}) is mixing with independent from $\eps$ rate. Once these properties are given, we do not more need (\ref{oHG1})-(\ref{oHG2}). 

More specifically, for Theorem  \nolinebreak\ref{otheo:introav} we need the first estimate from (\ref{oestimates}) to be held. For Theorems \nolinebreak \ref{otheo:smintro} and \ref{otheo:introF} we need (\ref{oestimates}) and uniqueness of the stationary measure for eq. (\ref{oini_er}). For Theorems \nolinebreak \ref{otheo:Iunifintro} and \ref{otheo:GKintro} we need the same, and additionally, we need eq.  (\ref{oini_er}) to be mixing with independent from $\eps$ rate. 

Note that the exponential estimates (\ref{oestimates}) are superfluous. It suffices to have polynomial estimates of sufficiently large degree. Moreover, if we have  only estimates which are not uniform in $\CC$ then all the results remain true except the uniformity of convergences in $\CC$. 
\end{rem}

{\bf Agreements.}
$1)$ We refer to item {\it (i)} of Theorem \ref{otheo:Iunifintro} as Theorem \ref{otheo:Iunifintro}.{\it i}, etc.

$2)$ By $C,C_1,C_2,\ldots$ we denote various positive constants and by $C(a),C_1(a),\ldots$~---  positive constants which depend on a parameter $a$. Sometimes we skip the dependence of the constants on parameters. However, unless otherwise stated, we always indicate if they depend on the choice of the set $\CC$, times $t,s,\tau,\ldots$, positions $j,k,l,m,\ldots\in\CC$ and small parameters $\eps, \la$. 
Constants $C,C(a),\ldots$ can change from formula to formula.

$3)$  We use notations $a \wedge b:=\min(a,b),\; a \vee b=\max(a,b)$.

$4)$ For $M\in\mN$ and vectors $a=(a_k),\, b=(b_k)\in\mC^M$,  by $a\cdot b$ we denote the Euclidean scalar product in 
$\mC^M\simeq\mR^{2M}$, and by $|a|$ the corresponding norm,
\footnote{We denote the $l_2$-norm in the space of complex vectors $\mC^M$ and the $l_1$-norm in the space of indices $\mZ^d$ by the same sign $|\cdot|$. This will not cause misunderstanding.}
$$a\cdot b:=\sum\limits_k(\Ree a_k \Ree b_k + \Imm a_k \Imm b_k )=\sum\limits_k\Ree a_k\ov b_k  
\quad\mbox{and}\quad |a|^2:=a\cdot a=\sum\limits_k|a_k|^2.$$

$5)$ By saying that a function $f:\mC^M\mapsto\mR$ is $C^n$-smooth we mean that it is $C^n$-smooth as a function $f:\mR^{2M}\mapsto\mR$.

$6)$
Let $X$ be some vector space and $f:(x_j)_{j\in\CC}\in X^{|\CC|}\mapsto \mR$. By $\Supp f$ we denote a minimal set $S\subseteq\CC$ such that the function $f$ is independent from $(x_j)_{j\notin S}$.

$7)$ For a metric space $X$ by $\LL_b(X)$ ($\LL_{loc}(X)$) we denote the space of bounded Lipschitz continuous (locally Lipschitz continuous) functions from $X$ to $\mR$.

$8)$ Assertions of the type "something is sufficiently small/sufficiently close to one/\ldots" always assume an estimate independent from the choice of the set $\CC$. 

$9)$ We assume $0<\eps\leq 1$ to be sufficiently small where it is needed. 

{\bf Acknowledgments.}
I am very grateful to  my PhD supervisors S. Kuksin and A. Shirikyan for formulation of the problem, guidance and encouragement.  
I would like to thank N. Cuneo, J.-P. Eckmann, V. Jaksic, J.L. Lebowitz and C. Liverani for useful discussions, and N. Cuneo and J.-P. Eckmann
for the excellent stay in Geneva. This work is supported by the Russian Science Foundation under grant 14-50-00005 and performed in Steklov Mathematical Institute of Russian Academy of Science.

\section{The limiting dynamics}
\label{osec:theory} 
In this section we study the limiting as $\eps\ra 0$ behaviour of solutions of equation (\ref{oini_er}). Most of results of this section are independent from the size of $\la$, where $0< \la\leq 1$.

\subsection{Estimates on solutions}
\label{osec:est}
	
It is well known that eq. (\ref{oini_er}) has a unique solution and this solution is defined for all $\tau\geq 0$ (see \cite{Khb}). In Appendix \nolinebreak\ref{oapp:mixing} we prove that eq. (\ref{oini_er})  is mixing. 
First we need to obtain estimates for its solution and stationary measure, which are unifrom in $\eps,\tau$ and $\CC$. 
\begin{lem}
\label{olem:est}
There exist constants $\al>0, \eps_0>0$ and $C>0$, independent from the choice of the set $\CC$, such that for all  $0<\eps<\eps_0,\,0<\la\leq 1, j\in\CC$ and $\tau\geq 0$  we have
\bee
\label{oestimates}
(i) \, \MO\max\limits_{s\in [\tau,\tau+1]} e^{\al(p_j^2(s)+q_j^2(s))} < C, \qquad\qquad 
(ii)\, \lan \mu,  e^{\al(p_j^2+q_j^2)}\ran < C,
\eee
where $(p,q)(\tau)$ is a solution of eq. (\ref{oini_er}), satisfying $\DD(p,q)(0)=\DD(p_0,q_0)$, and  $\mu$ is its unique stationary measure. 
\end{lem}
{\it Proof.}
Item {\it (ii)} follows from item {\it (i)}, the mixing property of eq. (\ref{oini_er}) and the Fatou lemma. 
\footnote{In fact, it can be obtained even without employing of the mixing property, in a slightly more complicated way (see Section \nolinebreak 2.5.2 from \cite{KuSh}). One should use that dynamics "forgets" initial conditions (see (\ref{ofog})), and the Fatou lemma.}
Let us now explain the idea of the proof of item {\it (i)}. 
Fix some  $1/2<\ga<1$, and denote
\bee\label{oUj}
U^j(p,q):=\sum\limits_{k\in\CC}\ga^{|k-j|} \big(\EE_k(p,q) +\frac{\eps}{2}p_k q_k\big),\quad j\in\mZ^d,
\eee
where we recall that the local energy $\EE_k$ is defined in (\ref{oH_j}). 
\footnote{For the brevity of notations here and further on in this proof we skip the upper index $\nu$.}
Estimate (\ref{oestimates}).{\it i} with the constant $C$ depending on the number of nodes $|\CC|$ can be obtained in a standard way, by applying the Ito formula to the function $e^{\al U^j}|_{\ga=1},$ see (\ref{oexp}). To derive it, one should crucially use that for $\ga=1$ the function  $\EE^j:=\sum\limits_{k\in\CC}\ga^{|k-j|}\EE_k(p,q)$ coincides with  the Hamiltonain  $H$, so that $\EE^j|_{\ga=1}$ is conserved by the Hamiltonian flow.
The obtained in such a way estimate is not uniform in $\CC$ because the r.h.s. of (\ref{oexp}) contains terms proportional to the sum $\sum\limits_{k\in\CC}\ga^{|j-k|}$, which equals to $|\CC|$ if $\ga=1$. To overcome this difficulty, we put $\ga<1$, so that the sum above is bounded uniformly in $\CC$. 
Despite that in this case the function $\EE^j$ is not a first integral of the Hamiltonain flow, the nearest-neighbour character of interaction between oscillators provides that if $\ga\volna 1$,  $\EE^j$ is its  "approximate integral", in the sense that $\eps^{-1}\{\EE^j,H\}\volna (1-\ga)$, see (\ref{oe2}). 
 This suffices to establish the desired estimate. 

Now we start the proof. First we will establish the following auxiliary proposition. 
Introduce the family of norms 
\bee\label{decrnorm}
\|x\|^2_j:= \sum\limits_{k\in\CC}\ga^{|j-k|}x_k^2, \quad\mbox{where}\quad j\in\mZ^d \quad\mbox{and}\quad x=(x_k)_{k\in\CC}\in\mR^{|\CC|}.  
\eee
\bpp\label{olem:est2}
There exists $\eps_1>0$ such that for any $1/2<\ga<1$, $j\in\mZ^d$ and $0<\eps<\eps_1$ we have 
$$(\|p\|_j^2+\|q\|^2_j)/4\leq U^j(p,q)\leq C(\|p\|_j^2+\|q\|^2_j)+C_1(\ga), \quad (p,q)\in\mR^{2|\CC|}.$$ 
\epp
{\it Proof.}
Let us prove the upper bound. Since (\ref{oHG1}) implies 
$
\ds{\EE_k\leq \frac{p_k^2}{2} + C\sum\limits_{l:|l-k|\leq 1} q_l^2 + C,}
$
we have
$$
U^j\leq \sum\limits_{k\in\CC} \ga^{|j-k|} \Big( \frac{p_k^2}{2}+ C\sum\limits_{l:|l-k|\leq 1} q_l^2 +C + \frac{p_k^2+q_k^2}{2} \Big) 
\leq	 \sum\limits_{k\in\CC} \ga^{|j-k|}  \big( p_k^2 + C_1 \sum\limits_{l:|l-k|\leq 1} q_l^2  +C).
$$  
Since $\ga>1/2$, we have 
$\ds{\sum\limits_{k\in\CC}\ga^{|j-k|} \sum\limits_{l:|l-k|\leq 1} q_l^2  \leq 2 \sum\limits_{k\in\CC} \sum\limits_{l:|l-k|\leq 1} \ga^{|j-l|} q_l^2 \leq C\|q\|^2_j}$. 
Moreover, 
$
\sum\limits_{k\in\CC}\ga^{|j-k|}  \leq \sum\limits_{m\in\mZ^d}\ga^{|m|} \leq C(\ga). 
$
Then we obtain the desired estimate. The lower bound follows from (\ref{oHG1}) and smallness of $\eps$ in a similar way.
\qed

Now let us formulate the following result, which we will establish after the end of the proof of the lemma.
\bl\label{olem:est1}
Let $1/2<\ga<1$ be sufficiently close to one and $0<\eps_0<\eps_1$ be sufficiently small, where $\eps_1$ is defined in Proposition \ref{olem:est2}. Fix any $\del_0>0$. Then for $0<\del<\del_0$ sufficiently small and any solution $(p,q)(\tau)$ of eq. (\ref{oini_er}), satisfying for some $j\in\mZ^d$ the estimate 
\bee\label{oinicond}
\MO e^{\del_0 U^j(p,q)(0)}<C,
\eee
we have
\bee\label{oestimates'}
\MO\max\limits_{s\in [\tau, \tau+1]} e^{\del U^j(p,q)(s)}<C_1,
\eee
for any $0<\eps<\eps_0$, $0<\la\leq 1$ and $\tau\geq 0,$ where $C_1=C_1(\ga,\eps_0,\del,C).$
\el 
We fix the parameters $\ga$ and $\eps_0$ as in Lemma \ref{olem:est1} and do not indicate dependence of constants from them.
Using first the upper bound from Proposition \ref{olem:est2} and then the Jensen inequality joined with ass. {\it HI}, we see that 
\bee\label{CCCdva}
\MO e^{\del_0  U^j(p_0,q_0)}\leq C_1 \MO e^{\del_0 C (\|p_0\|_j^2+\|q_0\|_j^2)} < C_2,
\eee
for $\del_0$ sufficiently small and any $j\in\mZ^d$.
Since $p_j^2+q_j^2\leq \|p\|_j^2+\|q\|_j^2$, the lower bound from Proposition \ref{olem:est2} joined with Lemma \ref{olem:est1} implies that 
\bee\label{CCCodin}
\MO\max\limits_{s\in [\tau, \tau+1]} e^{\del(p^2_j(s)+q^2_j(s))/4}\leq\MO\max\limits_{s\in [\tau, \tau+1]} e^{\del U^j(p,q)(s)}<C,
\eee
for $\del>0$ sufficiently small. Since the constant $C_2$ from (\ref{CCCdva}) is independent from $j$, the constant $C$ from (\ref{CCCodin}) also is, so that we get the desired estimate with $\al:=\del/4$.
\qed
\ssk

{\it Proof of Lemma \ref{olem:est1}}.
Without loss of generality we assume that $j=0$ and skip the index $j$ in the notations for $U^j$ and $\|\cdot\|_j$.
Moreover, for simplicity  we put $\la=1$: analysing the proof below it is easy to see that all estimates we obtain are uniform in $0<\la\leq 1.$

{\it Step 1.} First we will find $\ga, \eps_0$ and $\del_1>0$ such that for any $0<\eps<\eps_0$ and $s\geq 0$ we have
\bee
\label{o1est}
\MO e^{\del_1 U(s)}\leq C,
\eee 
where we have denoted $U(s):=U(p,q)(s)$. 
Further on we present only formal computation which could be justified by standard stopping-time arguments (see, e.g., \cite{KaSh}).
Applying  the Ito formula to $e^{\del_1 U(s)}$  we obtain
\begin{align}
\nonumber
\frac{d}{d s}e^{\del_1 U(s)} &=
\eps^{-1}\big\{e^{\del_1 U(s)},H\big\} - \sum\limits_{l\in\CC}p_l\chp_{p_l}e^{\del_1 U(s)} + \sum\limits_{l\in\CC}\TT_l\chp^2_{p_l^2}e^{\del_1 U(s)}  + \frac{d}{ds} M_{s_0,s} \\  \label{oexp}
&=\del_1 e^{\del_1 U(s)}(\YY_1+\YY_2 + \YY_3) +  \frac{d}{ds}M_{s_0,s}, 
\end{align}
where
$$
\YY_1=\eps^{-1}\{U,H\}, \quad
\YY_2=- \sum\limits_{l\in\CC}p_l\chp_{p_l}U, \quad
\YY_3=\sum\limits_{l\in\CC} \TT_l \big(\chp^2_{p_l^2}U+\del_1 (\chp_{p_l}U)^2\big),
$$
and the martingal $\ds{
M_{s_0,s}:=\del_1\int\limits_{s_0}^s e^{\del_1 U(s)}\sum\limits_{l\in\CC} \sqrt{2\TT_l}\chp_{p_l}U}\,d\beta_l$, where $s_0\leq s$.
Let us  estimate the terms $\YY_1,\YY_2,\YY_3$ separately. 

{\it Term $\YY_1$}. 
Writing $\YY_1$ in more details we obtain
\bee\label{oYg1}
\YY_1=\eps^{-1}\sum\limits_{k\in\CC}\ga^{|k|}\{\EE_k,H\} + \sum\limits_{k\in\CC}\frac{\ga^{|k|}}{2} \{p_kq_k, H\}. 
\eee
We have
\begin{align}\nonumber
\eps^{-1}\{\EE_k,H\} &= 
\Big\{\frac{p_k^2}{2}, \frac12 \sum\limits_{|l-m|=1}V(q_m,q_l) \Big\} + 
\Big\{\frac12 \sum\limits_{l:|l-k|=1}V(q_k,q_l), \sum\limits_{m\in\CC}\frac{p_m^2}{2} \Big\} \\ \nonumber
&= \Big\{\frac{p_k^2}{2}, \sum\limits_{l:|l-k|=1}V(q_k,q_l) \Big\} + 
\frac12 \sum\limits_{l:|l-k|=1} \Big\{V(q_k,q_l), \frac{p_k^2+p_l^2}{2} \Big\} \\ \label{186410683}
&= \sum\limits_{l:|l-k|=1} \Big\{\frac{p_k^2-p_l^2}{4}, V(q_k,q_l)\Big\}. 
\end{align}
Note that for any $(x_{kl})_{k,l\in\CC}\in\mR^{|\CC\times \CC|}$ we have 
$ \sum\limits_{k\in\CC}\sum\limits_{l:|l-k|=1}x_{kl}= \frac12\sum\limits_{|l-k|=1}(x_{kl}+x_{lk}).$ 
Using (\ref{186410683}) and applying the latter identity with 
$x_{kl}=\ds{\ga^{|k|}\Big\{\frac{p_k^2-p_l^2}{4}, V(q_k,q_l)\Big\}}$,
we find 
\begin{align}\nonumber
\eps^{-1}\sum\limits_{k\in\CC}\ga^{|k|}\{\EE_k,H\}&=
\sum\limits_{k\in\CC}\sum\limits_{l:|l-k|=1}\ga^{|k|}\Big\{\frac{p_k^2-p_l^2}{4}, V(q_k,q_l)\Big\} \\ \label{oe1}
&=\frac12\sum\limits_{|l-k|=1} (\ga^{|k|}-\ga^{|l|})\Big\{\frac{p_k^2-p_l^2}{4}, V(q_k,q_l)\Big\}.
\end{align}
Since for $k,l\in\CC$ satisfying $|l-k|=1$ we have $|l|-|k|=\pm 1$, we obtain 
$$
|\ga^{|k|}-\ga^{|l|}|=\ga^{|k|}|1-\ga^{|l|-|k|}|=
\left\{ 
\begin{array}{cl}
\ga^{|k|}(1-\ga), &\mbox{ if }|l|-|k|=1 \\
\ga^{|k|}\ga^{-1}(1-\ga), &\mbox{ if }|l|-|k|=-1
\end{array}  
\right.
\leq 2\ga^{|k|}(1-\ga),
$$
where we have used that $1/2<\ga<1$. In view of estimate (\ref{oHG2}), we find \\
$\ds{\Big|\Big\{\frac{p_k^2-p_l^2}{4}, V(q_k,q_l)\Big\}\Big|\leq C(p_k^2+p_l^2+q_k^2+q_l^2+1)}$. Then from (\ref{oe1}) we get
\bee\label{oe2}
\eps^{-1}\sum\limits_{k\in\CC}\ga^{|k|}\{\EE_k,H\}\leq (1-\ga)C \sum\limits_{|l-k|=1}\ga^{|k|}(p_k^2+p_l^2+q_k^2+q_l^2+1)
\leq (1-\ga)C_1 (\|p\|^2+\|q\|^2) + C_2(\ga).
\eee
Let us estimate the term $\ds{\sum\limits_{k\in\CC}\frac{\ga^{|k|}}{2}\{p_kq_k,H\}}$ from (\ref{oYg1}). We have 
$$
\{p_kq_k,H\} = p_k\chp_{p_k} H -q_k\chp_{q_k}H = p_k^2-q_k^2 - \eps\sum\limits_{l:|l-k|=1}q_k\chp_{q_k}V(q_k,q_l).
$$
Then, using that 
$|q_k\chp_{q_k}V(q_k,q_l)|\leq C(q_k^2+q_l^2+1)$, we find
\begin{align}\nonumber
\sum\limits_{k\in\CC}\frac{\ga^{|k|}}{2}\{p_kq_k,H\}
&\leq\frac12(\|p\|^2-\|q\|^2) + C\eps\sum\limits_{k\in\CC}\sum\limits_{l:|l-k|=1}\ga^{|k|}(q_k^2+q_l^2+1)\\ \label{oe3}
&\leq\|p\|^2/2+(-1/2+C_1\eps)\|q\|^2 + C_2(\ga).
\end{align}
Combining (\ref{oe2}) and (\ref{oe3}) we obtain
\bee\label{oe4}
\YY^1\leq \big((1-\ga)C+1/2\big) \|p\|^2 + \big((1-\ga+\eps)C-1/2\big) \|q\|^2+C_1(\ga).
\eee

{\it Term $\YY_2$.} Since
\bee\label{oe5}
\chp_{p_l}U
=\ga^{|l|}\Big(p_l+ \frac{\eps}{2} q_l\Big),
\eee
we find
\bee\label{oe6}
\YY_2= -\|p\|^2 -  \frac{\eps}{2}\sum\limits_{l\in\CC}\ga^{|l|}p_lq_l\leq (-1+\eps)\|p\|^2+ \eps\|q\|^2.
\eee

{\it Term $\YY_3$.} 
In view of  (\ref{oe5}) we have $\chp^2_{p^2_l}U=\ga^{|l|}$. Then, using that the temperatures $\TT_l$ are bounded  uniformly in $l\in\CC$ (see (\ref{otemp})),  we find
\bee\label{oe7}
\YY_3=\sum\limits_{l\in\CC}\TT_l\Big(\ga^{|l|}+\del_1 \ga^{2|l|}\big(p_l+ \frac{\eps}{2} q_l\big)^2\Big)
\leq C(\ga) + 2\del_1(\|p\|^2+\|q\|^2).
\eee
Substituting (\ref{oe4}),(\ref{oe6}) and (\ref{oe7}) to (\ref{oexp}), we get
\bee\label{oe8}
\frac{d}{d s}e^{\del_1 U}\leq \del_1 e^{\del_1 U} \Big(\Del \big(\|p\|^2+\|q\|^2\big) + C(\ga)\Big)+ \frac{d}{ds}M_{s_0,s},
\eee
where
$$\Del(\ga,\eps,\del_1):= \big((1-\ga)C+1/2-1+\eps+2\del_1\big)\vee \big((1-\ga+\eps)C-1/2+\eps + 2\del_1\big).$$ 
Fix $1/2<\ga<1$ sufficiently close to one, $\eps_0>0$ and $0<\del_1<\del_0$ sufficiently small, in such a way that 
$\Del(\ga,\eps,\del_1)<0$ for $\eps\leq\eps_0$. 
Then, using Proposition \ref{olem:est2}, for $\eps\leq\eps_0$ we obtain 
\bee\label{oe9}
\frac{d}{d s}e^{\del_1 U}\leq \del_1 e^{\del_1 U} \big(\Del_0 U + C\big)+ \frac{d}{ds}M_{s_0,s},
\eee
for some $\Del_0<0$, where we do not more indicate dependence of constants on the parameters $\ga,\eps_0$ and $\del_1$.  Then for an appropriate constant $C$ we get
\bee\label{oe10}
\frac{d}{d s}e^{\del_1 U}\leq - e^{\del_1 U} + C + \frac{d}{ds}M_{s_0,s}.
\eee
Fixing $s_0=0$, taking the expectation and applying the Gronwall inequality to (\ref{oe10}), we obtain
\bee\label{ofog}
\MO e^{\del_1 U(s)} \leq \MO e^{\del_1 U(0)} e^{-s} + C,
\eee
so that we have (\ref{o1est}).      
\ssk
 
{\it Step 2.} Now we will prove (\ref{oestimates'}) for $0<\del<\del_1/2$. 
Integrating inequality (\ref{oe10}) with $\del_1$ replaced by $\del$ over the interval $[\tau,s]$, where $\tau \leq s\leq \tau+1$, putting $s_0=\tau$  and taking $\MO\max\limits_{s\in[\tau,\tau+1]}$, we get
\bee
\label{oe11}
\MO \max\limits_{s\in[\tau,\tau+1]} e^{\del U(s)}\leq \MO e^{\del U(\tau)} + C+ \MO \max\limits_{s\in[\tau,\tau+1]} M_{\tau,s} \leq 
 C_1+  \MO \max\limits_{s\in[\tau,\tau+1]} M_{\tau,s}, 
\eee
where we have used (\ref{o1est}).
The Doob-Kolmogorov inequality ensures that 
\bee
\label{oDK}
\MO \max\limits_{s\in[\tau,\tau+1]} M_{\tau,s}\leq C \MO \sqrt{[M]_{\tau,\tau+1}} \leq C \sqrt {\MO [M]_{\tau,\tau+1}},
\eee
where $[M]_{\tau,s}$ denotes the quadratic variation of the martingale $M_{\tau,s}$. Due to (\ref{oe5}), 
\begin{align}
\nonumber
[M]_{\tau,\tau+1}
 &=\del^2 \int\limits_{\tau}^{\tau+1} e^{2\del U}\sum\limits_{l\in\CC} 2\TT_l\ga^{2|l|}\Big(p_l + \frac{\eps}{2} q_l\Big)^2\, ds
 \leq C\int\limits_{\tau}^{\tau+1} e^{2\del U} \sum\limits_{l\in\CC} \ga^{2|l|}(p_l^2 +  q_l^2) \, ds \\ \nonumber
 &\leq 4C \int\limits_{\tau}^{\tau+1} e^{2\del U} U \,ds,
\end{align}
in view of the lower bound from Proposition \ref{olem:est2} and the estimate  
$\|p\|^2+\|q\|^2 \geq \sum\limits_{l\in\CC} \ga^{2|l|}(p_l^2 +  q_l^2)$.
Since $2\del<\del_1$, estimate (\ref{o1est}) provides $\MO [M]_{\tau,\tau+1}\leq C$, so that (\ref{oe11}) joined with (\ref{oDK}) implies the desired relation (\ref{oestimates'}).
\qed

\subsection{a-variables and the effective equation}
\label{osec:eff}

Instead of studying eq. (\ref{oini_er}) in the real $(p,q)$-coordinates, 
it is convenient to pass to the complex variables 
$$
u=(u_j)_{j\in\CC}\in\mC^{|\CC|}, \quad  u_j(p,q)=p_j+iq_j, \quad j\in\CC.
$$
Then the actions and angles take the form
$$
I=(I_j)_{j\in\CC},\; I_j(u)=\frac{|u_j|^2}{2}\quad\mbox{and}\quad \ph=(\ph_j)_{j\in\CC},\; \ph_j(u)=\arg u_j.
$$
Denote 
$$
V_{jk}(u):=V(q_j(u),q_k(u))=V(\Imm u_j,\Imm u_k),
$$
where $j,k\in\CC$ satisfy $|k-j|=1$. 
Then in the $u$-variables eq. (\ref{oini_er}) takes the form 
\bee
\label{oslow}
\dot u_j = \eps^{-1}iu_j + P_j(u) + \sqrt {2 \TT_j}\dot\beta_j, 
\eee
where
\footnote{Note that the function $P_j$ is real valued.}
\bee\label{oPj}
P=(P_j)_{j\in\CC},\; P_j(u):=\la\sum\limits_{k \in\CC: |j-k|=1} i\nabla_j V_{jk}(u)-\Ree u_j,
\eee
and $\nabla_j:=2\chp_{\ov u_j}$ is the gradient with respect to the Euclidean scalar product in $\mC \simeq \mR^2$. The derivative $\chp_{\ov u_j}$ is understood formally, $\chp_{\ov u_j}:=(\chp_{p_j}+i\chp_{q_j})/2$.

Let $u^\eps(\tau)$ be a unique solution of eq. (\ref{oslow}) satisfying $\DD(u^\eps(0))=\DD(u_0),$ where 
$u_0:=p_0+iq_0.$
Our first aim is to study its limiting behaviour as $\eps\ra 0$. In the main order with  respect to $\eps$ eq. (\ref{oslow})  describes a system of fast synchronously rotating uncoupled oscillators (see also $\ph$-equation from (\ref{ointroaa})). Let us pass to a rotating coordinate system where the oscillators are immovable.  
Put 
\bee\label{oavar}
a^\eps=(a^\eps_j)_{j\in\CC},\quad a^\eps_j(\tau):=e^{-i\eps^{-1}\tau}u^\eps_j(\tau),\quad j\in\CC.
\eee
Obviously, we have 
\bee
\label{ov=a}
|u^\eps_j(\tau)|\equiv|a^\eps_j(\tau)|\quad\mbox{for all $j\in\CC$}, \quad\mbox{and}\quad I(u^\eps(\tau))\equiv I(a^\eps(\tau)).
\eee
The process $a^\eps(\tau)$ satisfies 
\begin{align}
\label{oa}
\dot a^{\eps}_j &= e^{-i\eps^{-1}\tau}P_j(e^{i\eps^{-1}\tau}a^\eps)  + \sqrt{2\TT_j} e^{-i\eps^{-1}\tau}\dot\beta_j, \quad j\in\CC,\\
\nonumber
\DD(a^\eps(0)) &=\DD(u_{0}),
\end{align}
where $e^{i\eps^{-1}\tau}a^\eps$ is a vector with components $\big(e^{i\eps^{-1}\tau}a^\eps\big)_k=e^{i\eps^{-1}\tau}a_k^\eps$, $k\in\CC$.
Regarding (\ref{oa}) as an equation in the extended phase space $\mC^{|\CC|}\times\mR_+\ni(a,\tht)$, we have
\bee
\label{oaex}
\dot a^{\eps}_j = e^{-i\tht}P_j(e^{i\tht}a^\eps)  + \sqrt{2\TT_j} e^{-i\tht}\dot\beta_j, \quad \dot\tht = \eps^{-1}.
\eee 
Eq. (\ref{oaex}) is a usual fast-slow system with the unique fast variable $\tht$. Applying the method of stochastic averaging due to R.Khas'minski, we will show that its limiting (as $\eps\ra 0$) dynamics is given by the $a$-equation, averaged with respect to $\tht$, see the next section. 
The latter has the form 
\bee
\label{oeff}
\dot u_j =\RR_j(u)+ \sqrt{\TT_j}\dot{\bm\beta_j}, \quad j\in\CC,
\eee
where 
\bee\label{oRR}
\RR_j(u):=\int\limits_{0}^{2\pi}e^{-i\tht} P_j(e^{i\tht}u)\,\dbar\tht, \quad \dbar\tht:=\frac{d\tht}{2\pi},
\eee
and $\bm\beta=(\bm\beta_j)_{j\in\CC}$ is a standard complex $|\CC|$-dimensional Brownian motion, so that $\bm\beta_j=\beta_j^1+i\beta_j^2$, 
where $(\beta_j^k)_{k=1,2,\, j\in\CC}$ are standard real independent Brownian motions. Following \cite{KM}, we call (\ref{oeff}) the {\it effective equation}. 

Let us emphasize that under the limit $\eps\ra 0$ the noise "bifurcates". Indeed, the rank of the dispersion matrix of the effective equation written in the real $(p,q)$-coordinates equals to $2|\CC|$ (so the diffusion matrix is non-degenerate), while that for eq. (\ref{oa}) equals to $|\CC|$. This happens because the fast rotation decomposes the noise into two independent components, see Proposition \ref{oprop:razdvwym}.

Let us calculate the drift $\RR_j(u)$. For this purpose we will  need to define the {\it resonant averaging}.
Take a continuous function $f:\,\mC^{|\CC|}\ra \mR$ and write it in the action-angle coordinates $(I,\ph)$. Then the resonant averaging $\lan f \ran_R$ is defined as the averaging of $f$ in angles in the direction  $\bm 1=(1,1\ldots,1)$ of their fast rotation 
(see $\ph$-equation in (\ref{ointroaa})). More precisely, 
\bee\label{oresavintro}
\lan f \ran_R(I,\ph): = \int\limits_{0}^{2\pi} f(I,\ph+\tht \bm 1)\, \dbar\tht.
\eee 
In the $u$-variables it takes the form
\bee\label{oresavv}
\lan f\ran_R(u)=\int\limits_{0}^{2\pi} f(e^{i\tht}u) \, \dbar\theta. 
\eee
Further on we will often  use properties of the resonant averaging given in Appendix \nolinebreak \ref{oapp:RA}.    
Denote by $H^{res}$ the resonant averaging of the interaction potential, 
\bee\label{hress}
H^{res}:=\frac12\sum\limits_{|m-k|=1} V_{mk}^{res},\quad \mbox{where}\quad  V_{mk}^{res}:=\lan  V_{mk} \ran_R.
\eee
We will call the function $ V_{mk}^{res}$ the {\it resonant potential } and $H^{res}$~--- the {\it resonant Hamiltonian}. 
In view of (\ref{oRR}) and (\ref{oPj}), we have 
\begin{align}\label{oRRR}
\RR_j(u)&= \la\int\limits_{0}^{2\pi}e^{-i\tht} \sum\limits_{k:|k-j|=1}i\nabla_j V_{jk}(e^{i\tht}u)\,\dbar\tht 
-\int\limits_{0}^{2\pi}e^{-i\tht} \Ree(e^{i\tht}u_j)\,\dbar\tht\\ \nonumber 
 &= \la i\nabla_j\int\limits_{0}^{2\pi}\frac12\sum\limits_{|m-k|=1} V_{mk}(e^{i\tht}u)\,\dbar\tht
 -\int\limits_{0}^{2\pi}e^{-i\tht} \frac{e^{i\tht}u_j + e^{-i\tht}\ov u_j}{2} \,\dbar\tht 
 = \la i\nabla_j H^{res}(u)-\frac{u_j}{2}.
\end{align}
Then the effective equation (\ref{oeff}) takes the form 
\bee\label{oeffH}
\dot u_j=\la i\nabla_j H^{res} - \frac{u_j}{2} + \sqrt{\TT_j}\dot{\bm\beta_j},\quad j\in\CC,
\eee
and describes dynamics of a Hamiltonian system given by the resonant Hamiltonian $\la H^{res}$, where each particle is coupled with its own stochastic Langevin-type thermostat. 
Note that eq. (\ref{oeffH}), written in the real $(p,q)$-coordinates, coincides with eq. (\ref{ointroeff}) from the introduction.

\subsection{Averaging theorems}
\label{sec:avth}

It is known that eq. (\ref{oeff}) has a unique solution, this solution is defined globally (see \cite{Khb}), and that eq. (\ref{oeff}) is mixing (see \cite{Khb,Ver87,Ver97})
\footnote{In order to establish the mixing property one could also use the method explained in Appendix \ref{oapp:mixing}.}.
 In Section \ref{osec:rape2} we will prove the following theorem, which jointly with  (\ref{ov=a}) immediately implies Theorem \nolinebreak\ref{otheo:introav}. 
\btt\label{otheo:a}
On the space $C([0,T],\mC^{|\CC|})$ we have the weak convergence of measures
\bee\label{oaconv}
\DD(a^\eps(\cdot))\raw \DD(u(\cdot)) \quad \mbox{as} \quad \eps\ra 0\quad\mbox{uniformly in $\CC$},
\eee
where $u(\tau)$ is a unique solution of the effective equation (\ref{oeff}) satisfying $\DD(u(0))=\DD(u_0)$.
\ett
Recall that the uniformity in $\CC$ of the weak convergences of measures through all the text is understood in the sense of {\it finite-dimensional projections}.
 For example, for convergence (\ref{oaconv}) it means that for any bounded set $\La\subset\mZ^d$ and   any  continuous bounded functional  $f:C([0,T],\mC^{|\CC|})\mapsto \mR$, satisfying
\footnote{Recall that $\Supp f$ is defined in {\it Agreements}.6.}
$\Supp f\subseteq\La$, we have
$$
\MO f\big(a^\eps(\cdot)\big)\ra \MO f\big(u(\cdot)\big) \ass\eps\ra 0 \quad\mbox{uniformly in }\CC,  
$$
where we assume that the set $\CC$ satisfies $\CC\supseteq\La$.

Convergence (\ref{oaconv}) transfers the estimates for solutions of eq. (\ref{oini_er}) obtained in Lemma \nolinebreak\ref{olem:est} to the effective equation. 
\bpp\label{olem:estef} Let $u(\tau)$ be the unique solution of the effective equation (\ref{oeff}) satisfying $\DD(u(0))=\DD(u_0)$, and $\mu$ be its unique stationary measure. Then for all $0<\la\leq 1$, $j\in\CC$ and $\tau\geq 0$  we have
\bee\nonumber
(i)\,\MO\max\limits_{s\in[\tau,\tau+1]} e^{\al |u_j(s)|^2}\leq C,\qquad  (ii)\,\lan \mu,e^{\al |u_j|^2}\ran \leq C,
\eee  
where the constants $C$ and $\al$ are the same that in Lemma \ref{olem:est}.
\epp
{\it Proof.} Let $u^\eps(\tau)$ be the unique solution of eq. (\ref{oslow}) satisfying $\DD(u^\eps(0))=\DD(u_0).$ Writing the first estimate from Lemma \ref{olem:est} in the complex variables we get $\MO\max\limits_{s\in[\tau,\tau+1]} e^{\al |u^\eps_j(s)|^2}<C$, for all $0<\eps<\eps_0$, $0<\la\leq 1,$ $j\in\CC$ and $\tau\geq 0$.
Then (\ref{ov=a}) joined with Theorem \nolinebreak\ref{otheo:a} and the Fatou lemma implies item {\it (i)}. Item {\it (ii)} follows from item {\it (i)}, the mixing property of the effective equation and the Fatou lemma. 
\qed
\ssk

Next we investigate the limiting (as $\eps\ra 0$) behaviour of the unique stationary measure $\mu^\eps$ of eq. (\ref{oslow}). 
Let us introduce an additional assumption. Consider an unbounded set $\CC_\infty\subseteq\mZ^d$. Take the inductive limit as $\CC\nearrow\CC_\infty$ of the effective equation,  i.e. eq. (\ref{oeff}) with the set $\CC$ replaced by $\CC_\infty$. We will call it the {\it $\CC_\infty$-effective equation}. 
\begin{defi}
{\it (i)} A weak solution $u^\infty(\tau)$ of the $\CC_\infty$-effective equation is called {\it regular} if there exists $\Del>0$ such that for any $T\geq 0$ we have
$$
\sup\limits_{j\in\CC_\infty} \MO \max\limits_{0\leq\tau\leq T} e^{\Del |u^\infty_j(\tau)|^2} < \infty.
$$ 
{\it (ii)} A probability measure $\mu^\infty$ defined on the Borel $\sigma$-algebra $\BB(\mC^{|\CC_\infty|})$
\footnote{The space $\mC^{|\CC_\infty|}$ is provided with the Tikhonov topology of pointwise convergence.} 
is called {\it regular stationary measure} of the $\CC_\infty$-effective equation if the latter admits a regular weak solution $u^\infty(\tau)$ satisfying  $\DD(u^\infty(\tau))\equiv\mu^\infty.$
\end{defi}
Sometimes we will assume 

{\bf H$\CC_\infty$}. {\it For each unbounded set $\CC_\infty\subseteq\mZ^d$ the corresponding $\CC_\infty$-effective equation  admits a unique regular stationary measure.}

Passing to the limit $\CC\nearrow\CC_\infty$ and using estimates of Proposition \ref{olem:estef}, we can prove that the $\CC_\infty$-effective equation has a regular stationary measure, but we do not know if it is unique (however, we believe that it is). The following result provides a sufficient condition for fulfillment of ass. {\it H$\CC_\infty$.} 
\bpp\label{oprop:ru} Assume that $\la$ is sufficiently small and second partial derivatives of the resonant potentials $V_{kj}^{res}$ are bounded. 
\footnote{If the second partial derivatives of the function $V_{kj}^{res}$ are bounded for one pair of indices  $k,j\in\CC$, then they are bounded by the {\it same} constant for all other $k,j\in\CC$. Indeed, this is evident since 
$V^{res}_{kj}(u)=\lan V(q_k(u),q_j(u))\ran_R.$}
Then assumption H$\CC_\infty$ is satisfied.
\epp
In particular, if the second partial derivatives of the interaction potential $V$ are bounded then those of $V_{kj}^{res}$ also are. Proposition \ref{oprop:ru} is a corollary of Proposition \ref{oprop:ru'}, where we show that under conditions of Proposition \ref{oprop:ru} the $\CC_\infty$-effective equation defines a mixing Markov process, in some appropriate space.  
Its proof is based on a simple observation that for $\la$ sufficiently small the $\CC_\infty$-effective equation is contracting there. See Section \nolinebreak \ref{osec:rape4}.
\btt
\label{otheo:sm} {\it (i)} 
Let $\mu^\eps$ and $\mu$ be  the unique stationary measures of equation (\ref{oslow}) and the effective equation (\ref{oeff}) correspondingly. Then $\mu^\eps\raw \mu$ as $\eps\ra 0$. 

{\it (ii)}
If assumption {\it H$\CC_\infty$} is satisfied then the convergence above holds uniformly in $\CC$.
\ett
The proof of the theorem is postponed to Section \nolinebreak\ref{osec:rape3}. It is based on a version of Theorem \nolinebreak\ref{otheo:a} for stationary solutions, and rotation invariance of the effective equation discussed below.

In Appendix \ref{oapp:mixing} it is proven that the rate of mixing of eq. (\ref{oslow}) is independent from $\eps$. Jointly with Theorems \ref{otheo:introav} and \ref{otheo:sm}, this implies the following result.  
\btt\label{otheo:Iunif} 
(i) The convergence $\DD\big(I(u^\eps(\tau))\big)\raw \DD\big(I(u(\tau))\big)$ as $\eps\ra 0$ provided by Theorem \nolinebreak \ref{otheo:introav} holds uniformly in $\tau\geq 0$ in 
the sense that for any continuous bounded function $f:\mR^{|\CC|}\mapsto \mR$ we have
\bee\label{oun1}
\sup\limits_{\tau\geq 0} \big|\MO f\big(I^\eps(\tau)\big)-\MO f\big(I(\tau)\big)\big| \ra 0 \ass \eps\ra 0,
\eee
where $I^\eps(\tau):=I(u^\eps(\tau))$ and $I(\tau):=I(u(\tau))$.
\footnote{We do not know if the convergence holds uniformly in both $\CC$ and $\tau$.}

(ii) We have
$$
\lim\limits_{\eps\ra 0} \lim\limits_{\tau\ra\infty} \DD\big(I^\eps(\tau)\big)=\lim\limits_{\tau\ra\infty}\lim\limits_{\eps\ra 0} \DD\big(I^\eps(\tau)\big)=
\lim\limits_{\substack{\eps\ra 0 \\ \tau\ra\infty}} \DD\big(I^\eps(\tau)\big)= \Pi_{I*} \mu,
$$
where $\mu$ is the unique stationary measure of the effective equation (\ref{oeff}), and $\Pi_I$ denotes the projection to the space of actions.
\ett

 {\it Proof.}
{\it (i)} 
Eq. (\ref{oslow}) and (\ref{oeff}) are mixing and the rate of mixing for eq. (\ref{oslow}) is uniform in $\eps< \eps_1$ (see Theorem \ref{otheo:mixing}). Then  for any $\del>0$ there exists $s>0$ such that for any $\eps$ satisfying $\eps<\eps_1$ 
 \bee\label{oun2}
\sup\limits_{\tau\geq s}\big|\MO f\big(I^\eps(\tau)\big)-\big\lan \Pi_{I*}\mu^\eps, f\big\ran\big|\leq \del/3,\quad
\sup\limits_{\tau\geq s}\big|\MO f\big(I(\tau)\big)-\big\lan \Pi_{I*}\mu, f\big\ran\big|\leq \del/3,
 \eee   
where $\mu^\eps$ is the unique stationary measure of eq. (\ref{oslow}).
In view of Theorem \ref{otheo:sm}.{\it i}, there exists $\eps_2$ such that for any $\eps< \eps_2$ we have
 \bee\label{oun3}
 \big|\big\lan \Pi_{I*}\mu^\eps, f \big\ran-\big\lan \Pi_{I*}\mu, f\big\ran\big|\leq \del/3.
 \eee 
 Inequalities (\ref{oun2}) and (\ref{oun3}) imply that for $\eps<\eps_1\wedge\eps_2$ we have
 \bee\label{oun4}
\sup\limits_{\tau\geq s} \big|\MO f\big(I^\eps(\tau)\big)-\MO f\big(I(\tau)\big)\big|\leq\del. 
 \eee
Applying Theorem \ref{otheo:introav} and using that the weak convergence of measures is equivalent to convergence in the dual-Lischitz norm (see \cite{Dud}, Theorem 11.3.3), it is not difficult to show that  
$
 \sup\limits_{0\leq\tau\leq s} \big|\MO f\big(I^\eps(\tau)\big)-\MO f\big(I(\tau)\big)\big|\ra 0  
$
as $\eps\ra 0$. Then
there exists $\eps_3$, such that for any $\eps<\eps_3$ we have
 $
 \sup\limits_{0\leq\tau\leq s} \big|\MO f\big(I^\eps(\tau)\big)-\MO f\big(I(\tau)\big)\big|\leq\del. 
 $
Jointly with (\ref{oun4}) this implies that for $\eps<\eps_1\wedge\eps_2\wedge\eps_3$ we have 
$\sup\limits_{\tau\geq 0} \big|\MO f\big(I^\eps(\tau)\big)-\MO f\big(I(\tau)\big)\big|\leq\del.$
 
{\it (ii)} The first two convergences follow from the mixing properties of eq. (\ref{oslow}) and (\ref{oeff}), and Theorems \ref{otheo:introav},\ref{otheo:sm}.{\it i}.
The last convergence follows from item {\it (i)} joined with the mixing property of eq. (\ref{oeff}).
 \qed

\subsection{The limiting behaviour of the local energy} 

Recall that the local energy $\EE^{\nu}_j(p,q)$ is defined in (\ref{oH_j}). We will consider it as a function of the complex variables, and abusing notations we will write  $\EE^{\nu}_j(u)$.
Define the vector of local energy 
$$\EE^{\nu}(u):=(\EE^{\nu}_j(u))_{j\in\CC}.$$
Let $u^\eps(\tau)$ be a solution of eq. (\ref{oslow}), either satisfying $\DD(u^\eps(0))=\DD(u_0)$ or a stationary one. 
Since $\EE^{\nu}=I+O(\nu)$, the limiting as $\eps\ra 0$ behaviour of the vector of local energy $\EE^{\nu}(u^\eps)$ coincides with that of the vector of actions $I(u^\eps)$, while the latter is governed by the effective equation, in the sense of Theorems \ref{otheo:introav},\ref{otheo:sm} and \ref{otheo:Iunif}. More precisely, we prove

\bpp\label{olem:introen} 
Let  $u^\eps(\tau)$ be a solution of eq. (\ref{oslow}) satisfying $\DD(u(0))=\DD(u_0)$, and $\wid u^\eps(\tau)$ be its stationary solution. Let $u(\tau)$ be a solution of the effective equation (\ref{oeff}) satisfying $\DD(u(0))=\DD(u_0)$, and $\mu$ be its unique stationary measure. 
Then, denoting 
$\EE^{\nu}(\tau):=\EE^{\nu}\big(u^\eps(\tau)\big),$ $\wid\EE^{\nu}(\tau):=\EE^{\nu}\big(\wid u^\eps(\tau)\big)$ and 
$I(\tau):=I(u(\tau))$, we have \\
$(i)\;\DD\big(\EE^{\nu}(\cdot)\big) \raw \DD\big(I(\cdot)\big)$ as  $\eps\ra 0$ on $C([0,T],\mR^{|\CC|})$ uniformly in $\CC$. \\
$(ii)\;\DD\big(\EE^{\nu}(\tau)\big) \raw \DD\big(I(\tau)\big)$ as  $\eps\ra 0$ uniformly in $\tau\geq 0$ in the sense explained in Theorem \nolinebreak\ref{otheo:Iunif}. \\
$(iii)\; \DD\big(\wid\EE^{\nu}(\tau)\big) \raw \Pi_{I*} \mu$ as $\eps\ra 0$ for any $\tau\geq 0$. If ass. {\it H$\CC_\infty$} is satisfied, this convergence holds uniformly in $\CC$. \\
$(iv)\;  \lim\limits_{\eps\ra 0} \lim\limits_{\tau\ra\infty} \DD\big(\EE^{\nu}(\tau)\big)=\lim\limits_{\tau\ra\infty} \lim\limits_{\eps\ra 0} \DD\big(\EE^{\nu}(\tau)\big)= 
\lim\limits_{\substack{\eps\ra 0 \\ \tau\ra\infty}} \DD\big(\EE^{\nu}(\tau)\big)=\Pi_{I*} \mu.$
\epp
{\it Proof.}
Since the interaction potential $V$ has at most a polynomial growth, the estimate {\it (i)} of Lemma \ref{olem:est} implies that the limit as $\eps\ra 0$ of $\DD\big(\EE^{\nu}(\cdot)\big)$ coincides  with that of $\DD\big(I(u^\eps(\cdot))\big)$ in the sense that if one limit exists then another one exists as well and the two are equal. Moreover, if one convergence is uniform in $\CC$ then another one also is. 
That is why item {\it (i)} follows from Theorem \nolinebreak\ref{otheo:introav}.
Items {\it (ii)-(iv)} can be obtained similarly, using Theorems \nolinebreak \ref{otheo:Iunif}.{\it i}, \ref{otheo:sm}  and \ref{otheo:Iunif}.{\it ii} correspondingly. 
 For the first limit from item {\it (iv)} one should additionally use that eq. (\ref{oslow}) is mixing.  
\qed 

\subsection{Rotation invariance of the effective equation}
\label{osec:preff}

Further on we will need the following results. Since the effective equation (\ref{oeff}) is obtained by the resonant averaging procedure, it is rotation invariant in the direction \nolinebreak $\bm 1$ of the fast rotation of angles:
\begin{prop}
\label{olem:vrawinvar}
If $u(\tau)$ is a weak solution of the effective equation (\ref{oeff}) then for any $\xi\in [0, 2\pi)$ the process $e^{i\xi} u(\tau)$ is also its weak solution. 
\end{prop}
{\it Proof.} Process $w(\tau)=e^{i\xi}u(\tau)$ is a weak solution of equation
$$
\dot w_j = \int\limits_{0}^{2\pi} e^{-i(\theta-\xi)}P_j(e^{i\theta} u)\, \dbar\theta + \sqrt{\TT_j}e^{i\xi}\dot{\bm\beta}_j, \quad j\in\CC.
$$
Changing the variable of integration $\wid\theta:=\theta-\xi$ and noting that $\{e^{i\xi}\bm\beta_j,\,j\in\CC\}$ is another set of standard complex independent Brownian motions, we obtain that the process $w(\tau)$ is a weak solution of the effective equation (\ref{oeff}).
\qed 

\bpp \label{olem:rotinv} 
Let $\mu$ be the unique stationary measure of the effective equation (\ref{oeff}). 
Then for any continuous function $f:\mC^{|\CC|}\mapsto \mR$ with at most a polynomial growth at infinity we have
$$
\lan \mu,f\ran = \big\lan\mu,\lan f\ran_R \big\ran.
$$
\epp
{\it Proof.}
Since the stationary measure $\mu$ of the effective equation is unique, Proposition \nolinebreak \ref{olem:vrawinvar} implies that it is rotation invariant in the direction \nolinebreak $\bm 1$: for any $ \xi\in[0,2\pi)$ we have
$
(e^{i\xi})_*\mu=\mu.
$ 
Then 
$$
\lan \mu,f\ran = \int\limits_0^{2\pi} \big\lan (e^{i\xi})_* \mu,f\big\ran \, \dbar \xi= \big\lan  \mu, \int\limits_0^{2\pi} f(e^{i\xi} \cdot) \, \dbar \xi\big\ran  =  \big\lan\mu,\lan f\ran_R \big\ran.
$$
\qed

\section{The energy transport}
\label{osec:app}

In this section we assume that $\la\ll 1$ and investigate the limiting (as $\eps\ra 0$) behaviour of the stationary Hamiltonian  energy flow for eq. (\ref{oslow}). We prove Theorems \ref{otheo:F} and \ref{otheo:GK} which are full versions of Theorems \ref{otheo:introF} and \ref{otheo:GKintro} given in the introduction. 

\subsection{The limiting behaviour of the energy flow}

Consider the Ornstein-Uhlenbeck process (\ref{oeffH})$|_{\la=0}$
\bee\label{oOU}
\dot u_m =- \frac{u_m}{2} + \sqrt{\TT_m}\dot{\bm\beta}_m, \quad m\in\CC.
\eee
It is well known that the unique stationary measure $\mu^0$ of eq. (\ref{oOU}) has the form 
\bee\label{omu0}
\mu^0=e^{-|u|_\TT^2}\,dm,
\eee
where $ |u|_\TT^2:=\sum\limits_{m\in\CC}(2\TT_m)^{-1}|u_m|^2 $ and $dm$ is a normalized  Lebesgue measure on $\mC^{|\CC|}\simeq\mR^{2|\CC|}$, $dm=d\mbox{Leb}/\big((2\pi)^{|\CC|}\Pi_{m\in\CC}\TT_m\big)$. 
The Ornstein-Uhlenbeck process (\ref{oOU}) is important for our study because of the following
\bpp\label{olem:lato0}
(i) Let $u^\la(\tau)$ and $u(\tau)$ be solutions of the effective equation (\ref{oeffH}) and eq. (\ref{oOU}) correspondingly, satisfying $\DD(u^\la(0))=\DD(u(0))=\DD(u_0)$. Then
$$
\DD(u^\la(\cdot))\raw \DD(u(\cdot)) \ass \la\ra 0 \quad\mbox{on}\quad C([0,T],\mC^{|\CC|})\quad\mbox{uniformly in $\CC$}.
$$
(ii) Let $\mu^\la$ be the unique stationary measure of eq. (\ref{oeffH}). Then $\mu^\la\raw\mu^0$ as $\la\ra 0$ uniformly in $\CC.$ 
\epp
{\it Proof.}  The required convergences can be established by arguments similar to those used in the proofs of Theorems \nolinebreak \ref{otheo:a} and \ref{otheo:sm}, but significantly simplified. The uniformity of convergence from item {\it (ii)}, similarly to that of Theorem  \ref{otheo:sm}.{\it ii}, follows from fulfillment of assumption analogous to {\it H$\CC_\infty$}. Namely, since equations (\ref{oOU}) are diagonal, for any unbounded set $\CC_\infty\subseteq\mZ^d$ a stationary measure of their inductive limit as $\CC\nearrow\CC_\infty$ (given by (\ref{oOU}) with $\CC=\CC_\infty$) is unique. 
\qed

For $k,j\in\mZ^d$ denote by $A^{kj}$ the generator of the process (\ref{oOU}) with $\CC$ replaced by the set $\{k,j\}$. Considering $A^{kj}$ as an operator on real valued functions, we have
\bee\label{oAkj}
A^{kj}=\sum\limits_{l=j,k} (2\TT_l\chp^2_{u_l\ov u_l} - \frac{u_l}{2}\cdot \nabla_l). 
\eee
We will need the following well known property of the Ornstein-Uhlenbeck process.
\bl\label{olem:PaVe}
{\it (i)} Fix $k,j\in\CC$. Let a measurable function $\xi:\mC^{|\CC|}\mapsto\mR$ has at most a polynomial growth at infinity, satisfies $\Supp\xi\subseteq \{k,j\}$ and
$\lan \mu^0, \xi \ran=0$. Then equation 
\bee\label{oA}
A^{kj}\eta=\xi
\eee
has a unique solution $\eta:\CCC\mapsto\mR$  in a class of functions 
satisfying the following properties $(a),(b),(c)$:  
$(a)\;\eta$ belongs to each Sobolev class $W^2_{p,loc}$, $p>1$, $(b)\; \eta$ has at most a polynomial growth at infinity, and $(c)\; \lan \mu^0, \eta \ran=0$.
The solution $\eta$ satisfies $\Supp\eta\subseteq \{j,k\}$ and has the form 
\bee\label{oeta}
\eta(u)=-\int\limits_0^\infty \fB_{\tau}^0\xi(u)\, d\tau,
\eee  
where $\fB_{\tau}^0$ denotes a Markov semigroup associated with the Ornstein-Uhlenbeck process (\ref{oOU}). 
Moreover, the gradient 
$\nabla\eta(u)$  also has at most a polynomial growth at infinity. 
 
{\it (ii)} If the function $\xi$ is $C^n$-smooth for some $n\in\mN$, then the solution $\eta$ is also $C^{n}$-smooth.
\el
{\it Proof.} {\it (i)} 
This is a particular case of Theorem \nolinebreak 1 from \cite{PaVe}. 

{\it (ii)} 
It follows in a standard way from smoothness of coefficients of the generator $A^{kj}$, $C^n$-smoothness of the  function $\xi$ and the Sobolev embedding theorem (see \cite{Tay}, Theorem \nolinebreak 5.11.1).
\qed
\ssk 

Recall that the energy flow $\JJ_{kj}(p,q)$ is defined in (\ref{oef}). Abusing notations, we put $\JJ_{kj}(u):=\JJ_{kj}(p(u),q(u))$. 
Introduce the {\it resonant energy flow} as
\bee\nonumber
\JJ_{kj}^{res}:=\lan \JJ_{kj}\ran_R.
\eee
Proposition \ref{oprop:RA}.{\it ii} joined with ass. {\it HV} implies that the function $\JJ_{kj}^{res}$ is $C^{r-1}$-smooth, and all its partial derivatives have at most a polynomial growth at infinity. 
\bpp\label{oprop:REF} We have
$$(i)\;  \JJ_{kj}^{res}=2\chp_{\ph_k} V^{res}_{kj}=-2\chp_{\ph_j} V^{res}_{kj};
\qquad
(ii) \; \lan\mu^0 , \JJ_{kj}^{res} \ran = 0.$$
\epp 
{\it Proof. (i)} 
The Poisson bracket written in the action angle coordinates  $(I,\ph)$ takes the form 
$\{h_1,h_2\}=\sum\limits_{k\in\CC}(\chp_{\ph_k}h_1\chp_{I_k}h_2  - \chp_{I_k}h_1\chp_{\ph_k}h_2)$. Then
 (\ref{oef}) implies
\bee\label{oEFla}
\JJ_{kj}=\{I_j-I_k, V_{kj}\}=(\chp_{\ph_k}V_{kj}-\chp_{\ph_j}V_{kj}).
\eee
Proposition \ref{oprop:RA}.{\it iii} provides that for $l\in\{j,k\}$ we have 
$\lan \chp_{\ph_l} V_{kj}\ran_R=\chp_{\ph_l} \lan V_{kj}\ran_R=\chp_{\ph_l} V_{kj}^{res}.$
It remains to take the resonant averaging of (\ref{oEFla}) and note that Proposition \ref{oprop:RA}.{\it iv} implies  
\bee\label{osymmetry}
\chp_{\ph_k} V^{res}_{kj}=-\chp_{\ph_j} V^{res}_{kj}.
\eee
{\it (ii)} It follows from item {\it (i)} and rotation invariance of the Gaussian measure $\mu^0$.  
\qed
\ssk

In view of item {\it (ii)} of Proposition \ref{oprop:REF}, the resonant energy flow $\JJ_{kj}^{res}$ satisfies conditions of Lemma \nolinebreak\ref{olem:PaVe}. Denote by $\eta_{kj}(u)$ obtained there solution of eq.
\bee\label{oeeeta}
A^{kj}\eta_{kj} = \JJ_{kj}^{res}.
\eee
\btt \label{otheo:F}
Let $\mu^{\eps,\la}$ be the unique stationary measure of eq. (\ref{oslow}).
Then for every $j,k\in\CC$, $|j-k|=1$, we have
\bee
\label{oF} 
\lan\mu^{\eps,\la},\JJ_{kj}\ran \ra \la\kappa(\TT_k,\TT_j)(\TT_{k}-\TT_{j}) + o(\la)\quad \mbox{as}\quad \eps\ra 0,
\eee
where $o(\la)/\la\ra 0$ as $\la\ra 0$ uniformly in $\CC$, and 
\bee\label{okappatheo}
\kappa(\TT_k,\TT_j)= 
-\frac{ \lan \mu^0,  \JJ_{kj}^{res}\eta_{kj}\ran}{\TT_k\TT_j}
=\frac{\lan\mu^0, \TT_k|\nabla_k\eta_{kj}|^2+\TT_j|\nabla_j\eta_{kj}|^2\ran}{2\TT_k\TT_j}.
\eee
If assumption {\it H$\CC_\infty$} is satisfied then convergence (\ref{oF}) holds uniformly in \nolinebreak$\CC$.

The function $\ka:\mR^2_+\mapsto\mR$ satisfies $\kappa(x,y)\equiv\kappa(y,x)$, is  $C^{r-1}$-smooth and nonnegative. If the resonant potential $V^{res}_{kj}$ is not independent from the angles $\ph$, then
$\kappa$ is strictly positive.   
\ett 
Emphasize that the function $\ka(\TT_k,\TT_j)$, which we call the {\it conductivity}, implicitly depends on the temperatures  through the measure $\mu^0$ and the function $\eta_{kj}$ (the latter depends on them since the operator $A^{kj}$ does).

Since $\eta_{kj}$ is the solution of eq. (\ref{oeeeta}) obtained in Lemma \ref{olem:PaVe}, formula (\ref{oeta}) joined with (\ref{okappatheo}) implies that the conductivity $\ka$  can  be also represented in the form 
$$
\ka(\TT_k,\TT_j)=\frac1{\TT_k\TT_j}\big\lan\mu^0,\JJ_{kj}^{res} \int\limits_0^\infty \fB_{\tau}^0\JJ^{res}_{kj}\,d\tau\big\ran.
$$
Note that the sufficient condition for fulfillment of ass. {\it H$\CC_\infty$} provided by Proposition \nolinebreak \ref{oprop:ru} is relevant for Theorem \ref{otheo:F} since convergence (\ref{oF}) gives some information only if $\la$ is small.

In order to prove Theorem \ref{otheo:F} we will need the following proposition, which can be obtained by simple integration by parts.
\bpp \label{olem:A*}
Let functions $\eta^1,\eta^2\in W^2_{2,loc}(\mC^{|\CC|}, \mR)$ with their gradients $\nabla \eta^1,\nabla \eta^2$ have at most a polynomial growth at infinity.  Then 
\bee\nonumber
\lan \mu^0,  \eta^2 A^{kj} \eta^1\ran =-\frac12 \lan \mu^0, \sum\limits_{l=k,j}\TT_l \nabla_l \eta^1 \cdot \nabla_l \eta^2 \ran= \lan \mu^0, \eta^1 A^{kj} \eta^2 \ran.
\eee
\epp
{\it Proof of  Theorem \ref{otheo:F}.}
Denote by $\mu^\la$ the unique stationary measure of the effective equation (\ref{oeffH}). Theorem \nolinebreak\ref{otheo:sm}.{\it i} implies
\bee\label{oconv}
\lan \mu^{\eps,\la}, \JJ_{kj} \ran \ra \lan \mu^\la, \JJ_{kj} \ran=\lan \mu^\la, \JJ^{res}_{kj} \ran  \quad\mbox{as}\quad \eps\ra 0,
\eee    
where the last equality follows from Proposition \nolinebreak\ref{olem:rotinv}.
If ass. {\it H$\CC_\infty$} is satisfied then Theorem \ref{otheo:sm}.{\it ii} implies that convergence (\ref{oconv}) holds uniformly in $\CC$.
Let us show that
\bee\label{oforma}
 \lan \mu^\la, \JJ^{res}_{kj} \ran=  \la\ka(\TT_k,\TT_j)(\TT_k-\TT_j)+o(\la),
\eee
where $o(\la)/\la\ra 0$ as $\la\ra 0$ uniformly in $\CC$. 
The key role will play the following
\bl\label{olem:razl}
 Let the function $\xi\in C^1(\mC^{|\CC|},\mR)$ satisfies $\Supp\xi\subseteq\{k,j\}$, has at most a polynomial growth at infinity and  $\lan \mu^0, \xi \ran=0$. Then
\bee\label{orazl}
\lan\mu^\la,\xi\ran = \la\lan \mu^0, f \xi \ran + o(\la),
\eee
where $o(\la)/\la\ra 0$ as $\la\ra 0$ uniformly in $\CC$, $f=\frac12\sum\limits_{|m-n|=1}f_{mn},$ where $f_{mn}=f_{nm}$, 
and $f_{mn}$ is the solution of eq. 
\bee\label{oAf}
A^{mn}f_{mn}=\frac{\TT_n-\TT_m}{\TT_m\TT_n} \JJ^{res}_{mn},
\eee
obtained in Lemma \ref{olem:PaVe}.
\footnote{Fix a  bounded set $\La\subset\mZ^d$. Almost without changes in the proof, a similar result can be established for a function $\xi$ satisfying $\Supp\xi\subseteq \La$. We restrict ourselves to the case $\La=\{k,j\}$ since we will apply the lemma only for a two-dimensional situation.} 
\el
Before proving the lemma we will finish the proof of the theorem. Applying Lemma \nolinebreak\ref{olem:razl} to the function $\xi=\JJ^{res}_{kj}$ we get 
\bee\label{oconv1}
\lan \mu^\la,\JJ_{kj}^{res}\ran =  \frac{\la}{2}\sum\limits_{|m-n|=1}\lan\mu^0,f_{mn} \JJ^{res}_{kj}\ran + o(\la).
\eee
\bpp\label{oprop:569} If the sets $\{m,n\}$ and $\{k,j\}$ do not coincide then 
\bee\label{o569}
\lan\mu^0,f_{mn} \JJ^{res}_{kj}\ran = 0.
\eee 
\epp
{\it Proof.}
Since $\Supp f_{mn}\subseteq \{m,n\}$,  if $m,n\neq k$ or $m,n\neq j$ then the function $f_{mn}$ does not depend on the angle $\ph_k$ or $\ph_j$ correspondingly. Then (\ref{o569}) follows from Proposition \nolinebreak\ref{oprop:REF}.{\it i} and rotation invariance of the Gaussian measure $\mu^0$. 
\qed

Since $f_{jk}=f_{kj}$, (\ref{oconv1}) joined with Proposition \ref{oprop:569} implies
\bee\label{3243861}
\lan \mu^\la,\JJ_{kj}^{res}\ran=\la \lan\mu^0,f_{kj} \JJ^{res}_{kj}\ran + o(\la).
\eee
Let $\eta_{kj}$ be the solution of equation (\ref{oeeeta}),
obtained in Lemma \ref{olem:PaVe}. Eq. (\ref{oAf}) implies  
\bee\label{oetaaa}
f_{kj}=\frac{\TT_j-\TT_k}{\TT_k\TT_j} \eta_{kj}, 
\quad\mbox{so that}\quad 
\lan\mu^0,f_{kj}\JJ^{res}_{kj}\ran = \hat\kappa_{kj}(\TT)(\TT_k-\TT_j),
\eee
where by $\TT$ we have denoted the temperature profile $(\TT_j)_{j\in\CC}$, and
$\ds{
\hat\kappa_{kj}(\TT):=-\frac{\lan\mu^0,\eta_{kj}\JJ^{res}_{kj}\ran}{\TT_k\TT_j}.
}$
Using that the function $\JJ_{kj}^{res}$ does not depend on the temperatures $\TT$, $\eta_{kj}$ depends on them only through the components $\TT_k,\TT_j$, that $\Supp \JJ_{kj}^{res}$, $\Supp \eta_{kj}\subseteq \{j,k\}$ and $\mu^0$ is a product measure, we see that the function $\hat\kappa_{kj}$ depends on $\TT$ only through    $\TT_k,\TT_j$, so that $\hat\kappa_{kj}(\TT)= \kappa_{kj}(\TT_k,\TT_j)$, for some function $\ka_{kj}$. Obviously, the latter  does not depend on the choice of the indices $k,j$, so we skip them and get
\bee \label{okkkapa}  
\ka(\TT_k,\TT_j)=-\frac{\lan\mu^0,\eta_{kj}\JJ^{res}_{kj}\ran}{\TT_k\TT_j}.
\eee
Now (\ref{oetaaa})  joined with  (\ref{3243861}) and (\ref{oconv}) implies the desired convergence (\ref{oF}). 
To obtain the second representation from (\ref{okappatheo}) for the conductivity $\kappa$,  
first we use eq. (\ref{oeeeta}) and then  Proposition \nolinebreak\ref{olem:A*}:
$$
\lan\mu^0,\eta_{kj} \JJ^{res}_{kj}\ran=\lan\mu^0, \eta_{kj} A^{kj}\eta_{kj}\ran= -\frac12\lan\mu^0, \sum\limits_{l=k,j}\TT_l|\nabla_l \eta_{kj}|^2\ran,
$$ 
so that we get 
$
\ka(\TT_k,\TT_j)={\ds\frac{\big\lan\mu_0, \sum\limits_{l=k,j}\TT_l|\nabla_l \eta_{kj}|^2\big\ran}{2\TT_k\TT_j}}.
$
In particular, this implies that the conductivity $\kappa$ is nonnegative.
Moreover, if the resonant potential $V^{res}_{kj}$ is not independent from the angles $\ph$, then in view of Proposition \ref{oprop:REF}.{\it i} the function $\JJ^{res}_{kj}$ does not vanish. Consequently, $\eta_{kj}$ is not a constant, so  $\ka$ is strictly positive for any $\TT_k,\TT_j>0$. 

Let us show that $\ka(\TT_k,\TT_j)=\ka(\TT_j,\TT_k)$. Since $\JJ^{res}_{kj}=-\JJ^{res}_{jk}$ and $A^{kj}=A^{jk}$, we have $\eta_{kj}=-\eta_{jk}$. Then, due to (\ref{okkkapa}), 
$\ka(\TT_k,\TT_j)=-(\TT_k\TT_j)^{-1}\lan\mu^0,\eta_{jk}\JJ^{res}_{jk}\ran=\ka(\TT_j,\TT_k)$.
 
It remains to prove that the function $\ka$ is $C^{r-1}$-smooth.  
It is convenient to change the variables: let $v_j=\TT_j^{-1/2}u_j,\,j\in\CC$. Then, due to (\ref{okkkapa}), we have 
$$
 \ka(\TT_k,\TT_j) =  -(\TT_k \TT_j)^{-1}\lan \wid \mu^0, \wid \JJ^{res}_{kj}\wid \eta_{kj}\ran,
$$
where
$
\wid\mu^0:=e^{-|v|^2/{2^{|\CC|}}}\,d\wid m,\; d\wid m:=d\mbox{Leb}/(2\pi)^{|\CC|}, \; \wid \JJ^{res}_{kj}(v):=\JJ^{res}_{kj}(u(v))$ and $\wid \eta_{kj}(v):=\eta_{kj}(u(v))$. 
\bpp\label{oprop:dteta}
For every $v\in\mC^{|\CC|}$ the functions $\wid\JJ_{kj}^{res}(v)=\wid\JJ_{kj}^{res}(v,\TT_k,\TT_j)$ and $\wid\eta_{kj}(v)=\wid\eta_{kj}(v,\TT_k,\TT_j)$ are $C^{r-1}$-smooth with respect to the temperatures $(\TT_k,\TT_j)\in\mR^2_+$. The corresponding partial derivatives till the  order  $r-1$ inclusive are continuous in $(v,\TT_k,\TT_j)\in\mC^{|\CC|}\times\mR^2_+$ and have at most a polynomial growth at infinity in $v$, which is uniform with respect to the temperatures $(\TT_k,\TT_j)$ if they belong to a compact set in $\mR^{2}_+$.
\epp
Since the function $\JJ^{res}_{kj}$ is independent from the temperatures $\TT_k,\TT_j$, assertion of Proposition \ref{oprop:dteta} concerning the function $\wid\JJ^{res}_{kj}$ is obvious. That, concerning $\wid\eta_{kj}$  is not so clear. It is not implied by standard arguments of elliptic theory because the generator 
$A^{kj}$ has growing coefficients. Its proof is based on simple analysis of formula (\ref{oeta}) and is postponed to the end of the section.

Since the measure $\wid\mu^0$  has an exponentially decaying density, Proposition \ref{oprop:dteta} implies that the function $\lan \wid \mu^0, \wid \JJ^{res}_{kj}\wid \eta_{kj}\ran$ is $C^{r-1}$-smooth, so that the conductivity $\ka$ also is.
\qed
\ssk

{\it Proof of Lemma \ref{olem:razl}.} 
It is known (see \cite{BKR}) that the unique stationary measure  $\mu^\la$ of the effective  equation (\ref{oeffH}) has a continuous density $\rho^\la$ with respect to the normalized Lebesgue measure $dm$, so that $d\mu^\la=\rho^\la\, dm$. 
Then, in view of (\ref{omu0}), we have
\bee\label{omumu}
d\mu^\la=h^\la\,d\mu^0,
\eee 
where the density $h^\la$ satisfies
\bee\label{orhop}
\rho^\la=e^{-|u|^2_\TT}h^\la.
\eee
The proof of the lemma is based on a decomposition of the density $h^\la$ in the small parameter $\la$. Let us start with writing down an equation for $h^\la$.
Denote by $A$ the generator of the Ornstein-Uhlenbeck process (\ref{oOU}),
\bee\label{oAAAAAAA}
A=\Del^\TT - \frac{u}{2}\cdot\nabla, \quad\mbox{where}\quad \Del^\TT:=2\sum\limits_{m\in\CC} \TT_m\chp^2_{u_m\ov u_m}.
\footnote{In the real $(p,q)$-coordinates the operator $\Del^\TT$ has the form of the weighted laplacian $\Del^\TT=\frac12\sum\limits_{m\in\CC}\TT_m(\chp^2_{p^2_m}+\chp^2_{q^2_m})$.}
\eee
Then the generator of the effective equation (\ref{oeffH}) takes the form $A+\la B$, where 
$$B:=g\cdot\nabla 
\quad \mbox{and}\quad 
g=(g_m)_{m\in\CC}, 
\quad g_m:=i\nabla_m H^{res}=i\sum\limits_{n:|n-m|=1}\nabla_m V^{res}_{mn}.$$ 
The density $\rho^\la$ satisfies in the weak sense the inverse Kolmogorov equation
\bee\label{oKol}
(A^*+ \la B^*)\rho^\la =0, 
\eee
where 
$$
A^*:= \Del^\TT +\frac{u}{2}\cdot\nabla + |\CC| \quad \mbox{and}\quad B^*= - g\cdot\nabla.
$$
In order to calculate the operator $B^*$, we have used the equality $\nabla\cdot g=0$, following from the identity $\nabla_m\cdot(i\nabla_m\psi)\equiv 0$, which holds for any function $\psi\in C^2(\CCC,\mR)$.
Applying the operator $A^*$ to the both sides of (\ref{orhop}), by the direct computation we find
 $A^*\rho^\la =e^{-|u|^2_\TT} Ah^\la$. Then, substituting (\ref{orhop}) in (\ref{oKol}) and multiplying the resulting equation by 
$e^{|u|_\TT^2}$ we get
\bee\label{oAB}
(A+\la \wid B^*) h^\la =0 , \quad\mbox{where}\quad \wid B^*h^\la:= e^{|u|^2_\TT} B^*(e^{-|u|^2_\TT}h^\la).
\eee
Now we will decompose equation (\ref{oAB}) in the small parameter $\la$, then find a solution of the obtained system till order $\la$ and show that the rest is small.  Let us start with a formal computation. 
Put
\bee\label{of=}
h^\la=1+\la f + \la \wid f^\la, 
\eee 
where
$f$ does not depend on $\la$ and $\wid f^\la=o(1)$ as $\la\ra 0$.
Substituting (\ref{of=}) in (\ref{oAB}), in order $\la^0$ we have the identity $A1=0$, while in the order $\la$ we obtain
\bee\label{oAf1}
Af=-\wid B^*1.
\eee 
To find the function $\wid f_\la$, we solve eq. (\ref{oAf1}) and put 
\bee\label{owidf}
\wid f^\la:= \la^{-1}(h^\la-1-\la f),
\eee
so that (\ref{of=}) becomes an identity.
\bpp\label{oprop:just}
(i) Eq. (\ref{oAf1}) has a solution $f$ of the form claimed in the formulation of the lemma. 

(ii) The function $\wid f^\la$, defined in (\ref{owidf}) with $f$ from item {\it (i)}, satisfies
\bee\label{of2}
 \lan \mu^0, \wid f^\la \xi\ran \ra 0 \quad \mbox{as}\quad \la\ra 0\quad \mbox{uniformly in } \CC,
\eee
where $\xi$ is a function satisfying the conditions of the lemma. 
\epp
Take $f$ and $\wid f^\la$ as in Proposition \ref{oprop:just}. Then, using that $\lan \mu^0,\xi\ran=0$, (\ref{omumu}) and (\ref{of=}), we obtain
\bee\label{omuh}
\lan\mu^\la,\xi\ran= \lan \mu^0, h^\la\xi\ran= \la\lan \mu^0, f\xi\ran + o(\la), 
\eee 
where $o(\la)=\la \lan \mu^0, \wid f^\la \xi\ran,$ so that we get the desired decomposition (\ref{orazl}).
\qed
\ssk

{\it Proof of Proposition \ref{oprop:just}.}
{\it Item (i).} 
Using that for any function $\psi(u)\in C^1(\CCC,\mR)$ we have $(i\nabla_m\psi)\cdot u_m=-\chp_{\ph_m}\psi$, and the symmetry (\ref{osymmetry}),  we calculate
\begin{align}\nonumber
\wid B^* 1&=2\sum_{m\in\CC} \TT_m^{-1}g_m\cdot u_m = 2\sum_{m\in\CC} \TT_m^{-1}\sum\limits_{n:|n-m|=1}i\nabla_m V_{mn}^{res}\cdot u_m  
= -2\sum\limits_{|n-m|=1}\TT_m^{-1}\chp_{\ph_m}V^{res}_{mn} \\ \label{oB1}
&= \sum\limits_{|n-m|=1}(\TT_n^{-1}-\TT_m^{-1})\chp_{\ph_m}V^{res}_{mn}=\frac12 \sum\limits_{|n-m|=1} (\TT_n^{-1}-\TT_m^{-1})\JJ^{res}_{mn},
\end{align}  
where in the last equality we have employed Proposition \ref{oprop:REF}.{\it i}.
Take the solution $f_{mn}$ of 
\bee\label{offkj}
 A^{mn}f_{mn}=(\TT^{-1}_m-\TT^{-1}_n)\JJ^{res}_{mn}={\frac{\TT_n-\TT_m}{\TT_n\TT_m}}\JJ^{res}_{mn},
\eee
obtained in Lemma \ref{olem:PaVe}. 
Put 
$$f:=\frac12\sum\limits_{|m-n|=1} f_{mn}.$$
Since $\Supp f_{mn}\subseteq\{m,n\}$, we obviously have $Af_{mn}=A^{mn}f_{mn}$. Then (\ref{offkj}) joined with (\ref{oB1}) implies (\ref{oAf1}).
Since $\JJ^{res}_{mn}=-\JJ^{res}_{nm}$ and
 $A^{mn}=A^{nm}$, equations for $f_{mn}$ and $f_{nm}$ coincide, so that we get $f_{mn}=f_{nm}$. 

{\it Item (ii).} We will need the following two identities. Take any function $\eta\in W^2_{2,loc}(\mC^{|\CC|},\mR)$ such that $\eta$ and $\nabla \eta$ have at most a polynomial growth. 
 
1) Since $\mu^\la$ is a stationary measure of the effective equation (\ref{oeffH}) and $A+\la B$ is the generator of the latter, we have $(A+\la B)^*\mu^\la =0$. So that, $\lan\mu^\la,A\eta\ran=-\la\lan\mu^\la,B\eta\ran.$ 

2) The integration by parts implies $\lan\mu^0,\wid B^*1\eta\ran=\lan\mu^0, B\eta \ran$.

Also let us note that Proposition \ref{olem:A*} still holds if replace the operator $A^{kj}$ by $A$ and the sum over $l\in\{k,j\}$ by the sum over $l\in\CC$.

Now let us take the solution $\eta$ of equation $A^{kj}\eta =\xi$, obtained in Lemma \ref{olem:PaVe}. Since $\Supp\eta\subseteq\{k,j\}$, we have $A\eta=A^{kj}\eta=\xi$. 
Using (\ref{owidf}), (\ref{omumu}), the version of Proposition \ref{olem:A*} above,  (\ref{oAf1}) and  the identities $1),2)$ above,  we get
\begin{align}\nonumber
\lan\mu^0, \wid f^\la \xi\ran&=\lan\mu^0, \wid f^\la A\eta\ran= \la^{-1} \big( \lan\mu^0, h^\la A\eta \ran - \lan\mu^0, (1+\la f) A\eta\ran \big) 
\\ \nonumber
&=\la^{-1}\big(\lan \mu^\la, A\eta \ran - \lan\mu^0, A (1+\la f) \eta \ran \big) 
= -\lan \mu^\la,  B\eta\ran   +  \lan\mu^0, \wid B^*1 \eta\ran =-\lan \mu^\la-\mu^0, B\eta\ran.
\end{align}
Since the function $\xi$ is $C^1$-smooth, Lemma \ref{olem:PaVe}.{\it ii} implies that the solution $\eta$ also is. So that, the function $B\eta$ is continuous. Then Proposition \ref{olem:lato0}.{\it ii} implies
\bee\label{o684}
\lan \mu^\la-\mu^0, B\eta\ran\ra 0 \quad\mbox{as}\quad \la\ra 0.
\eee
Recall that the uniformity in $\CC$ of convergence in Proposition \ref{olem:lato0}.{\it ii} is understood, as usual, in the sense of finite-dimensional projections. Then, in order to prove that convergence (\ref{o684}) holds uniformly in $\CC$, it suffices to find a bounded set $\La\subset\mZ^d$, independent from the choice of the set $\CC$, such that 
$\Supp B\eta\subseteq \La$. Since $\Supp\eta\subseteq\{k,j\}$, we have  $B\eta=g_j\cdot\nabla_j \eta + g_k\cdot\nabla_k \eta$. Since for any $m$ we have $\Supp g_m\subseteq\{n:|m-n|\leq 1\}$, the set $\La$ can be chosen 
as $\La=\{n\in\mZ^d:|n-j|\wedge |n-k|\leq 1 \}$.
\qed
\ssk

{\it Proof of Proposition \ref{oprop:dteta}}.
In this proof we accept the following agreement: saying that something is "uniform in $\TT$", we mean the uniformity in the temperatures $(\TT_k,\TT_j)$,  if they belong to a compact subset of $\mR^{2}_+$.

Since the resonant energy flow $\JJ^{res}_{kj}$ is independent from the temperatures, $\wid\JJ^{res}_{kj}$ depends on them only through the function $u(v)$, so the desired properties of the function $\wid\JJ^{res}_{kj}$ follow.
Let us turn to the function $\wid\eta_{kj}$. Since $\eta_{kj}$ is a solution of eq. (\ref{oeeeta}) obtained in Lemma \nolinebreak \ref{olem:PaVe}, due to formula (\ref{oeta}) we have
\bee\label{ofhyjyrg}
\eta_{kj}(u)=-\int\limits_0^\infty \fB_{\tau}^0\JJ^{res}_{kj}(u)\,d\tau.
\eee   
Since the change of variables $u\mapsto v$ transforms the Ornstein-Uhlenbeck process (\ref{oOU}) to 
\bee\label{oOUu}
\dot v_j=-\frac{v_j}{2}+\dot{\bm\beta}_j, \quad j\in\CC,
\eee
relation (\ref{ofhyjyrg}) implies
\bee\label{owideta}
\wid\eta_{kj}(v)=-\int\limits_0^\infty \wid\fB^0_{\tau}\wid\JJ^{res}_{kj}(v) \,d\tau,
\eee
where $\wid\fB^0_{\tau}$ is a Markov semigroup associated with the process (\ref{oOUu}). Note that the operator $\wid\fB^0_{\tau}$  does not depend on the temperatures.
Put $\chp_\TT:=\chp_{\TT_k^{l_1}\TT_j^{l_2}}^{l_1+l_2}$, where $0\leq l_1+l_2\leq r-1$. 
To prove that the function $\wid\eta_{kj}$ is $(r-1)$-time differentiable in $\TT_k,\TT_j$, we will just show that 
\bee\label{oobmen}
\chp_{\TT}\wid\eta_{kj}(v)=-\int\limits_0^\infty \wid\fB^0_{\tau}\chp_\TT\wid\JJ^{res}_{kj}(v) \,d\tau, 
\eee
where the last integral converges.
Changing the variables $u\mapsto v$ in the identity of Proposition \ref{oprop:REF}.{\it ii}, we get 
$\lan \wid \mu^0, \wid \JJ^{res}_{kj}\ran =0$.
This holds for any $\TT_k,\TT_j>0$, so that we have  $\lan \wid\mu^0, \chp_\TT \wid\JJ^{res}_{kj}\ran =0$. Since $\mu^0$ is a unique stationary measure of the Ornstein-Uhlenbeck process (\ref{oOU}), the measure $\wid\mu^0$ is that for eq. (\ref{oOUu}). The mixing property of the latter implies 
\bee\label{oexpconv}
\big|\wid\fB^0_{\tau}\chp_\TT \wid\JJ^{res}_{kj}(v) \big|= \big|\wid\fB^0_{\tau}\chp_\TT\wid\JJ^{res}_{kj}(v)- \lan\wid\mu^0,  \chp_\TT\wid\JJ^{res}_{kj} \ran \big| \leq C(1+|v|^m)e^{-b\tau},
\eee
for any $\tau\geq 0$ and some constants $C,m,b>0$. It is possible to show that the constants $C,m$ and $b$ may be chosen uniformly in $\TT.$
It follows that the integral from the r.h.s. of (\ref{oobmen}) converges uniformly in $\TT$. Then it is not difficult to show that
$\int\limits_0^\infty \wid\fB^0_{\tau}\chp_\TT\wid\JJ^{res}_{kj} \,d\tau=\chp_\TT\int\limits_0^\infty \wid\fB^0_{\tau}\wid\JJ^{res}_{kj} \,d\tau$. So that, applying $\chp_\TT$ to the both sides of (\ref{owideta}), we get (\ref{oobmen}). 

Relation  (\ref{oobmen})  joined with (\ref{oexpconv}) implies that the function $\chp_{\TT}\wid\eta_{kj}(v)$ has at most a polynomial growth in $v$, which is uniform in $\TT$. The proof of the fact that the function $\chp_\TT\wid\eta_{kj}$ is continuous with respect to $(v,\TT_k,\TT_j)$ is not complicated. Using that solutions of eq. (\ref{oOUu}) depend on initial conditions in a continuous way, one should show that the r.h.s. of (\ref{oobmen}) with the time integral $\int\limits_0^\infty\ldots\,d\tau$ replaced by $\int\limits_0^T\ldots\,d\tau$ is continuous in  $(v,\TT_k,\TT_j)$, and then observe that the rest $\int\limits_T^\infty\ldots\,d\tau$  is small, in view of (\ref{oexpconv}).  
\qed

\subsection{Examples}
\label{osec:ex}

{\bf 1.} Consider the quadratic interaction potential  
$$
V(q_j,q_k)=(q_j-q_k)^2.
$$
It turns out that in this case the conductivity $\ka(\TT_j,\TT_k)$ from Theorem \ref{otheo:F}  does not depend on the temperatures, so is a positive constant.
We skip the proof since next we will consider a similar but more complicated situation.
\\
{\bf 2.} Let us now calculate the conductivity for the interaction potential 
\bee\label{oexpot}
V(q_j,q_{k})=(q_j-q_k)^4.
\eee 
The potential $V$ has growth of power four, so assumption {\it HV} is violated. Despite this we have the following result.
\begin{prop}
\label{oprop:eest}
(i)  Eq. (\ref{oslow}) is exponentially mixing with uniform in $\eps$ rate (in the sense as in Theorem  \nolinebreak \ref{otheo:mixing}). 

(ii) 
There exists $\eps_0$ such that for any $0<\eps<\eps_0$, $0<\la\leq 1$, $\tau\geq 0$ and $m\geq 0 $ we have
$$\MO\max\limits_{s\in[\tau,\tau+1]}(|p(s)|^2+|q(s)|^2)^{m} \leq C(|\CC|,m), \quad\quad  \big\lan\mu,(|p|^2+|q|^2)^{m}\big\ran  \leq C(|\CC|,m),$$
where $(p,q)(\tau)$ is a solution of eq. (\ref{oslow}) satisfying $\DD(p,q)(0)=\DD(p_0,q_0)$, and $\mu$ is its unique stationary measure. 
\end{prop}
{\it Proof.} 
{\it (i)}  By the same arguments that was used in the proof of Theorem \ref{otheo:mixing}.

{\it (ii)} By applying the Ito formula to the function $(H^\nu+\frac{\eps}{2}\sum\limits_{j\in\CC}p_jq_j)^m$. 
\footnote{For the both items one should use that 
 $\{\frac12\sum\limits_{j\in\CC}p_jq_j, \frac12\sum\limits_{|k-j|=1}(q_k-q_j)^4\}=-\sum\limits_{|k-j|=1}(q_k-q_j)^4.$}   
\qed

As it was explained in Remark \ref{orem:HG1}, Proposition \ref{oprop:eest} implies that all theorems proved in this and previous chapters remain valid, except the uniformity of convergences in $\CC$ which fails.
\begin{prop}
\label{oprop:conddd}
 The conductivity $\ka$ has the form
$$\kappa(\TT_j,\TT_k)=C(\TT_j+\TT_{k})^2,$$
where the constant $C$ is independent from the temperatures.
\end{prop} 
{\it Proof.}
By the direct computation we get
\bee\nonumber
V^{res}_{kj}(u) = \frac{3}{8} |u_j-u_k|^4.
\eee 
To calculate the conductivity $\ka$, defined by formula (\ref{okappatheo}), we need to find the solution $\eta_{kj}$ of eq. (\ref{oeeeta}), satisfying properties $(a)$-$(c)$ from Lemma \ref{olem:PaVe}. Writing the operator $A^{kj}$ in the polar coordinates, we see that it commutes with the operator $\chp_{\ph_l}$, for any $l\in\CC$.  Then, in view of the relation $\JJ^{res}_{kj}=2\chp_{\ph_k}V^{res}_{kj}$ established in Proposition \ref{oprop:REF}{\it (i)}, it suffices to find a solution $\phi_{kj}$ of equation 
\bee\label{ophi}
A^{kj}\phi_{kj}= 2V_{kj}^{res}+C(\TT_k,\TT_j),
\eee
for some constant $C(\TT_k,\TT_j)$ (possibly depending on the temperatures $\TT_j,\TT_k$), such that the function $\eta_{kj}:=\chp_{\ph_k}\phi_{kj}$ satisfies the conditions $(a)$-$(c)$ from Lemma \nolinebreak\ref{olem:PaVe}. By the direct computation we get that the function
$$
\phi_{kj}=-\frac{3}{16} |u_j-u_k |^4 - \frac{3}{2}(\TT_j+\TT_k)|u_j-u_k|^2 
$$
satisfies (\ref{ophi}). Using that  $\chp_{\ph_k}(|u_j-u_k|^2)=-2(iu_k)\cdot u_j$, we find 
\bee\label{o763}
\eta_{kj}=\chp_{\ph_k}\phi_{kj}= \frac{3}{4}(iu_k)\cdot u_j\big(|u_j-u_k|^2 + 4 (\TT_j+\TT_k)\big).
\eee
It is not difficult to see that the function $\eta_{kj}$ satisfies the required conditions. Note also that
\bee\label{oxie}
\JJ^{res}_{kj}=2\chp_{\ph_k}V_{kj}^{res}= -3(iu_k)\cdot u_j|u_j-u_k|^2.
\eee
Substituting (\ref{o763}) and (\ref{oxie}) to  
the first representation for the conductivity $\ka$ from (\ref{okappatheo}), passing to the coordinates $r_j:=|u_j|,\,r_k:=|u_k|,\, \psi=\ph_j-\ph_k$, and using that
$$
|u_j-u_k|^2= r_k^2+r_j^2 -2r_kr_j\cos\psi \quad\mbox{and}\quad (iu_k)\cdot u_j =r_kr_j\sin\psi, 
$$
we get the desired result by the direct computation.
\qed

\subsection{Space-time correlations of the energy flow}
\label{osec:GK}

In this section we study stationary space-time correlations of the energy flow under the limits $\eps,\la\ra 0$ and investigate their connection with the conductivity $\ka$.
Since till the end of the section we will not talk about the uniformity of convergences with respect to the choice of the set $\CC$, further on constants are permitted to depend on it.
\begin{theo}\label{otheo:GK} 
Let $\fB_\tau^{\eps,\la}$ be a Markov semigroup associated with eq. (\ref{oslow}) and $\mu^{\eps,\la}$ be its unique stationary measure. Take any $k,j,m,l\in\CC$ satisfying $|k-j|=|m-l|=1$ and denote
$$
\YY^{\eps,\la}:=\int_0^\infty \lan\mu^{\eps,\la}, \JJ_{kj} \fB_\tau^{\eps,\la}\JJ_{ml}\ran 
- \lan \mu^{\eps,\la}, \JJ_{kj}\ran\lan \mu^{\eps,\la}, \JJ_{ml}\ran\,d\tau. 
$$
Then
\bee\label{oGK}
\lim\limits_{\la\ra 0}\lim\limits_{\eps\ra 0} \YY^{\eps,\la}
=\left\{
\begin{array}{cl}
\TT_j\TT_k\ka(\TT_j,\TT_k) &\quad\mbox{if}\quad m=k,l=j, \\
-\TT_j\TT_k\ka(\TT_j,\TT_k) &\quad\mbox{if}\quad m=j,l=k,\\
0 &\quad\mbox{if}\quad \{k,j\}\neq\{m,l\}.
\end{array}
\right.
\eee
\end{theo}
\begin{cor}\label{ocor:GK}
Assertion of Theorem \ref{otheo:GKintro} is satisfied. 
\end{cor}
{\it Proof of Corollary \ref{ocor:GK}.} Let us first note that for the constant temperature profile $\TT_n=\hat\TT$ $\forall n\in\CC$ we have
$\lan \mu^{\eps,\la}, \JJ_{kj}\ran=0$, 
for any 
$k,j\in\CC$, $|k-j|=1$.
Indeed, since in the present case the measure $\mu^{\eps,\la}$ is just the Gibbs measure,
$\mu^{\eps,\la}=Z^{-1}(\hat\TT)e^{-H^\nu/\hat\TT} \,d\mbox{Leb}$, 
this follows from evenness of the Hamiltonian $H^\nu$ with respect to the variable $p$, and oddness of the energy flow $\JJ_{kj}$ with respect to it.

Now let us rewrite the r.h.s. of (\ref{o8345}) as   
\bee\nonumber
\frac{1}{\hat\TT^2 N}\lim\limits_{\la\ra 0}\lim\limits_{\eps\ra 0} \Big(\sum\limits_{j=0}^{N-1}
\int\limits_{0}^\infty \big\lan\mu^{\eps,\la},  \JJ_{jj+1} \fB_\tau^{\eps,\la} \JJ_{jj+1}\big\ran\,d\tau 
+ 
\sum\limits_{0\leq j,k\leq N-1,\,j\neq k}
\int\limits_{0}^\infty \big\lan\mu^{\eps,\la},  \JJ_{jj+1} \fB_\tau^{\eps,\la} \JJ_{kk+1}\big\ran\,d\tau\Big).
\eee
Due to Theorem \ref{otheo:GK}, the limits of the first summand in the brackets equal to $N\hat\TT^2\hat\ka(\hat\TT)$, while the limits of the second one vanish.
\qed

{\it Proof of Theorem \ref{otheo:GK}.} The proof of the theorem is based on the following auxiliary lemma, which we will establish in the end of the section.
Denote by $\fB_{\tau}^\la$ and $\fB_{\tau}^0$ the Markov semigroups associated with the effective equation (\ref{oeffH}) and the Ornstein-Uhlenbeck process (\ref{oOU}) correspondingly.
Let

$
\ds{\eta_{ml}^{\eps,\la}(u):=\int\limits_0^\infty \fB_\tau^{\eps,\la}\JJ_{ml}(u)- \lan \mu^{\eps,\la}, \JJ_{ml}\ran\,d\tau,}
$
$\ds{\eta_{ml}^\la(u):=\int\limits_0^\infty \fB_{\tau}^\la\JJ_{ml}^{res}(u) -\lan \mu^{\la}, \JJ^{res}_{ml}\ran\,d\tau},$
$$
\eta_{ml}(u):=-\int\limits_0^\infty \fB_{\tau}^0\JJ_{ml}^{res}(u)\,d\tau,
\footnote{Recall that $\lan \mu^{0}, \JJ^{res}_{ml}\ran=0$, due to Proposition \ref{oprop:REF}.{\it ii}, where $\mu^0$ is the unique stationary measure of the Ornstein-Uhlenbeck process (\ref{oOU}).}
$$
where $\mu^\la$ is the unique stationary measure of eq. (\ref{oeffH}). Note that, due to (\ref{oeta}), the function $\eta_{ml}$ coincides with that from Theorem \ref{otheo:F}.
Put also 
$$\hat\eta^{\eps,\la}_{ml}(u):=\int\limits_0^\infty |\fB_\tau^{\eps,\la}\JJ_{ml}(u) - \lan \mu^{\eps,\la}, \JJ_{ml}\ran|\,d\tau,$$ 
and define $\hat\eta_{ml}^\la(u)$ and $\hat\eta_{ml}(u)$ in a similar way.
\bl\label{olem:GK}
(i) There exist constants $C,n>0$, independent from $\eps$ and $\la$, such that for any $u\in\mC^{|\CC|}$ we have
$$
|\eta^{\eps,\la}_{ml}(u)|,\hat\eta^{\eps,\la}_{ml}(u),|\eta_{ml}^\la(u)|,\hat\eta_{ml}^\la(u),|\eta_{ml}(u)|,\hat\eta_{ml}(u)\leq C(1+|u|^n).
$$
(ii) For any $u\in\CCC$  we have  
$$
\eta_{ml}^{\eps,\la}(u)\ra \eta_{ml}^\la(u) \ass \eps\ra 0\quad\mbox{and}\quad \eta^\la_{ml}(u)\ra-\eta_{ml}(u) \ass \la\ra 0,
$$
where the both convergences hold uniformly in $u\in K$, if $K\subset\CCC$ is a compact set. 

(iii) For any $\tht\in[0,2\pi)$   we have $\eta_{ml}^\la(e^{i\tht}u)\equiv \eta_{ml}^\la(u)$.
\el   
Let us first show that the integral $\YY^{\eps,\la}$ converges. Due to Fubini's theorem, 
for this purpose it suffices to establish that 
\bee\label{oGK2}
\big\lan\mu^{\eps,\la}, \int\limits_0^\infty |\JJ_{kj}(\fB_\tau^{\eps,\la} \JJ_{ml} -\lan \mu^{\eps,\la}, \JJ_{ml}\ran)|\, d\tau \big\ran <\infty.
\eee
Note that the l.h.s. of (\ref{oGK2}) equals to 
$\lan \mu^{\eps,\la}, |\JJ_{kj}| \hat\eta_{ml}^{\eps,\la}\ran$.
Since the energy flow $\JJ_{kj}$ and the function  $\hat\eta_{ml}^{\eps,\la}$  have at most a polynomial growth, inequality (\ref{oGK2}) follows from the estimate for the stationary measure $\mu^{\eps,\la}$ provided by Lemma \nolinebreak\ref{olem:est}.{\it ii}.
Now let us pass to the limit $\eps\ra 0$.  Fubini's theorem implies that 
\bee\label{yyy1}
\YY^{\eps,\la}=\lan \mu^{\eps,\la}, \JJ_{kj} \eta^{\eps,\la}_{ml}\ran.
\eee 
Theorem \ref{otheo:sm}.{\it i} joined with Lemma \ref{olem:GK}.{\it i,ii}  implies 
\bee\label{yyy2}
\lan \mu^{\eps,\la},\JJ_{kj} \eta_{ml}^{\eps,\la} \ran \ra \lan \mu^\la, \JJ_{kj} \eta_{ml}^\la \ran \ass \eps\ra 0.
\eee
In view of Proposition \ref{olem:rotinv} and Lemma \ref{olem:GK}.{\it iii}, we have
\bee\label{yyy3}
\lan \mu^\la, \JJ_{kj} \eta_{ml}^\la \ran=\big\lan \mu^\la, \lan\JJ_{kj}\eta_{ml}^\la\ran_R  \big\ran
=\big\lan \mu^\la, \lan \JJ_{kj} \ran_R \eta_{ml}^\la \big\ran=\lan \mu^\la, \JJ_{kj}^{res} \eta_{ml}^\la \ran.
\eee
Now we pass to the limit $\la\ra 0$. Proposition \ref{olem:lato0}.{\it ii} joined with Lemma \ref{olem:GK}.{\it ii} implies
\bee\label{yyy4}
\lan \mu^\la, \JJ_{kj}^{res}\eta_{ml}^\la \ran \ra -\lan \mu^0,\JJ_{kj}^{res} \eta_{ml} \ran \ass\la\ra 0.
\eee
Combining (\ref{yyy1})-(\ref{yyy4}), we see that it remains to establish the identity
\bee\label{resst}
-\lan \mu^0,\JJ_{kj}^{res} \eta_{ml} \ran=
\left\{\begin{array}{cl}
\TT_j\TT_k\ka(\TT_j,\TT_k) &\quad\mbox{if}\quad m=k,l=j, \\
-\TT_j\TT_k\ka(\TT_j,\TT_k) &\quad\mbox{if}\quad m=j,l=k,\\
0 &\quad\mbox{if}\quad \{m,l\}\neq\{k,j\}.
\end{array}
\right.
\eee
The first equality of (\ref{resst}) immediately follows from the formula (\ref{okappatheo}) while the second one from the relation $\eta_{jk}=-\eta_{kj}$. The third equality follows from the rotation invariance of the Gaussian measure $\mu^0$ and can be obtained by the argument used in the proof of Proposition \ref{oprop:569}.
\qed
\msk

{\it Proof of Lemma \ref{olem:GK}}. 
Let us start with several auxiliary propositions.
\bpp\label{oprop:launif}
The effective equation (\ref{oeffH}) is exponentially mixing with the rate which is uniform in $0\leq\la\leq 1$  (in the sense as in Theorem \nolinebreak \ref{otheo:mixing}.)
\epp
{\it Proof.} It can be shown by an argument similar to that used in the proof of Theorem \nolinebreak\ref{otheo:mixing}, but simplified. Note only that in the proof of the reccurence one should apply the Ito formula to the function 
$e^{\del \tau}\sum\limits_{j\in\CC} I_j$.
\footnote{This is related to the fact that the sum of actions $\sum\limits_{j\in\CC} I_j$ is a first integral of the resonant Hamiltonian $H^{res}$. The latter follows from the relation 
$\{\sum\limits_{j\in\CC} I_j, H^{res}\}=-\sum\limits_{j\in\CC}\chp_{\ph_j}H^{res}
=-\sum\limits_{j\in\CC}\sum\limits_{k:|j-k|=1}\chp_{\ph_j}V_{jk}^{res}=-\sum\limits_{|j-k|=1}\chp_{\ph_j}V_{jk}^{res}$, 
and Proposition \nolinebreak \ref{oprop:RA}.{\it iv}.}
\qed

\bpp\label{oprop:GK2} 
For any compact set $K\subset\CCC$,
the time-integral from the definition of the function $\eta_{ml}^{\eps,\la}$ converges uniformly in $u\in K$ and $\eps$, while that from the definition of the function $\eta_{ml}^\la$ converges uniformly in $u\in K$ and $\la$.
\epp
{\it Proof.} 
Recall that in Appendix \ref{oapp:mixing} it is proven that eq. (\ref{oslow}) is exponentially mixing with the rate which is uniform in $\eps$. Then it can be shown that for all $u\in\mC^{|\CC|}$  we have
\bee\label{opavepr}
|\fB^{\eps,\la}_\tau \JJ_{ml}(u)- \lan \mu^{\eps,\la}, \JJ_{ml}\ran |
\leq C(1+|u|^n)e^{-b\tau},
\eee
where the constants $C,n,b>0$ are independent from $\eps$ (see e.g. a proof of Theorem 1 from \cite{PaVe}).
It follows that the integral $\int\limits_0^\infty \fB^{\eps,\la}_\tau \JJ_{ml}(u) - \lan \mu^{\eps,\la}, \JJ_{ml}\ran\, d\tau$ converges uniformly in $\eps$ and $u\in K$. To prove that the integral 
$\int\limits_0^\infty \fB_{\tau}^\la \JJ_{ml}^{res}(u) - \lan \mu^{\la}, \JJ^{res}_{ml}\ran\, d\tau$ converges uniformly in $\la$ and $u\in K$, we use Proposition \ref{oprop:launif} and argue similarly.
\qed
\ssk

Now we turn to the proof of the lemma.

{\it Item (i)}. 
The estimates  $|\eta_{ml}^{\eps,\la}|,\hat\eta_{ml}^{\eps,\la}\leq C(1+|u|^n)$ follow from (\ref{opavepr}). The other estimates can be obtained similarly.
 
{\it Item (ii)}. 
Let us first show that $\eta_{ml}^{\eps,\la}(u)\ra \eta_{ml}^\la(u)$ as $\eps\ra 0$ uniformly in $u\in K$.
Due to Proposition \nolinebreak\ref{oprop:GK2}, it suffices to prove that for any $T>0$ we have
\bee\nonumber
\int\limits_0^T \fB^{\eps,\la}_\tau \JJ_{ml}(u) - \lan \mu^{\eps,\la}, \JJ_{ml}\ran\, d\tau \ra  \int\limits_0^T \fB_{\tau}^\la \JJ_{ml}^{res}(u) - \lan \mu^{\la}, \JJ^{res}_{ml}\ran\, d\tau 
\ass \eps\ra 0, 
\eee
uniformly in $u\in K.$ Since, due to Theorem \ref{otheo:sm}.{\it i} joined with Proposition \ref{olem:rotinv}, we have $\lan \mu^{\eps,\la}, \JJ_{ml}\ran\ra \lan \mu^{\la}, \JJ^{res}_{ml}\ran$ as $\eps\ra 0$, it suffices to establish that
\bee\label{ogkl1}
\int\limits_0^T \fB^{\eps,\la}_\tau \JJ_{ml}(u)\, d\tau \ra  \int\limits_0^T \fB_{\tau}^\la \JJ_{ml}^{res}(u)\, d\tau 
\ass \eps\ra 0 \quad\mbox{uniformly in $u\in K$}. 
\eee
Let us rewrite the l.h.s. of  (\ref{ogkl1}) as $\int\limits_0^T \MO \JJ_{ml}(u^{\eps,\la}(\tau))\, d\tau$, 
where $u^{\eps,\la}(\tau)$ is a solution of eq. (\ref{oslow}) satisfying  $u^{\eps,\la}(0)=u$.
Denote by $a^{\eps,\la}$ the process $u^{\eps,\la}$ written in the $a$-variables.
\begin{rem}\label{orem:123}
Simple analysis of proofs of  Lemma \nolinebreak \ref{olem:lem} and Theorem \ref{otheo:a} shows that convergences obtained there for the process $a^{\eps,\la}$ hold uniformly in $u\in K$, if the set $\CC$ is fixed.
\end{rem}
Applying Lemma \ref{olem:lem} to the function $h(u,\tht):=\JJ_{ml}(e^{i\tht}u)$, in view of Remark \nolinebreak\ref{orem:123} we obtain that the integral above is uniformly in $u\in K$ close to 
$\int\limits_0^T \MO \JJ_{ml}^{res}(a^{\eps,\la}(\tau))\, d\tau.$
Due to Theorem \ref{otheo:a}, the latter integral is uniformly in $u\in K$ close to 
$$\int\limits_0^T \MO \JJ_{ml}^{res}(u^\la(\tau))\, d\tau=\int\limits_0^T \fB_{\tau}^\la \JJ_{ml}^{res}(u)\, d\tau,$$ 
where $u^\la(\tau)$ is a solution of the effective equation (\ref{oeffH}) satisfying $u^\la(0)=u$.

The convergence $\eta_{ml}^\la(u)\ra -\eta_{ml}(u)$ as $\la\ra 0$ can be established with help of Proposition \nolinebreak\ref{olem:lato0} in a similar way. 

{\it Item (iii)}. Let $u^\la(\tau)$ be a solution of the effective equation (\ref{oeffH}) satisfying $u^\la(0)=u$.
The rotation invariance of the effective equation (see Proposition \nolinebreak\ref{olem:vrawinvar}) implies that $e^{i\tht}u^\la(\tau)$ is its weak solution. Since it satisfies $e^{i\tht}u^\la(0)=e^{i\tht}u$, we have 
$$
\eta_{ml}^\la(e^{i\tht}u)=
\int\limits_0^\infty \MO \JJ_{ml}^{res}(e^{i\tht}u^\la(\tau))-\lan\mu^\la, \JJ_{ml}^{res}\ran\,d\tau=
\int\limits_0^\infty \MO \JJ_{ml}^{res}(u^\la(\tau))-\lan\mu^\la, \JJ_{ml}^{res}\ran\,d\tau=\eta_{ml}^\la(u),
$$
where we have used Proposition \ref{oprop:RA}.{\it i}.
\qed

\section{Proofs of the averaging theorems}
\label{osec:rape}

Here we prove Theorems \ref{otheo:a}, \ref{otheo:sm} and Proposition \ref{oprop:ru'} which implies Proposition \ref{oprop:ru}.

\subsection{Averaging lemma}
\label{osec:rape1}

We start with the following averaging lemma which is the main tools in the proofs of Theorems \ref{otheo:a} and \ref{otheo:sm}.  Let $u^\eps(\tau)$ be a solution of eq. (\ref{oslow}), either satisfying $\DD(u^\eps(0))=\DD(u_0)$ or a stationary one. Let $a^\eps(\tau)$ be the corresponding $a$-variables, given by (\ref{oavar}). Fix a bounded set $\La\subset\mZ^d$ and assume that the set $\CC$ satisfies $\CC\supseteq\La$. 
Take a function $h(u,\tht)\in\LL_{loc}(\mC^{|\CC|}\times [0,2\pi))$ such that for each $\tht$ the function $h(\cdot,\tht)$ satisfies $\Supp h(\cdot,\tht)\subseteq\La$ and has at most a polynomial growth at infinity which is uniform in $\tht$.
Then there exist non-decreasing functions $c,c_1:\mR\mapsto\mR$ such that 
\bee\label{occ_1}
|h(u_1,\tht)|\leq c(\max\limits_{j\in\La}|u_{1j}|),\;|h(u_1,\tht)-h(u_2,\tht)|\leq c_1(\max\limits_{k=1,2,\,j\in\La}|u_{kj}|)\sum\limits_{j\in\La}|u_{1j}-u_{2j}|,
\eee
for any $u_1,u_2\in\CCC$, $\tht\in[0,2\pi)$, where the function $c$ has at most a polynomial growth at infinity. Denote
$$\lan h\ran_\tht (u):=\int\limits_0^{2\pi} h(u, \tht) \, \dbar\tht.$$
\begin{lem}
\label{olem:lem}
Take a function $h$ as above. Then 
\begin{equation}
\nonumber
\MO \max\limits_{0\leq\tau\leq T} \left| \int\limits_0^\tau \big( h(a^{\eps}(s), \eps^{-1} s) - \lan h\ran_\tht (a^\eps(s))\big) \, ds \right| \rightarrow 0 \mbox{ as } \eps \rightarrow 0,
\end{equation}
uniformly in $\CC$ and with respect to the choice of the function $h$ satisfying (\ref{occ_1}) for fixed $c,c_1$. 
\end{lem}
{\it Proof.}  For purposes of the proof we first introduce some notations. 
For an event $\Gamma$ and a random variable $\xi$ we denote 
$$
\mbox{\bf E}_{\Gamma}\,\xi:=\MO(\xi\mI_{\ov\Gamma}).
$$
By $\ae(r)$ we denote various functions of $r\in\mR$ such that $\ae(r)\rightarrow 0$ as $r\rightarrow\infty$ uniformly in $\CC$ and $h$, satisfying (\ref{occ_1}) for fixed $c,c_1$.  By $\ae_\infty(r)$ we denote functions $\ae(r)$ such that $\ae(r)=o(r^{-m})$ uniformly in $\CC$ and $h$, for each $m>0$. We write $\ae(r)=\ae(r;a)$ to indicate that $\ae(r)$ depends on a parameter $a$.

Denote by $U_1(\La)$ the  neighbourhood of radius $1$ of $\La$ in $\CC$, i.e.
$$U_1(\La):=\{n\in\CC\big| \, \mbox{there exists } k\in\La \mbox{ satisfying } |n-k|\leq 1\}.$$ 
Fix $R>0$. Set
\bee
\label{o4etirep}
\Omega_R = \{ \max\limits_{k\in U_1(\La)}\max\limits_{0\leq \tau \leq T} |a^\eps_k(\tau)| \geq R\}.
\eee
Due to the estimate {\it (i)} of Lemma \ref{olem:est}, we have
 \begin{equation}
\label{oodin}
\PR(\Omega_R)\leq \ae_\infty(R).
\end{equation}
In view of the polynomial growth of the function $h$,
$$
\MO_{\ov\Om_R} \max\limits_{0\leq \tau \leq T}  \left| \int\limits_0^\tau h(a^\eps(s),\eps^{-1}s)\,ds \right| \leq \ae_\infty(R),
$$
and the function $\lan h \ran_{\tht} (a^\eps(s))$ satisfies a similar relation.  
Then, it is sufficient to show that for any $R\geq 0$
$$
\UU^\eps:= \MOOM \max\limits_{0\leq \tau \leq T} \left| \int\limits_0^\tau h(a^\eps(s),\eps^{-1}s) - \lan h \ran_{\tht}(a^\eps(s)) \, ds \right|\ra 0 \quad\mbox{as}\quad\eps\ra 0\quad\mbox{uniformly in }\CC,h.
$$
For this purpose we consider a partition of the interval $[0,T]$ to subintervals of the length $\del=\sqrt\eps$ by the points 
$
\tau_l= l \del,$ $ 0\leq l \leq L$, where $L=[T/\del].$
Denote 
$$
\eta_l=\int\limits_{\tau_l}^{\tau_{l+1}} h(a^\eps(s),\eps^{-1}s) - \lan h\ran_{\tht}(a^\eps(s))   \, ds.
$$
Then
\bee\label{oueps}
\UU^\eps\leq  \MOOM \sum\limits_{l=0}^{L-1} |\eta_l| + 2\del c(R).
\eee
We have
\begin{align}
\nonumber
|\eta_l| &\leq \left|\int\limits_{\tau_l}^{\tau_{l+1}} h(a^\eps(s),\eps^{-1}s) - h(a^\eps(\tau_l),\eps^{-1}s)\, ds \right|  +  \left|\int\limits_{\tau_l}^{\tau_{l+1}} h(a^\eps(\tau_l),\eps^{-1}s)  - \lan h\ran_{\tht}(a^\eps(\tau_l)) \, ds \right|  \\
\nonumber
 &+ \left|\int\limits_{\tau_l}^{\tau_{l+1}}  \lan h\ran_{\tht}(a^\eps(\tau_l)) -  \lan h\ran_{\tht}(a^\eps(s)) \, ds \right| =: \YY_l^1 + \YY_l^2 + \YY_l^3.  
\end{align}
{\it Terms $\YY_l^1,\YY_l^3$.} Since the function $h$ is locally Lipschitz, 
\bee
\label{oy1,3}
\MOOM (\YY_l^1+\YY_l^3)\leq 2\del c_1(R) \sum\limits_{j\in \La} \MOOM \max\limits_{\tau_l\leq s\leq \tau_{l+1}} |a_j^\eps(s)-a_j^\eps(\tau_l)| \leq \del \ae(\del^{-1};R),
\eee
where the last inequality follows from eq. (\ref{oa}).  

{\it Term $\YY_l^2$.} Changing the variable of integration $\hat s:= \eps^{-1}s$,  we obtain
\bee\label{oy2}
\MOOM\YY_2= \eps  \MOOM\left|\int\limits_{\eps^{-1}\tau_l}^{\eps^{-1}\tau_{l+1}} h(a^\eps(\tau_l),\hat s)  - \lan h\ran_{\tht}(a^\eps(\tau_l)) \, d\hat s \right|\leq 2\eps c(R),
\eee 
due to the definition of the function $\lan h\ran_{\tht}$.

In view of the identity $\eps=\del^2$, estimates (\ref{oy1,3}) and (\ref{oy2}) imply 
$$
 \MOOM |\eta_l|\leq \del\big(\ae(\del^{-1};R)+2\del c(R)\big).
$$
Noting that $L\leq T\del^{-1}$, from (\ref{oueps})  we get 
$$
\UU^\eps\leq T\big(\ae(\del^{-1};R)+2\del c(R)\big)+ 2\del c(R)\ra 0 \quad \mbox{as}\quad \eps\ra 0 \quad\mbox{uniformly in $\CC,h$.}
$$
\qed
\ssk

Let now $u^\eps(\tau)$ be a stationary solution of eq. (\ref{oslow}). 
\begin{cor}\label{olem:lem'}
Let a function $h\in\LL_{loc}(\mC^{|\CC|})$ satisfies $\Supp h\subseteq\La$ and has at most a polynomial growth at infinity. Then for any $\tau\geq 0$ we have
\bee\nonumber
\MO h(u^\eps(\tau))-\MO h(a^\eps(\tau)) \ra 0 \quad\mbox{as}\quad \eps\ra 0\quad\mbox{ uniformly in }\CC.
\eee
\end{cor} 
{\it Proof.} 
There exist functions $c,c_1:\mR\mapsto\mR$, where $c$ has at most a polynomial growth at infinity, such that
\bee\label{occ_1'}
|h(u_1)|\leq c(\max\limits_{j\in\La}|u_{1j}|),\;|h(u_1)-h(u_2)|\leq c_1(\max\limits_{k=1,2,\,j\in\La}|u_{kj}|)\sum\limits_{j\in\La}|u_{1j}-u_{2j}|,
\eee
for any $u_1,u_2\in\CCC$. Consider the function $\hat h(u,\tht):=h(e^{i\tht}u)$. Clearly, it satisfies (\ref{occ_1}) with $c,c_1$ from (\ref{occ_1'}). 
Applying Lemma \ref{olem:lem}, we get 
\bee\label{o352}
 \int\limits_0^T \MO h(u^\eps(\tau))-\MO \lan h\ran_R(a^\eps(\tau))\, d\tau \ra 0 \quad\mbox{as}\quad \eps\ra 0,
\eee
uniformly in $\CC$ and with respect to the choice of the function $h$ satisfying (\ref{occ_1'}). 
In view of Proposition \ref{oprop:RA}.{\it i}, we have  
$\lan h\ran_R(a^\eps(\tau)) \equiv\lan h\ran_R(u^\eps(\tau))$.
Then stationarity of the process $u^\eps$ implies that the function under the integral in (\ref{o352}) is independent from time, so that for any $\tau\geq 0$ we have
\bee\label{o352'}
\MO h(u^\eps(\tau))-\MO \lan h\ran_R(a^\eps(\tau))\ra 0 \quad\mbox{as}\quad \eps\ra 0 \quad\mbox{uniformly in $\CC$ and $h$.}
\eee 
Let $h^{\eps,\tau}(u):=h(e^{-i\eps^{-1}\tau}u).$ Since the functions $h^{\eps,\tau}$ satisfy (\ref{occ_1'}) for any $\eps$ and $\tau$, we have
$$
\MO h^{\eps,\tau}(u^\eps(\tau))-\MO \lan h^{\eps,\tau}\ran_R(a^\eps(\tau)) \ra 0 \quad\mbox{as}\quad \eps\ra 0 \quad\mbox{uniformly in $\CC$.} 
$$
Since $h^{\eps,\tau}(u^\eps(\tau))\equiv h(a^\eps(\tau))$ and $\lan h^{\eps,\tau}\ran_R= \lan 
h\ran_R$, we get
$$
\MO h(a^\eps(\tau))-\MO \lan h\ran_R(a^\eps(\tau)) \ra 0 \quad\mbox{as}\quad \eps\ra 0 \quad\mbox{uniformly in $\CC$.} 
$$
Jointly with (\ref{o352'}) this implies the desired convergence.
\qed

\subsection{Proof of Theorem \ref{otheo:a}} 
\label{osec:rape2}

To prove the theorem it suffices to find weak solutions $a^\eps(\tau)$ and $u(\tau)$
of eq. (\ref{oa}) and the effective equation (\ref{oeff}) correspondingly, defined on the same probability space and  satisfying $\DD(a^\eps(0))=\DD(u(0))=\DD(u_0)$,  such that  for every $j\in\CC$ we have
\bee\label{obut}
\MO \max\limits_{0\leq\tau\leq T} |w^\eps_j(\tau)|\ra 0 \assN,
\eee 
where $w^\eps(\tau):=a^\eps(\tau)-u(\tau).$
Let us start with constructing suitable Brownian motions $\beta$ and $\bm\beta$. Denote 
$$
\bm\beta^\eps=(\bm\beta_j^\eps)_{j\in\CC},\quad \bm\beta_j^\eps(\tau):=\sqrt{2}\int\limits_0^\tau e^{-i\eps^{-1}s}\, d\beta_j.
$$
\bpp\label{oprop:razdvwym}
We have $\DD(\bm\beta^\eps(\cdot))\raw\DD(\bm\beta(\cdot))$ as $\eps\ra 0$ in $C([0,T],\mC^{|\CC|})$.
\epp
Proof of the proposition is given after the end of the proof of the theorem.
Due to Proposition \ref{oprop:razdvwym} and Skorokhod theorem, we can find random processes $\wid{\bm\beta}^\eps(\tau)$ and $\wid{\bm\beta}(\tau)$, defined on the same probability space and satisfying 
$
\DD(\wid{\bm\beta}^\eps(\cdot)) = \DD(\bm\beta^\eps(\cdot))$, $\DD(\wid{\bm\beta}(\cdot))= \DD(\bm\beta(\cdot)),  
$
such that 
\bee\label{orrw}
\wid{\bm\beta}^\eps \ra \wid{\bm\beta}\ass \eps\ra 0\quad \mbox{in}\quad C([0,T],\mC^{|\CC|}) \quad\mbox{a.s.}
\eee
Take solutions $a^\eps(\tau)$ and $u(\tau)$ of eq. (\ref{oa}) and the effective equation (\ref{oeff})  on the probability space above (extended if needed), corresponding to the Brownian motions 
$\ds{\wid\beta(\tau):=\frac{1}{\sqrt{2}}\int\limits_0^\tau e^{i\eps^{-1}s} \, d\wid{\bm\beta}^\eps}$ and $\wid{\bm\beta}$, and having the same initial conditions, distributed as $u_0$.
Let us show that the process $w^\eps=a^\eps-u$ satisfies (\ref{obut}).
Due to eq. (\ref{oa}) and (\ref{oeff}),
\begin{align}\nonumber
|w_j^\eps(\tau)| &\leq\Big|\int\limits_0^\tau e^{-i\eps^{-1}s}P_j(e^{i\eps^{-1}s}a^\eps(s))-\RR_j(a^\eps(s))\, ds\Big| 
+ \Big|\int\limits_0^\tau \RR_j(a^\eps(s))-\RR_j(u(s))\, ds\Big|  \\
\label{o375}
&+ \sqrt{\TT_j}\big|\wid{\bm\beta}^\eps_j(\tau)-\wid{\bm\beta}_j(\tau) \big|  =: \YY^\eps_{1j}(\tau)+\YY^\eps_{2j}(\tau) + \YY^\eps_{3j}(\tau).
\end{align}
First we will estimate terms $\YY^\eps_{1j},\YY^\eps_{2j}$ and $\YY^\eps_{3j}$ separately.

{\it Terms $ \YY^\eps_{1j}$ and $ \YY^\eps_{3j}$.}
Using that $\MO \max\limits_{0\leq\tau\leq T} |\wid{\bm\beta}_j^\eps(\tau)|^2\vee |\wid{\bm\beta}_j(\tau)|^2<C$, it is not difficult to show that (\ref{orrw}) implies 
\bee\nonumber
\MO \max\limits_{0\leq\tau\leq T} \YY^\eps_{3j}(\tau) \ra 0 \ass \eps\ra 0.
\eee
Since each of the vectors $\bm\beta^\eps$ and $\bm\beta$ has independent components, the processes $\wid{\bm\beta}_j^\eps,\wid{\bm\beta}_j$ can be chosen independently from the choice of the set $\CC$, so that the convergence above holds uniformly in $\CC$.

Applying Lemma \ref{olem:lem} to the function $h(u,\tht)=e^{-i\tht}P_j(e^{i\tht}u)$, we have
\bee\nonumber
\MO\max\limits_{0\leq\tau\leq T} \YY^\eps_{1j}(\tau)\ra 0 \ass \eps\ra 0 \quad\mbox{uniformly in }\CC.
\eee
Denote $\del^\eps=(\del^\eps_j)_{j\in\CC}$, where
$
\del_j^\eps:=\MO\max\limits_{0\leq\tau\leq T} (\YY^\eps_{1j}(\tau) + \YY^\eps_{3j}(\tau)),
$
so that 
\bee\label{o3854}
\del_j^\eps\ra 0 \assN.
\eee

{\it Term $ \YY^\eps_{2j}$.} Since, by assumption {\it HV}, the second derivatives of the potential $V$ have at most a quadratic growth, the same holds for the first derivatives of the functions $\RR_j$. Then for each $j\in\CC$ and any $\tau\geq 0$ we have 
\bee\label{odif}
\YY^{\eps}_{2j} \leq C\int\limits_{0}^\tau \big(1+\sum\limits_{k:|k-j|\leq 1}(|a^\eps_k|^2+|u_k|^2)\big) \sum\limits_{k:|k-j|\leq 1} |w^\eps_k| \, ds. 
\eee

Now we go back to the proof of (\ref{obut}).
Define for $R>0$ a stopping time
\footnote{We can not define a stopping time in a standard way 
$\hat\tau_R=\inf\{\tau\geq 0: \, \max\limits_{m\in\CC}|a^\eps_m(\tau)|^2\vee |u_m(\tau)|^2 \geq R\}$ 
since $\PR(\hat\tau_R\leq T)\ra 0$ as $R\ra\infty$ non-uniformly in $\CC$. To overcome this difficulty we introduce a stopping time $\tau_R$ which admits a slow growth of the absolute values $|a^\eps_m|,|u_m|$ with respect to $|m|$. 
}
\bee\label{ostop}
\tau_R=\inf\{\tau\geq 0:\,\exists m\in\CC\mbox{ satisfying } |a^\eps_m(\tau)|^2\vee |u_m(\tau)|^2\geq R(|m|^{1/2}+1)\}.
\eee
 For $L\in\mN$ denote $|w^\eps|_L:=\sum\limits_{|j|\leq L} e^{-|j|} |w^\eps_j|$.  Since for any $s\leq\tau_R$ and $k\in\CC$ satisfying $|k|\leq L+1$ we have 
\bee\label{46832186}
|a^\eps_k(s)|^2\vee|u_k(s)|^2\leq R((L+1)^{1/2} + 1) \leq  C(R)\sqrt{L},
\eee 
estimates (\ref{o375}) and (\ref{odif}) imply that for $0\leq \tau\leq T$ we have
\begin{align}\nonumber
\MO |w^\eps(\tau\wedge\tau_R)|_L &\leq |\del^\eps|_L + C \sum\limits_{|j|\leq L}e^{-|j|}\MO\int\limits_0^{\tau\wedge\tau_R}\Big(1+\sum\limits_{k:|k-j|\leq 1}(|a^\eps_k|^2+|u_k|^2)\Big) \sum\limits_{k:|k-j|\leq 1} |w^\eps_k|\, ds \\ 
\nonumber
&\leq |\del^\eps|_L+ C_1(R) \sqrt{L}\,\MO\int\limits_0^{\tau\wedge\tau_R} \sum\limits_{|j|\leq L} \sum\limits_{k:|k-j|\leq 1} e^{-|j|} |w^\eps_k|\,ds \\ 
\nonumber
&\leq |\del^\eps|_L+ C_2(R) \sqrt{L}\,\MO\int\limits_0^{\tau\wedge\tau_R} \Big( |w^\eps|_L  + 
 e^{-L}\sum\limits_{|k|= L+1} |w^\eps_k|\Big)  \, ds \\ 
\label{o6854}
&\leq |\del^\eps|_L +  C_2(R) \sqrt{L} \int\limits_0^{\tau}\Big( \MO |w^\eps(s\wedge\tau_R)|_L  + e^{-L}\sum\limits_{|k|= L+1} |w^\eps_k(s\wedge\tau_R)|\Big) \, ds.
\end{align}
Due to (\ref{46832186}), for $k\in\CC$ satisfying $|k|=L+1$ we have
$|w^\eps_k(s\wedge\tau_R)|\leq|a^\eps_k(s\wedge\tau_R)|+|u_k(s\wedge\tau_R)|\leq C(R)L^{1/4}.$
Since the cardinality of the set $\{k:|k|=L+1\}$ is bounded by $CL^{d-1}$, we get
$$
\sum\limits_{|k|= L+1} |w^\eps_k(s\wedge\tau_R)|\leq C(R)L^{d-3/4}.
$$
By the Gronwall lemma, relation (\ref{o6854}) joined with the inequality above implies  
$$
\MO |w^\eps(\tau\wedge\tau_R)|_L\leq (|\del^\eps|_L + C(R)e^{-L}L^{d-3/4})e^{C(R)\sqrt{L}\tau}.
$$
In view of estimate (\ref{o3854}), we have $|\del^\eps|_L\ra 0$ as $\eps\ra 0$ uniformly in $\CC$, for each $L\in\mN$. Letting $L\ra \infty$ and $\eps\ra 0$ in such a way that $|\del^\eps|_Le^{C(R)\sqrt{L}\tau}\ra 0$, we obtain that $\MO |w^\eps(\tau\wedge\tau_R)|_L\ra 0$ uniformly in $\CC$. 
In particular, for each $k\in\CC$ and any $0\leq \tau\leq T$ we have
\bee\label{o4856}
\MO |w^\eps_k(\tau\wedge\tau_R)|\ra 0 \assN.
\eee
Inequality (\ref{o375}) jointly with estimate (\ref{odif}) implies 
\bee\label{o35235}
\MO \max\limits_{0\leq\tau\leq T}|w_j^\eps(\tau\wedge\tau_R)| \leq \del_j^\eps+C(R)(\sqrt{|j|}+1)\int\limits_0^{T} \MO \sum\limits_{k:|k-j|\leq 1} |w^\eps_k(s\wedge \tau_R)|\, ds.
\eee
Since the function under the time integral in (\ref{o35235}) is bounded by a constant $C(R,j)$, in view of (\ref{o4856}) the dominated convergence theorem implies that this integral tends to zero as $\eps\ra 0$, uniformly in $\CC$. Then 
in view of (\ref{o3854}) for every $j\in\CC$ we have 
\bee\label{o5743}
\MO \max\limits_{0\leq\tau\leq T}|w_j^\eps(\tau\wedge\tau_R)| \ra 0 \assN.
\eee
To recover convergence (\ref{obut}) from (\ref{o5743}) we will need the following proposition. 
\bpp\label{oprop:tau_R}
(i) $\PR(\tau_R\leq T)\ra 0$ as $R\ra \infty$ uniformly in $\CC$. 

(ii) For all $j\in\CC$ we have 
$\MO \max\limits_{0\leq\tau\leq T} e^{\al |u_j(\tau)|^2}\leq C,$
where $\al,C>0$ are the constants from Lemma \nolinebreak \ref{olem:est}. 
\epp
 Using the Cauchy-Schwarz inequality, we find 
\begin{align}
\nonumber
\MO \max\limits_{0\leq\tau\leq T}|w_j^\eps(\tau)| &=
\MO \mI_{(\tau_R\geq T)}\max\limits_{0\leq\tau\leq T}|w_j^\eps(\tau\wedge\tau_R)| + 
\MO \mI_{(\tau_R\leq T)}\max\limits_{0\leq\tau\leq T}|w_j^\eps(\tau)| \\ \nonumber
&\leq \MO \max\limits_{0\leq\tau\leq T} |w_j^\eps(\tau\wedge\tau_R)| 
+\big(\PR(\tau_R\leq T)\big)^{1/2} \big(\MO \max\limits_{0\leq\tau\leq T} |w_j^\eps(\tau)|^2\big)^{1/2}.
\end {align}
Proposition \ref{oprop:tau_R}.{\it ii} joined with the estimate of Lemma \ref{olem:est}.{\it i} implies that $\MO \max\limits_{0\leq\tau\leq T} |w_j^\eps(\tau)|^2 < \nolinebreak C$. Then, in view of (\ref{o5743})  and Proposition \ref{oprop:tau_R}.{\it i}, we arrive at (\ref{obut}).
\qed
\ssk

{\it Proof of Proposition \ref{oprop:tau_R}.}
{\it (i)} 
We have 
\begin{align}
\nonumber
\PR(\tau_R\leq T)&=\PR\big(\exists j\in\CC:\;\max\limits_{0\leq \tau \leq T}|a^\eps_j(\tau)|^2\vee |u_j(\tau)|^2\geq R(|j|^{1/2}+1)\big) \\\nonumber
&=\PR\big(\exists j\in\CC:\; \max\limits_{0\leq \tau \leq T}|a^\eps_j(\tau\wedge\tau_R)|^2\vee |u_j(\tau\wedge\tau_R)|^2\geq R(|j|^{1/2}+1)\big) \\\nonumber
&\leq \sum\limits_{j\in\CC}\PR\big(\max\limits_{0\leq \tau\leq T }|a^{\eps}_j(\tau\wedge\tau_R)|^2\geq R(|j|^{1/2}+1)\big) \\
\nonumber
&+  \sum\limits_{j\in\CC}\PR\big(\max\limits_{0\leq \tau\leq T }  |u_j(\tau\wedge\tau_R)|^2\geq R(|j|^{1/2}+1)\big).
\end{align}
In view of the estimate 
$\MO \max\limits_{0\leq\tau\leq T} e^{\al |a_j^{\eps}(\tau\wedge \tau_R)|^2}\leq \MO \max\limits_{0\leq\tau\leq T} e^{\al |a_j^{\eps}(\tau)|^2}<C$ provided by Lemma \nolinebreak\ref{olem:est}, convergence (\ref{o5743}) 
joined with the Fatou lemma implies
\bee\label{ovRR}
\MO \max\limits_{0\leq\tau\leq T} e^{\al |u_j(\tau\wedge\tau_R)|^2}\leq C, \quad j\in\CC.
\eee
 Then, using the Chebyshev inequality, we obtain
\bee\nonumber
\PR(\tau_R\leq T)\leq C\sum\limits_{j\in\CC}e^{-\al R (|j|^{1/2}+1)} \leq C\sum\limits_{j\in\mZ^d}e^{-\al R (|j|^{1/2}+1)} \ra 0 \quad\mbox{as}\quad R\ra\infty \quad\mbox{uniformly in $\CC$.}
\eee
{\it (ii)}
The desired estimate follows from item {\it (i)} joined with (\ref{ovRR}) and the Fatou lemma.
\qed
\ssk

{\it Proof of Proposition \ref{oprop:razdvwym}.} Since components of each of the vectors $\bm\beta$ and $\bm\beta^\eps$ are independent, it suffices to prove that 
$\DD(\bm\beta_j^\eps(\cdot))\raw\DD(\bm\beta_j(\cdot))$ as $\eps\ra 0$ in $C([0,T],\mC)$ for each $j\in\CC$.
In a standard way it can be shown that the set of measures $\{\DD(\bm\beta_j^\eps(\cdot)), \, 0<\eps\leq 1\}$ is tight.  Let $Q$ be its limiting point as $\eps_k\ra 0$. Take a process $\hat{\bm\beta_j}(\tau)$ such that $\DD(\hat{\bm\beta_j}(\cdot))=Q$. 
We need to show that $\hat{\bm\beta_j}$ is a standard complex Brownian motion. 
Since the processes $\bm\beta_j^\eps$ are uniformly in $\eps$ square-integrable martingales,  the process $\hat{\bm\beta_j}$ also is. Then it suffices to establish that 
$$
[\hat\beta_1]_\tau=[\hat\beta_2]_\tau=\tau \quad\mbox{and}\quad [\hat\beta_1,\hat\beta_2]_\tau\equiv 0,
$$
where $\hat\beta_1=\Ree\hat{\bm\beta_j}$, $\hat\beta_2=\Imm\hat{\bm\beta_j}$, $[\cdot]_\tau$ denotes the quadratic variation and $[\cdot,\cdot]_\tau$ stands for the cross-variation. We will only show that $[\hat\beta_1]_\tau=\tau$, proofs of the other assertions are similar. Due to the definition of the quadratic variation, for this purpose it suffices to prove that the process $\ga(\tau):=\big(\hat\beta_1(\tau)\big)^2-\tau$ is a martingale. Denote 
$\ga^{\eps_k}(\tau):=\big(\beta_1^{\eps_k}(\tau)\big)^2-\tau$, where $\beta_1^{\eps_k}:= \Ree\bm\beta_j^{\eps_k}$. 
Clearly, we have
\bee\label{orw1}
\DD(\ga^{\eps_k}(\cdot)) \raw \DD (\ga(\cdot)) \quad\mbox{as}\quad k\ra\infty.
\eee
On the other hand, since $[\beta_1^{\eps_k}]_\tau=2\int\limits_0^\tau \cos^2(\eps_k^{-1}s) \, ds $, 
\bee\label{orw2}
\ga^{\eps_k}(\tau)= M^{\eps_k}(\tau) + 2\int\limits_0^\tau \cos^2(\eps_k^{-1}s) \, ds-\tau,
\eee
where $M^{\eps_k}(\tau)$ is a square-integrable martingale.
Since $\max\limits_{0\leq\tau\leq T} \big|2\int\limits_0^\tau \cos^2(\eps_k^{-1}s) \, ds-\tau \big|\ra 0$ as $k\ra\infty$, (\ref{orw2}) jointly with (\ref{orw1}) implies that
$$
\DD(M^{\eps_k}(\cdot))\raw  \DD (\ga(\cdot))  \quad\mbox{as}\quad k\ra\infty,
$$
so that $\ga(\tau)$ is a square-integrable martingale as well.
\qed
\msk

\subsection{Proof of Theorem \ref{otheo:sm}.}
\label{osec:rape3} 

{\it Item (i)}. Let $u^\eps(\tau)$ and $u(\tau)$  be stationary solutions of eq. (\ref{oslow}) and the effective equation (\ref{oeff}) correspondingly, $\DD(u^\eps(\tau))\equiv \mu^\eps$ and $\DD(u(\tau))\equiv \mu$. 
Take a function $h\in\LL_b(\mC^{|\CC|})$. It suffices to show that 
\bee\label{ostres1}
\MO h(u^\eps(\tau))\ra\MO h(u(\tau))\quad \mbox{as} \quad \eps\ra 0.
\eee
Let us pass to the $a$-variables $a^\eps(\tau)$, corresponding to the process $u^\eps(\tau)$.
Consider the space $C([0,\infty),\mC^{|\CC|})$ provided with the topology of uniform convergence on finite time intervals, which is given by the metrics
\bee\label{ometric}
\rho(v,w)=\sum\limits_{K=1}^\infty2^{-K}\sup\limits_{0\leq\tau\leq K} |v(\tau)-w(\tau)|\wedge 1.
\eee
\begin{prop}
\label{olem:tight} 
The set of laws $\DD(a^\eps(\cdot))$, $0<\eps< \eps_0$, 
\footnote{Recall that $\eps_0$ is defined in Lemma \ref{olem:est}.}
is tight in $C([0,\infty),\mC^{|\CC|})$. 
\end{prop}
{\it Proof.} 
It suffices to establish that the set of measures $\DD(a^{\eps}(\cdot))$ is tight in  $C([0,T],\mC^{|\CC|})$ for any $T\geq 0$.
The latter follows  in a standard way from the estimates of Lemma \ref{olem:est}, equation (\ref{oa}) and Arzela-Ascoli theorem. 
\qed

Consider some limiting point
\bee
\label{oQ}
\DD(a^{\eps_k}(\cdot))\raw Q \quad \mbox{as} \quad \eps_k\ra 0. 
\eee
It turns out that the measure $Q$ is a law of a weak solution of the effective equation:
\bpp\label{oprop:Q}
The measure $Q$ coincides with the law $\DD(u^0(\cdot))$ in the space  $C([0,\infty),\mC^{|\CC|})$, where $u^0(\tau)$ is a weak solution of the effective equation (\ref{oeff}). 
\epp
Proposition \ref{oprop:Q} can be established by argument similar to that used in the proof of Theorem \ref{otheo:a}.
\footnote{Note that this argument does not lead to the uniformity in $\CC$ of convergence (\ref{oQ}). Indeed, in the present case the distributions of initial conditions $\DD(a^{\eps_k}(0))$ are different for different $\eps_k$, so we should add in (\ref{o6854}) the term $\MO |w^{\eps_k}(0)|_L$. We do not know if it convergences to zero  as $\eps_k\ra 0$ uniformly in $\CC$. }
We do not prove it here since below we will establish an analogous result in a more complicated, infinite-dimensional setting (see Proposition \ref{oprop:Qinfty}). 
Proposition \ref{oprop:Q} provides that for any $\tau\geq 0$ we have
$\MO h  (a^{\eps_k}(\tau)) \ra \MO h  ( u^{0}(\tau))$ as $\eps_k\ra 0.$
Jointly with Corollary \ref{olem:lem'} this implies 
\bee\label{ovkv0}
\MO h(u^{\eps_k}(\tau))\ra \MO  h (u^0(\tau))\quad \mbox{as} \quad \eps_k\ra 0.
\eee
Since the process $u^\eps$ is stationary, (\ref{ovkv0}) implies that the process $u^0(\tau)$ also is, so that
$\DD(u^0(\tau))\equiv\DD(u(\tau))\equiv\mu$. Thus, we get (\ref{ostres1}).

{\it Item (ii)}. Assume that condition {\it H$\CC_\infty$} is satisfied. Recall that the uniformity in $\CC$ of the weak convergences of measures through all the text is understood in the sense of finite-dimensional projections. That is, we need to prove that for any  bounded set $\La\subset \mZ^d$ and a function $h\in\LL_b(\mC^{|\CC|})$ satisfying $\Supp h\subseteq\La$, convergence (\ref{ostres1}) holds uniformly in $\CC$ satisfying $\CC\supseteq\La$. For this purpose it suffices to show that for any sequence of bounded sets $(\CC_n)_{n\in\mN}$, $\CC_n\subset \mZ^d$ satisfying $\CC_n\supseteq\La$ for each $n\in\mN$, there exists a subsequence $(\hat\CC_{n})_{n\in\mN}$ such that the desired convergence holds uniformly in $\CC\in(\hat\CC_{n})_{n\in\mN}$. 

The proof follows a scheme below. As $(\hat\CC_n)_{n\in\mN}$ we take an arbitrary subsequence which has a "limit" as $n\ra\infty$, say $\hat\CC_\infty$. Ass. {\it H$\CC_\infty$} provides that the corresponding $\hat\CC_\infty$-effective equation has a unique stationary measure in $\MM^{\hat\CC_\infty}$. Then, arguing as in item {\it (i)}, we establish convergence (\ref{o1456}), where  $u^{\eps,\hat\CC_n}$ denotes a stationary solution of eq. (\ref{oslow}) with $\CC=\hat\CC_n$, while $u^\infty$ denotes that of the $\hat\CC_\infty$-effective equation. This provides control of the expectations $\MO h(u^{\eps,\hat\CC_n}(\tau))$ for large $n$ and small $\eps$, sufficient to prove the desired uniformity of convergence.  
Now let us present the rigorous proof. 
Denote by $B_N$ a ball in $(\mZ^d,|\cdot|)$ of the radius $N$ centred at zero, where $|\cdot|$ denotes the $l_1$-norm in $\mZ^d$.
\bpp\label{oprop:Cn}
Let $(\CC_n)_{n\in\mN}$ be  a sequence of bounded sets in $\mZ^d$. Then there exists a subsequence $(\hat\CC_{n})_{n\in\mN}$  and a set $\hat\CC_\infty\subseteq\mZ^d$ such that $\hat\CC_{n}\ra\hat\CC_\infty$ as $n\ra\infty$ in the sense that for any $N\in\mN$ and all 
$n\geq N$ we have 
$\hat\CC_n\cap B_N=\hat\CC_\infty\cap B_N$.
\epp
{\it Proof.}
Since the number of sites in $B_1$ is finite, there exists a subsequence $(\CC_{n_{n_1}})_{n_1\in\mN}$ such that for every $j\in B_1$ we have either $j\in\CC_{n_{n_1}}$ or $j\notin\CC_{n_{n_1}}$  at the same time for all $n_1$. Then the set $\CC_{n_{n_1}}\cap B_1$ is independent from $n_1$. Similarly, we find a subsequence $(\CC_{n_{{n_1}_{n_2}}})_{n_2\in\mN}$  such that the set $\CC_{n_{{n_1}_{n_2}}}\cap B_2$ is independent from $n_2$. We continue the procedure and apply the diagonal process to choose a subsequence 
$(\hat\CC_n)_{n\in\mN}=\CC_{n_1},\CC_{n_{{n_1}_2}},\CC_{n_{{n_1}_{{n_2}_3}}},\ldots$.
Then for any $N\in\mN$ the sets $\hat\CC_n\cap B_N$ coincide for all $n\geq N$. Let 
$\hat\CC_\infty:=\cup_{n\in\mN}(\hat\CC_n\cap B_n).$ Clearly, the pair $(\hat\CC_n)_{n\in\mN}$ and $\hat\CC_\infty$ 
satisfies the assertion of the proposition.
\qed
\ssk

Choose a subsequence $(\hat\CC_n)_{n\in\mN}$ of the sequence $(\CC_n)_{n\in\mN}$ as in Proposition \ref{oprop:Cn}. 
Let $u^{\eps, \hat\CC_n}(\tau)$ be a stationary solution of eq. (\ref{oslow}) with $\CC=\hat\CC_n$, and $u^{\hat\CC_n}(\tau)$ be that of the effective equation (\ref{oeff}).
We need to prove that convergence (\ref{ostres1}) holds uniformly in $\CC\in (\hat\CC_n)_{n\in\mN}$, i.e. to show that for any $\del>0$ there exists $\hat\eps>0$ such that for any $\eps<\hat\eps$ and all $n\in\mN$ we have 
\bee
\label{oras}
|\MO h(u^{\eps,\hat\CC_n}(\tau))-\MO h(u^{\hat\CC_n}(\tau))|< \delta. 
\eee
Denote $u^{\eps,n}=(u^{\eps,n}_j)_{j\in\hat\CC_\infty}$, where 
\bee
\label{oprodolzhenie}
u_j^{\eps,n}:=
\left\{
\begin{array}{cl}
u_j^{\eps,\hat\CC_n},\quad &\mbox{if}\quad j\in\hat\CC_\infty\cap\hat\CC_n, \\
0, \quad &\mbox{if}\quad j\in\hat\CC_\infty\sm\hat\CC_n.
\end{array}
\right.
\eee 
Let $a^{\eps,n}=(a^{\eps,n}_j)_{j\in\hat\CC_\infty}$ be the corresponding $a$-variables.
Consider the space $C([0,\infty),\mC^{|\hat\CC_\infty|})$ provided with the topology of uniform convergence on finite time intervals  which is given by the metrics (\ref{ometric}), where the distance $|v(\tau)-w(\tau)|$ is replaced by 
$$
\rho_{\mC^{|\hat\CC_\infty|}}(v(\tau),w(\tau))=\sum\limits_{j\in\hat\CC_\infty}2^{-|j|}|v_j(\tau)-w_j(\tau)|\wedge 1.
$$
Using the uniformity in $\CC$ of estimates from Lemma \ref{olem:est} we get that the set of laws $\{\DD(a^{\eps,n}(\cdot)),\; 0<\eps\leq 1,\, n\in\mN\}$ is tight in the space $C([0,\infty),\mC^{|\hat\CC_\infty|})$. Let $Q^{\infty}$ be its limiting point as $\eps_k\ra 0,\,n_k\ra\infty.$  
\bpp\label{oprop:Qinfty}
In the space  $C([0,\infty),\mC^{|\hat\CC_\infty|})$ the measure $Q^\infty$ coincides with the law $\DD(u^{\infty}(\cdot))$, where $u^{\infty}(\tau)$ is a weak solution of the $\hat\CC_\infty$-effective equation. For any $j\in\hat\CC_\infty$, $0<\la\leq 1$ and $\tau\geq 0$ it satisfies 
\bee\label{nwoigwigr}
\MO \max\limits_{s\in[\tau,\tau+1]}e^{\al|u_j^\infty(s)|}\leq C.
\eee
\epp
Before presenting the proof of the proposition, let us finish the proof of the theorem.
Since $\La\subseteq\hat\CC_n$ for all $n\in\mN$, we have $\La\subseteq\hat\CC_\infty$. Then, due to the inclusion $\Supp h\subseteq\La$, the function $h$ can be considered as $h:\mC^{|\hat\CC_\infty|}\mapsto\mR$. 
Proposition \ref{oprop:Qinfty} implies that $\MO h(a^{\eps_k,n_k}(\tau))\ra\MO h(u^\infty(\tau))$ as $\eps_k\ra 0, n_k\ra\infty$. Since, due to Corollary \ref{olem:lem'}, we have $\MO h(a^{\eps_k,n_k}(\tau))-\MO h(u^{\eps_k,n_k}(\tau))\ra 0$ as $\eps_k\ra 0, n_k\ra\infty$, we get
\bee\label{o385}
\MO h(u^{\eps_k,n_k}(\tau))\ra\MO h(u^\infty(\tau))\quad \mbox{as} \quad \eps_k\ra 0, n_k\ra\infty,
\eee
so that $u^\infty$ is a stationary process. Due to (\ref{nwoigwigr}), the stationary distribution $\DD(u^\infty(\tau))$
belongs to $\MM^{\hat\CC_\infty}$.  
Then, in view of ass. {\it H$\CC_\infty$}, it is defined uniquely, so that the limit in (\ref{o385}) does not depend on the sequence $(\eps_k,n_k)_{k\in\mN}$ and holds as $\eps\ra 0$, $n\ra\infty$. 
Since  $\Supp h \subseteq\La\subseteq \hat\CC_\infty\cap \hat \CC_n$ for all $n\in\mN$, we have $h(u^{\eps,n}(\tau))\equiv h(u^{\eps,\hat\CC_n}(\tau))$, so that
\bee\label{o1456}
\MO h(u^{\eps,\hat\CC_n}(\tau))\ra\MO h(u^\infty(\tau))\quad \mbox{as} \quad \eps\ra 0, n\ra\infty.
\eee
Due to item {\it (i)} of the theorem, for each $n\in\mN$ we also have 
\bee\label{o6838}
\MO h(u^{\eps,\hat\CC_n}(\tau))\ra\MO h(u^{\hat\CC_n}(\tau)) \ass \eps\ra 0. 
\eee
Jointly with (\ref{o1456}) this implies that
$\MO h(u^{\hat\CC_n}(\tau))\ra \MO h(u^{\infty}(\tau))$  when $n\ra\infty$.
Consequently, for any $\delta>0$ there exist $N\in\mN$ and $\eps_1>0,$ such that for every $n\geq N$ and $\, 0<\eps < \eps_1$, we have 
$$
|\MO h(u^{\eps,\hat\CC_n}(\tau))-\MO h(u^{\infty}(\tau))|,\;|\MO h(u^{\hat\CC_n}(\tau))-\MO h(u^{\infty}(\tau))|< \delta/2.
$$  
Then for $n$ and $\eps$ as above, we get (\ref{oras}).
In view of (\ref{o6838}), we can choose $\eps_2>0$ such that (\ref{oras}) is also satisfied for every $0<\eps<\eps_2$ and $n<N$.
Then it holds for all $n\in\mN$ and $\eps<\eps_1\wedge\eps_2$.
\qed
\ssk

{\it Proof of Proposition \ref{oprop:Qinfty}.} 
Before starting the proof let us make the following remark. Since 
$\Supp P_j,\Supp R_j\subseteq \{k\in\mZ^d: |k-j|\leq 1\} \cap \{k\in\CC\}$,  the functions
\footnote{We recall that the functions $P_j$ and $\RR_j$ are defined in  (\ref{oPj}) and  (\ref{oRR}) correspondingly}
 $P_j$ and $\RR_j$ depend on the choice  of the set $\CC$. 
To indicate this, we denote  those corresponding to the set $\hat\CC_{n_k}$  as $P_j^k$, $\RR_j^k$ and  those corresponding to the set $\hat\CC_\infty$ as $P_j^\infty$, $\RR_j^\infty$. 
Now we start the proof. For shortness we write $a^k:=a^{\eps_k,n_k}$.
Consider the process $b^k=(b^k_j)_{j\in\hat\CC_\infty}$, where 
\bee\nonumber
b_j^{k}(\tau)= a_j^{k}(\tau) -a_j^{k}(0) - \int\limits_0^\tau \RR^\infty_{j}(a^{k}(s))  \, ds, \quad j\in\hat\CC_\infty.
\eee 
Consider also a process $u^\infty(\tau)$ satisfying $\DD(u^\infty(\cdot))=Q^\infty$, and put $b=(b_j)_{j\in\hat\CC_\infty}$, where
$$
b_j(\tau)= u^\infty_j(\tau) -u^\infty_j(0) - \int\limits_0^\tau \RR^\infty_{j}(u^\infty(s))  \, ds, \quad j\in\hat\CC_\infty.
$$
The convergence $\DD(a^k(\cdot))\raw Q^\infty$ as $k\ra\infty$ implies
\bee\label{obb465}
\DD(b^k(\cdot))\raw \DD(b(\cdot)) \ass k\ra\infty \quad \mbox{on}\quad C([0,\infty),\mC^{|\hat\CC_\infty|}).
\eee
On the other hand, according to (\ref{oprodolzhenie}) and eq. (\ref{oa}), for  $j\in\hat\CC_\infty\cap\hat\CC_{n_k}$ we have
\bee\label{objktau}
b_j^{k}(\tau) = \Theta_j^{k}(\tau)+\hat{\bm\beta}^k_j(\tau),
\eee
where $\hat{\bm\beta^k}=(\hat{\bm\beta}^k_j)_{j\in\hat\CC_\infty}$, $\hat{\bm\beta}^k_j(\tau):=\sqrt{2\TT_j}\int\limits_0^\tau e^{-i\eps_k^{-1}s}\, d\beta_j$, 
$(\beta_j)_{j\in \hat\CC_\infty}$ are standard real independent Brownian motions, and
$$
\Theta_j^{k}(\tau)= \int\limits_0^\tau \big( e^{-i\eps_k^{-1}s}P^k_j(e^{i\eps_k^{-1}s} a^{\eps_k,\hat\CC_{n_k}}(s)) - \RR^\infty_{j}(a^{k}(s)) \big) \,ds,
$$
where $a^{\eps_k,\hat\CC_{n_k}}$ denotes the $a$-variables corresponding to the process $u^{\eps_k,\hat\CC_{n_k}}$.
Since  $\hat\CC_{n_k}\ra\hat\CC_\infty$ as $k\ra\infty$,
for sufficiently large $k$ (depending on $j$) we have $P_j^k=P_j^\infty$ and
$\RR_j^\infty(a^k(s))\equiv \RR_j^\infty(a^{\eps_k,\hat\CC_{n_k}}(s))$.  
Then, applying Lemma \ref{olem:lem} to the function $h(u,\tht)=e^{-i\tht}P^\infty_j(e^{i\tht} u)$, we get
\bee\nonumber
\MO\max\limits_{0\leq \tau\leq T} |\Theta_{j}^{k}(\tau) |\rightarrow 0 \quad \mbox{as}\quad k\ra\infty\quad\mbox{for every }T\geq 0.
\eee
Moreover, Proposition \ref{oprop:razdvwym} implies that $\DD(\hat{\bm\beta}^k(\cdot))\raw \DD(\hat{\bm\beta}(\cdot))$ as $k\ra\infty$ in $C([0,\infty),\mC^{|\hat\CC_\infty|})$, where $\hat{\bm\beta}=(\hat{\bm\beta}_j)_{j\in\hat\CC_\infty}$, 
$\hat{\bm\beta}_j=\sqrt{\TT_j}\bm\beta_j$, and $(\bm\beta_j)_{j\in\hat\CC_\infty}$ are standard complex independent Brownian motions.
Then from (\ref{objktau}) we obtain
$$
\DD(b^k(\cdot))\raw \DD(\hat{\bm\beta}(\cdot)) \ass k\ra\infty \quad \mbox{on}\quad C([0,\infty),\mC^{|\hat\CC_\infty|}).
$$ 
Thus, due to (\ref{obb465}), we have
$
\DD(b(\cdot))=\DD(\hat{\bm\beta}(\cdot)), 
$
so that $u^\infty(\tau)$ is a weak solution of the $\hat\CC_\infty$-effective equation.
Estimate (\ref{nwoigwigr}) follows from the Fatou lemma and Lemma \ref{olem:est}.
\qed
\ssk

\subsection{The $\CC_\infty$-effective equation} 
\label{osec:rape4}

In this section we prove that under conditions of Proposition \ref{oprop:ru} the $\CC_\infty$-effective equation defines a Markov process and is mixing, in a suitable space. In particular, this will imply Proposition \ref{oprop:ru}. Denote
$$
\|u\|_0^2:=\sum\limits_{k\in\wid\CC}\ga^{|k|}|u_k|^2,
$$
where $u\in\mC^{|\wid\CC|}$ and $\wid\CC=\CC$ or $\wid\CC=\CC_\infty$ (the choice will be clear from the context). Fix the constant $1/2< \ga<1$ as in Lemma \ref{olem:est1}. Consider the space 
$$
\XX:=\{u\in\mC^{|\CC_\infty|}:\, \|u\|_0<\infty\},
$$
and denote by $\PP(\XX)$ a space of Borel probability measures on $\XX$. 
\bpp\label{oprop:ru'}
Assume that conditions of Proposition \ref{oprop:ru} are fulfilled. 
Then for any unbounded set $\CC_\infty\subseteq \mZ^d$ the corresponding $\CC_\infty$-effective equation defines a Markov process in $\XX$ and is mixing. That is,  it has a unique stationary measure $\mu^\infty$ in the class of measures $\PP(\XX)$, and for any its solution $u(\tau)$ satisfying $\PR(u(\tau)\in\XX \mbox{ for all }\tau\geq 0)=1$ we have the convergence $\DD(u(\tau))\raw\mu^\infty$ as $\tau\ra\infty$.
\epp
In order to prove Proposition \ref{oprop:ru'} we will need the following
\bpp\label{oprop:est}
Let $u(\tau)$ be a solution of the effective equation (\ref{oeffH}) satisfying \\
$\MO e^{\del_0\|u(0)\|_0^2}<C$ for some $\del_0>0$. Then there exists $\del>0$ such that for any $0<\la\leq 1$ and $\tau\geq 0$ we have 
\bee\nonumber
\MO\max\limits_{s\in[\tau,\tau + 1]} e^{\del \|u(s)\|_0^2}\leq C_1(\ga,\del,C).
\eee
\epp
{\it Proof.} Note that the norm $\|\cdot\|_0$ restricted to $\mR^{|\CC|}$ coincides with the norm from (\ref{decrnorm}) with $j=0$. Let $u=p+iq$.  Since $\|u\|_0^2=\|p\|_0^2+\|q\|_0^2$, the upper bound from Proposition \nolinebreak \ref{olem:est2} implies that  
$\MO e^{\hat\del_0 U^0(u(0))}<C$ for some $\hat\del_0>0$, where $U^0(u):=U^0(p(u),q(u)).$ 
\footnote{See (\ref{oUj}) for the definition of $U^0.$}
Consider a solution $u^\eps(\tau)$ of eq. (\ref{oslow}) satisfying $\DD(u^\eps(0))=\DD(u(0))$. 
Due to Lemma \ref{olem:est1}, there exists $\hat\del>0$ such that
$$
\MO\max\limits_{s\in[\tau,\tau + 1]} e^{\hat\del U^0(u^\eps(s))}<C_1,
$$
for any $0<\eps<\eps_0$, $0<\la\leq 1$ and $\tau\geq 0$. 
Due to the lower bound from Proposition \ref{olem:est2}, we have $U^0(u)\geq \|u\|_0^2/4$, so that 
$$\MO\max\limits_{s\in[\tau,\tau + 1]} e^{\del \|u^\eps(s)\|_0^2}\leq C_1,$$
where $\del:=\hat\del/4.$ Then Theorem \ref{otheo:introav} joined with the Fatou lemma implies the desired estimate.
\qed

{\it Proof of Proposition \ref{oprop:ru'}.}
The proof is divided into three steps. In the first one we show that if $\la$ is sufficiently small, then  the $\CC_\infty$-effective equation is contracting in the space $\XX$. In the second step we prove that the $\CC_\infty$-effective equation is well-posed and defines a Markov process in $\XX$. In the third one we show that the $\CC_\infty$-effective equation has a stationary measure from $\PP(\XX)$, prove that the latter is unique, and obtain the desired convergence.
  
{\it Step 1.} Let $u^1(\tau)$ and $u^2(\tau)$ be two solutions of the $\CC_\infty$-effective equation having deterministic initial conditions $u_{0}^1, u_{0}^2\in\XX$ and corresponding to the same Brownian motion $(\bm\beta_j)_{j\in\CC_\infty}$.
Let $w:=u^1-u^2$. 
Since the second partial derivatives of the potential $V_{kj}^{res}$ are bounded, for all $j\in\CC_\infty$ we have
\bee\label{o8632}
\frac{d}{d\tau}\frac{|w_j|^2}{2}\leq \la C\sum\limits_{k:|k-j|\leq 1} |w_k||w_j| - \frac{|w_j|^2}{2},  \quad\mbox{a.s.,}
\eee
see (\ref{oeffH}).
Multiplying the both sides of (\ref{o8632}) by $\ga^{|j|}$ and summing over $j\in\CC_\infty$, we find
$$
\frac{d}{d\tau}\|w\|_0^2\leq (\la C -1) \|w\|_0^2, \quad\mbox{a.s.}
$$
Applying the Gronwall inequality, we get 
\bee\label{ocontra}
\|w(\tau)\|_0^2 \leq\|w(0)\|_0^2 e^{(\la C -1)\tau}, \quad\mbox{a.s.}
\eee
Assume $\la<1/C$. Then  
\bee\label{o5468}
\|w(\tau)\|_0\ra 0 \ass \tau\ra\infty, \quad\mbox{a.s.}
\eee

{\it Step 2.} Let us show first that the $\CC_\infty$-effective equation admits a weak solution in the space $\XX$. Let $u^\infty_0\in\XX$ be a deterministic initial data. Take a sequence of sets $\CC_n:=\CC_\infty\cap\{j\in\mZ^d: |j|\leq n\}$. Let $u^{\CC_n}(\tau)$ be a solution of the effective equation (\ref{oeffH}) with $\CC=\CC_n$ and initial conditions 
$u_0^{\CC_n}\in\mC^{|\CC_n|}$
satisfying $u_{0j}^{\CC_n}=u_{0j}^{\infty}$ for every $j\in\CC_n$. 
We define the process $u^{n}=(u^{n}_j)_{j\in\CC_\infty}$, where  $u_j^n=u_j^{\CC_n},$ if $j\in\CC_n$, and $u_j^n=0$, if $j\in\CC_\infty\sm\CC_n$. 
Proposition \ref{oprop:est} joined with the Arzela-Ascoli theorem implies
\bpp\nonumber
The set of measures $\{\DD(u^n(\cdot)),\;n\in\mN\}$ is tight in $C([0,\infty),\mC^{|\CC_\infty|})$. 
\epp
Let 
\bee\label{oignoswg}
\DD(u^{n_k}(\cdot))\raw Q^\infty \ass k\ra\infty.
\eee
Take a process $u^\infty(\tau)$ satisfying $\DD(u^\infty(\cdot))=Q^\infty$.
It is possible to show that $u^\infty(\tau)$ is a weak solution of the $\CC_\infty$-effective equation. Obviously, it satisfies $u^\infty(0)=u_0^\infty$.
Moreover,  Proposition \ref{oprop:est} joined with convergence (\ref{oignoswg}) and Fatou's lemma implies 
that $\MO\max\limits_{s\in [\tau,\tau+1]} e^{\del \|u^\infty(s)\|_0^2}\leq C$, so $\PR(u^\infty(\tau)\in\XX \mbox{ for all }\tau\geq 0)=1$. 

Estimate (\ref{ocontra}) implies the pathwise uniqueness in $\XX$ of solutions for the $\CC_\infty$-effective equation. Jointly with  existence of a weak solution,  
by Yamada-Watanabe arguments (see \cite{Yor,RSZ}), the pathwise uniqueness  implies  existence of a strong solution. 
Then, using arguments from Chapter 7 of \cite{Oks}, it can be shown that the set of solutions corresponding to all possible initial data from $\XX$  form a Markov family.

{\it Step 3}. Let $\msP_\tau(u,\cdot)$ be the corresponding transition function. Convergence (\ref{o5468}) implies that for any $u_0^1,u_0^2\in\XX$ we have 
\bee\nonumber
\|\msP_\tau(u_0^1,\cdot)-\msP_\tau(u_0^2,\cdot)\|_{Lip}^*\ra 0 \ass \tau\ra\infty,
\eee 
where $\|\cdot\|_{Lip}^*$ denotes the dual-Lipschitz norm in $\PP(\XX)$. It follows that, if the $\CC_\infty$-effective equation has a stationary measure $\mu^\infty\in\PP(\XX)$, then it is unique, and for any solution $u(\tau)$ of the $\CC_\infty$-effective equation satisfying $\PR(u(\tau)\in\XX \mbox{ for all }\tau\geq 0)=1$ we have  $\DD(u(\tau))\raw \mu^\infty$ (see {\it Step 1} form the proof of Theorem \nolinebreak 3.1.3 in \cite{KuSh}). To show that the stationary measure exists, we employ an argument similar to that used in the beginning of {\it Step 2} joined with the estimate of Proposition \nolinebreak\ref{olem:estef}.{\it ii}. 
\qed
\ssk

Now let us show that Proposition \ref{oprop:ru'} implies Proposition \ref{oprop:ru}.

{\it Proof of Proposition  \ref{oprop:ru}}. {\it Existence}. Estimates of Proposition \nolinebreak\ref{olem:estef} imply that the stationary measure $\mu^\infty$ of the $\CC_\infty$-effective equation constructed in the proof of Proposition \nolinebreak \ref{oprop:ru'} is regular.

{\it Uniqueness.} Let $\mu^\infty$ be a regular stationary measure of the $\CC_\infty$-effective equation. Then there exists a regular weak solution $u^\infty(\tau)$ of the latter satisfying $\DD(u(\tau))\equiv \mu^\infty$. Clearly, the regularity implies that  $\PR(u^\infty(\tau)\in\XX \mbox{ for all }\tau\geq 0)=1$, so that $\mu^\infty$ is a stationary measure of the Markov process in the space $\XX$, given by the $\CC_\infty$-effective equation. Proposition \ref{oprop:ru'} provides that such measure is unique.
\qed

\appendix
\numberwithin{equation}{section}

\section{Appendix: Uniformity of mixing}
\label{oapp:mixing}

In this appendix we show that eq. (\ref{oini_er}) is exponentially mixing with uniform in $\eps$ rate. Note that for fixed $\eps$ the mixing property of eq. (\ref{oini_er}) is well understood, see e.g. \cite{MSH}, where it is proven for the case of smooth interaction potential $V$.
It is convenient to work in the complex variables $u=p+iq\in\mC^{|\CC|}$, so that instead of  (\ref{oini_er}) we deal with eq. (\ref{oslow}). The proof is based on the fact that the latter, written in the $a$-variables, has uniformly in $\eps$ bounded coefficients (see eq. (\ref{oa})). 

We do not follow dependence of the rate of mixing on the choice of the set $\CC$, so in this section the constants $C,C_1,\ldots$ are permitted to depend on  $\CC$.
\footnote{We do not know if the rate of mixing is uniform with respect to the choice of the set $\CC$. If it is, it seems to be a very complicated problem.  Note that we do not need this uniformity to get in Theorem \ref{otheo:sm} the uniformity of convergence in $\CC$. Indeed, in the proof of the latter we do not use the mixing property but we work directly with stationary solutions.}
Moreover, for simplicity of notations we put $\la=1$.
Let $\fB_{\tau}^{\eps}$ be a Markov semigroup associated with eq. (\ref{oslow}).
\btt\label{otheo:mixing}
Equation (\ref{oslow}) has a unique stationary measure $\mu^\eps$. 
There exist constants $\eps_1>0$ and $C,b>0$ such that for all $0<\eps<\eps_1$ and any Borel probability measure $\rho$ satisfying $\lan\rho, |u|^2 \ran < \infty$ we have
\bee\label{ou1}
\|\fB_{\tau*}^{\eps}\rho-\mu^\eps\|_{var}\leq C\big(1+\lan \rho, |u|^2\ran \big)e^{-b\tau},
\eee
where $\|\cdot\|_{var}$ denotes the variational norm.
\ett
{\it Proof.} In this proof by $B_R$ we denote a closed ball in $\mC^{|\CC|}$ of radius $R$, centred at zero. By $\msP_{\tau}^{\eps}(u,\cdot)$ we denote the transition function of the Markov process (\ref{oslow}).
\ssk

A standard Bogolioubov-Krylov argument implies existence of a stationary measure $\mu^\eps$ for eq. (\ref{oslow}). 
Assume that for any $u_1, u_2\in\CCC$, $0<\eps<\eps_1$ and $\tau\geq 0$ we have
\bee\label{osq}
\|\msP_\tau^{\eps}(u_1,\cdot)-\msP_\tau^{\eps}(u_2,\cdot)\|_{var} \leq C(1+|u_1|^2+|u_2|^2) e^{-b\tau}.
\eee
Using arguments from the introduction to Section 3, and of Sections 3.1.1,3.1.2 from \cite{KuSh}, it can be shown that (\ref{osq}) implies the assertion of the theorem. 
Coupling argument from the introduction to Section 3 of \cite{KuSh} implies that to prove (\ref{osq}) it suffices to show that the Markov process given by eq. (\ref{oslow}) satisfies the following two properties.

{\it Recurrence.} Let $u^\eps(\tau)$ be a solution of eq. (\ref{oslow}) satisfying $u^\eps(0)=u_1$, where $u_1\in\mC^{|\CC|}$. Put 
$$
\tau_R^\eps:=\inf\{\tau\geq 0: u^\eps(\tau)\in B_R\}. 
$$
Then there exist $R,\del>0$ and $\eps_2>0$ such that for any $0<\eps<\eps_2$  we have 
\bee\label{ou2}
\MO e^{\del\tau^{\eps}_R}\leq C(1+|u_1|^2).
\eee

{\it Squeezing.} For any $R>0$ there exists constants $0<\vartheta\leq 1$ and $\eps_3>0$ such that for any  $u_1,u_2\in B_R$ and $0<\eps<\eps_3$ we have
\bee\label{ou3}
\|\msP_1^{\eps}(u_1,\cdot)-\msP_1^{\eps}(u_2,\cdot)\|_{var} \leq 1-\vartheta.
\eee 
We start with the proof of the reccurence.

{\it Proof of the recurrence.} For simplicity of notations we will skip the upper index $\eps$.
Let 
$$
U(u)=H(u)+\frac{\eps}{2}\sum\limits_{j\in\CC} p_jq_j,
$$
where $u=p+iq$, $p=(p_j)_{j\in\CC},\,q=(q_j)_{j\in\CC}$.
Applying Ito's formula to the function $e^{\del\tau}U(u(\tau))$ with $0<\del<1$, we obtain 
\footnote{Here it is simpler to look at eq. (\ref{oini_er}) than at (\ref{oslow}).}
\begin{align}\nonumber
e^{\del \tau}U(u(\tau))&= U(u_1) + 
\int_0^\tau e^{\del s}(\del U+ \eps^{-1}\{U,H\}-\sum\limits_{j\in\CC}p_j \chp_{p_j}U+ \sum\limits_{j\in\CC} \TT_j\chp^2_{p_j^2}U) \,ds 
\\
\label{ou4} 
&+ 
\int_0^\tau \sum\limits_{j\in\CC} e^{\del s}  \sqrt{2\TT_j} \chp_{p_j} U\,d\beta_j.
\end{align}
Using estimate (\ref{oHG2}), we get
\begin{align}\nonumber
\eps^{-1}\{U,H\}&=
\Big\{\frac12 \sum\limits_{j\in\CC} p_jq_j,\sum\limits_{k\in\CC}\frac{p_k^2+q_k^2}{2} \Big\}+
\Big\{\frac12 \sum\limits_{j\in\CC} p_jq_j, \frac{\eps}{2}\sum\limits_{|k-m|= 1} V(q_m,q_k)\Big\} \\ \label{ou4,5}
&\leq
\frac12(|p|^2-|q|^2)+C_1\eps(1+|q|^2).
\end{align}
Moreover, for $\eps$ sufficiently small,
\begin{align}\nonumber
|u|^2/4\leq U(u)\leq C_2(1+|u|^2), &\quad 
-\sum\limits_{j\in\CC}  p_j\chp_{p_j} U  \leq (-1+\eps)|p|^2+ \eps |q|^2, \\
\label{ou5}
 \sum\limits_{j\in\CC} \TT_j\chp^2_{p_j^2}U & =\sum\limits_{j\in\CC}\TT_j\leq C_3.
\end{align}
For $K>0$ put $\xi:=\tau_R^\eps\wedge K$. Then (\ref{ou4}) joined with (\ref{ou4,5}) and (\ref{ou5}) implies  
\bee\label{ou6}
\MO e^{\del\xi} U(u(\xi)) \leq U(u_1)
+\MO\int\limits_0^\xi e^{\del s}(\Del|u|^2+ C_4)\, ds,
\eee
where 
$$
\Del:=\big[(\del C_2+1/2-1+\eps)\vee(\del C_2-1/2+C_1\eps+\eps)\big]=[(-1/2+\del C_2+\eps)\vee(-1/2+\del C_2+\eps(C_1+1))],
$$
and $C_4:=C_2 + C_1+ C_3$. 
Choose $\eps_2$ and $\del$ in such a way that for $\eps\leq\eps_2$ we have $\Del(\eps,\del)<0$, and fix $R\geq\sqrt{-C_4/\Del(\eps_2,\del)}$. Let $s\leq\xi$. Then $|u(s)|\geq R$, so that
$\Del|u(s)|^2+ C_4\leq 0$, for any $\eps\leq\eps_2$.
Thus, (\ref{ou6}) implies 
$$
\MO e^{\del\xi} U(u(\xi)) \leq U(u_1).
$$
Since, in view of the first estimate from (\ref{ou5}), we have $U(u(\xi))\geq |u(\xi)|^2/4\geq R^2/4$, we obtain 
$$\MO e^{\del\xi}\leq 4 U(u_1)/R^2\leq 4C_2(1+|u_1|^2)/R^2.$$
Letting $K\ra\infty$ and applying Fatou's lemma we get (\ref{ou2}).
\qed
\ssk

{\it Proof of the squeezing.}
Denote by $\msP^{a,\eps}(s,\tau;u,\cdot)$ the transition function of the Markov process (\ref{oa}), corresponding to initial conditions $u$ and initial time $s$. Let 
$\msP_{\tau}^{a,\eps}(u,\cdot):=\msP^{a,\eps}(0,\tau;u,\cdot)$.  
In view of the definition of the $a$-variables (\ref{oavar}), we obviously have
\bee\label{os1}
\|\msP_1^{\eps}(u_1,\cdot)-\msP_1^{\eps}(u_2,\cdot)\|_{var}=\|\msP_1^{a,\eps}(u_1,\cdot)-\msP_1^{a,\eps}(u_2,\cdot)\|_{var}.
\eee
Let $\GG\in\BB(\CCC)$. First we will obtain uniform in $\eps$ and $u\in B_R$ estimates from below for the transition probability $\msP_1^{a,\eps}(u,\GG)$. For this purpose we will use Girsanov's theorem (see \cite{Oks}). Let $a^\eps(\tau)$ be a solution of eq. (\ref{oa}) satisfying $a^\eps(0)=u$, where $u\in B_R$, and 
$$
\xi_K^\eps=\inf\{\tau\geq 0:\, |a^\eps(\tau)|\geq K\}. 
$$ 
In order to satisfy Novikov's condition needed for application of Girsanov's theorem (see below), instead of considering the process $a^\eps(\tau)$ we will consider the process $a^{\eps,K}(\tau)$ which coincides with $a^\eps(\tau)$ for $\tau\leq\xi_K^\eps$ and for $\tau\geq\xi_K^\eps$ satisfies
\bee\nonumber
\dot a^{\eps,K}_j=\sqrt{2\TT_j}e^{-i\eps^{-1}\tau}\,\dot\beta_j,\quad j\in\CC.
\eee
Writing eq. (\ref{oa}) in the real coordinates and using that $P_j$ are real valued functions, we see that 
Girsanov's theorem implies that for all $0 \leq\tau\leq 1$ the process $a^{\eps,K}(\tau)$ satisfies the equation 
\bee\label{os32}
\dot a^{\eps,K}_j=\sqrt{2\TT_j}e^{-i\eps^{-1}\tau}\,\dot\beta_j^{\eps,K},\quad a^{\eps,K}_j(0)=u_j,\quad j\in\CC,
\eee 
where $(\beta_j^{\eps,K})_{j\in\CC}$ is a standard $|\CC|$-dimensional real Brownian motion with respect to the measure $\PRQ^{\eps,K}$ given by
\bee\label{os2}
d \PRQ^{\eps,K}(\om)=M^{\eps,K}(\om) d\PR(\om)
\eee
with 
$$
M^{\eps,K}=\exp\Big(-\int\limits_0^{1\wedge\xi_K^\eps}\sum\limits_{j\in\CC}\PP^\eps_j(\tau,a^{\eps,K}(\tau)) d\beta_j 
- \frac12\int\limits_0^{1\wedge\xi_K^\eps}|\PP^\eps(\tau,a^{\eps,K}(\tau))|^2\, d\tau \Big),
$$
where by $\PP^\eps=(\PP_j^\eps)_{j\in\CC}$ we have denoted 
$$\PP_j^\eps(\tau,u):=(2\TT_j)^{-1/2}P_j(e^{i\eps^{-1}\tau}u).$$
Note that Novikov's condition 
$
\ds{\MO \exp\Big(\frac12\int\limits_0^{1\wedge\xi_K^\eps}|\PP^\eps(\tau,a^{\eps,K}(\tau))|^2\, d\tau \Big) <\infty},
$
required for the application of Girsanov's theorem, is satisfied since for $\tau\leq\xi_K^\eps$ we have $|a^{\eps,K}(\tau)|\leq K$. 

Now let us estimate the transition probability $\msP_1^{a,\eps}(u,\GG)=\PR\big(a^\eps(1)\in\GG\big)$. We have
\begin{align}\label{os3}
\PR\big(a^\eps(1)\in\GG\big)&\geq \PR\big(a^\eps(1)\in\GG, \max\limits_{0\leq\tau\leq 1} |a^{\eps}(\tau)|< K\big)\\
\nonumber
& = \PR\big(a^{\eps,K}(1)\in\GG, \max\limits_{0\leq\tau\leq 1} |a^{\eps}(\tau)|< K\big) \geq  \PR\big(a^{\eps,K}(1)\in\GG\big)-\ae(K),
\end{align}
where Lemma \ref{olem:est}.{\it i} joined with (\ref{ov=a}) implies that $\ae(K)\ra 0$ as $K\ra\infty$ uniformly in $\eps$ sufficiently small.
Due to (\ref{os2}), for any $L>0$ we have
\begin{align}\nonumber
\PR\big(a^{\eps,K}(1)\in\GG\big)&=\int\limits_{\{a^{\eps,K}(1)\in\GG\}} (M^{\eps,K})^{-1}\, d\PRQ^{\eps,K}\geq
e^{-L} \PRQ^{\eps,K}\big(a^{\eps,K}(1)\in\GG, M^{\eps,K}\leq e^L\big) \\
\label{os4}
&\geq e^{-L} \big(\PRQ^{\eps,K}\big(a^{\eps,K}(1)\in\GG\big) - e^{-L}\big),
\end{align}
where we have employed the exponential supermartingale inequality.
In view of (\ref{os32}),  the random variable $a^{\eps,K}(1)$  has the Gaussian distribution with mean $u$, with respect to the measure $\PRQ^{\eps,K}$. Its covariance matrix written with respect to the real coordinates has the form 
$\ds{D=\diag\begin{pmatrix} D^{11}_j & D^{12}_j \\ D^{12}_j & D^{22}_j \end{pmatrix}_{j\in\CC}}$, where
\begin{align}\nonumber
D_{j}^{11}&=2\TT_j\int\limits_0^1\cos^2(\eps^{-1}\tau)\,d\tau = \TT_j (1+\frac{\eps}{2}\sin (2\eps^{-1})),   
\\ \nonumber
D_{j}^{22}&=2\TT_j\int\limits_0^1\sin^2(\eps^{-1}\tau)\,d\tau = \TT_j (1-\frac{\eps}{2}\sin (2\eps^{-1})), 
\\ \nonumber
D_{j}^{12}&=2\TT_j\int\limits_0^1\sin(\eps^{-1}\tau)\cos(\eps^{-1}\tau)\,d\tau 
= \frac{\eps\TT_j }{2}(1-\cos (2\eps^{-1})).
\end{align} 
Since for $\eps$ sufficiently small we have 
$\ds{D\geq \frac12 \diag\begin{pmatrix} \TT_j & 0 \\ 0 & \TT_j \end{pmatrix}_{j\in\CC}}$, we find that
\bee\label{os5}
\PRQ^{\eps,K}\big(a^{\eps,K}(1)\in\GG\big)\geq \PRQ^{\eps,K}\big(a^{\eps,K}(1)\in\GG\cap B_1\big)\geq C(u) \mbox{Leb} (\GG\cap B_1),
\eee
where $\mbox{Leb}(\GG\cap B_1)$ denotes the Lebesgue measure of the set $\GG\cap B_1$. 
Obviously,
\bee\label{os6}
\hat C:=\min\limits_{u\in B_R} C(u)>0.
\eee
Combining estimates (\ref{os3})-(\ref{os6}), for any $u\in B_R$ we get
\bee\label{os7}
\msP_1^{a,\eps}(u,\GG)\geq e^{-L}\big(\hat C\, \mbox{Leb}(\GG\cap B_1) - e^{-L}\big) - \ae(K).
\eee
Now we are able to estimate the variational norm (\ref{os1}).
Without loss of generality we assume 
$\msP_1^{a,\eps}(u_1,\GG)\geq \msP_1^{a,\eps}(u_2,\GG)$. Then, in view of (\ref{os7}),
\begin{align}\nonumber
|\msP_1^{a,\eps}(u_1,\GG) - \msP_1^{a,\eps}(u_2,\GG) |&= 1-\msP_1^{a,\eps}(u_1,\ov \GG)- \msP_1^{a,\eps}(u_2,\GG) \\
\nonumber
&\leq 1-e^{-L}\big(\hat C \mbox{Leb}(B_1) - 2e^{-L}\big) + 2\ae(K).
\end{align}
Choosing first $L$ and then  $K$ sufficiently large, we obtain 
$$
|\msP_1^{a,\eps}(u_1,\GG) - \msP_1^{a,\eps}(u_2,\GG) |\leq 1- \vartheta, \quad\mbox{where}\quad \vartheta:= e^{-L}\big(\hat C \mbox{Leb}(B_1) - 2e^{-L}\big) - 2\ae(K)>0.
$$
It remains to note that the constant $\vartheta$ is independent from $\eps$, $u_1,u_2\in B_R$ and $\GG\in\BB(\CCC)$. Then,  in view of (\ref{os1}), we arrive at (\ref{ou3}).
\qed

\section{Appendix: Low temperature regime}
\label{oapp:lt}

Let $(\wid p,\wid q)=(\wid p_j,\wid q_j)_{j\in\CC}\in\mR^{2|\CC|}$. Consider the equation
\bee\nonumber
\frac{d}{dt}\wid q_j=\wid p_j,\quad 
\frac{d}{dt}\wid p_j=-\wid q_j - \eps\sum\limits_{k:|k-j|=1}\chp_{\wid q_j}V(\wid q_j,\wid q_k) - \eps \wid p_j + \sqrt{2\eps\del\TT_j}\frac{d}{dt}\beta_j, \quad j\in\CC,
\eee
where $\eps,\del\ll 1$. 
It describes a system of weakly coupled oscillators, weakly interacting with thermostats of temperatures $\del\TT_j$, $j\in\CC$. Let us put 
$$p_j:=\sqrt\del\wid p_j, \quad q_j:=\sqrt\del\wid q_j,$$
 and assume that the interaction potential $V$ is a homogeneous polynom of degree $m\geq 3$. Then 
\bee\nonumber
\frac{d}{dt}q_j=p_j,\quad 
\frac{d}{dt}p_j=-q_j - \eps\del^{(m-2)/2}\sum\limits_{k:|k-j|=1}\chp_{q_j}V(q_j,q_k) - \eps p_j + \sqrt{2\eps\TT_j}\frac{d}{dt}\beta_j, \quad j\in\CC.
\eee
Putting $\la:=\del^{(m-2)/2}$, we arrive at equation (\ref{oini_er}), written in  the fast time $t$.

At the physical level of rigor the condition above for the form of the interaction potential can be weaken. Since we study small amplitude solutions, the interaction potential $V(\wid q_j,\wid q_k)$ can be replaced by a leading order term of its Taylor series. Thus, it suffices to assume that the latter is a homogeneous polynom of degree $m\geq 3$.

\section{Appendix: Resonant averaging}
\label{oapp:RA}
Here we discuss some properties of the resonant averaging, given by formulas
(\ref{oresavintro}) and (\ref{oresavv}). 
Since we will use the derivatives with respect to the angles $\ph=(\ph_j)_{j\in\CC}$, let us note that if a function
$f:\mC^{|\CC|}\mapsto\mR$ is $C^1$-smooth, then $f$ is continuously  differentiable with respect to the angles $\ph$. Indeed, this follows from the formula 
$$
\chp_{\ph_j}f=i u_j\chp_{u_j}f- i \ov u_j\chp_{\ov u_j}f.
$$
\bpp\label{oprop:RA}
 Let $f:\mC^{|\CC|}\ra\mR$ be a continuous function. Then

(i) For any $\xi\in[0,2\pi)$ we have $\lan f \ran_R (e^{i\xi}u) \equiv \lan f\ran_R (u)$. 

(ii) Let $f\in\LL_b(\mC^{|\CC|}), \LL_{loc}(\mC^{|\CC|})$ or $C^n(\mC^{|\CC|})$, where $n\in\mN$. Then $\lan f \ran_R\in\LL_b(\mC^{|\CC|}), \LL_{loc}(\mC^{|\CC|})$ or $C^n(\mC^{|\CC|})$ correspondingly. If partial derivatives of the function $f$ have at a most polynomial growth at infinity, then those of $\lan f\ran_R$ also have at most the polynomial growth. 

(iii) Let $f\in C^1(\mC^{|\CC|}).$ Then for every $j\in\CC$ we have $\lan\chp_{\ph_j} f\ran_R\equiv \chp_{\ph_j}\lan f\ran_R$. 

(iv) Let $f\in C^1(\mC^{|\CC|})$ be such that $\Supp f\subseteq \{j,k\}.$ Then $ \chp_{\ph_j}\lan f\ran_R=- \chp_{\ph_k}\lan f\ran_R.$ 
\epp
{\it Proof.} Items {\it (i)} and {\it (ii)} follow from formula (\ref{oresavv}). Item {\it (iii)} follows from (\ref{oresavintro}).  To prove item {\it (iv)} it suffices to note that, in view of (\ref{oresavintro}), the resonant averaging $\lan f\ran_R(I,\ph)$ depends on the angles $\ph_j,\ph_k$ only through their difference $\ph_j-\ph_k$.
\qed


\begin{thebibliography}{00}

\bibitem[AKN]{AKN} V.Arnold, V. V. Kozlov and A.I. Neistadt, {\it Mathematical Aspects of Classical and Celestial Mechanics}, 3rd ed., Springer, Berlin (2006).
\bibitem[BaBeO]{BBO} G. Basile, C. Bernardin, S. Olla, {\it Thermal conductivity for a momentum conservative model}, Commun. Math. Phys., {\bf 287} (2009), 67-98.
\bibitem[BaOS]{BOS} G. Basile, S. Olla, H. Spohn, {\it Energy transport in stochastically perturbed lattice dynamics}, Arch. Rat. Mech. Anal., {\bf 195} (2010), 171-203.
\bibitem[BeHu]{BeHu} C. Bernardin, F. Huveneers, {\it Small perturbation of a disordered harmonic chain by a noise and an anharmonic potential}, Probab. Theory Relat. Fields, {\bf 157} (2013), 301-331.
\bibitem[BeHuLeLO]{BeHuLeLO} C. Bernardin, F. Huveneers,  J.L. Lebowitz, C. Liverani, S. Olla, {\it Green-Kubo formula for weakly coupled system with dynamical noise}, arXiv: 1311.7384v1.
\bibitem[BeKLeLu]{BeKLeLu} C. Bernardin, V. Kannan, J.L. Lebowitz, J. Lukkarinen, {\it Harmonic Systems with Bulk Noises}, Journal of Statistical Physics, {\bf 146} (2012),  800-831.
\bibitem[BeO05]{BO} C. Bernardin, S. Olla, {\it Fourier's law for a microscopic model of heat conduction},
Journal of Statistical Physics, {\bf 118} (2005), 271-289.
\bibitem[BeO11]{BO1} C. Bernardin, S. Olla, {\it Transport properties of a chain of anharmonic oscillators with random flip of velocities}, 
Journal of Statistical Physics, {\bf 145} (2011), 1224-1255.
\bibitem[BKR]{BKR} V.I. Bogachev, N.V. Krylov, M. Rockner, {\it On regularity of transition probabilities and invariant measures of singular diffusions under minimal conditions}, Communications in Partial Differential Equations, {\bf 26} (2001), 2037-2080.
\bibitem[BoLeLu]{BLL} F. Bonetto, J. L. Lebowitz, J. Lukkarinen, {\it Fourier's Law for a Harmonic Crystal with Self-Consistent Stochastic Reservoirs}, Journal of Statistical Physics, {\bf 116} (2004),  783-813.
\bibitem[BoLeLuO]{BLLO} F. Bonetto, J. L. Lebowitz, J. Lukkarinen, S. Olla, {\it Heat Conduction and Entropy Production in Anharmonic Crystals with Self-Consistent Stochastic Reservoirs}, Journal of Statistical Physics, {\bf 134} (2009), 1097-1119.  
\bibitem[BoLeR]{Leb} F. Bonetto, J.L. Lebowitz, L. Rey-Bellet, {\it Fourier's law: a challenge to theorists}, Mathematical
Physics 2000, Imp. Coll. Press, London, (2000), 128-15.
\bibitem[Car]{Car} P. Carmona, {\it Existence and uniqueness of an invariant measure for a chain of oscillators in contact with two heat baths}, Stochastic Process. Appl., {\bf 117} (2007), 1076-1092.
\bibitem[CE]{CE} N. Cuneo, J.-P. Eckmann, {\it Non-equilibrium steady states for chain of four rotors}, arXiv:1504.04964.
\bibitem[CEPo]{CEPo} N. Cuneo, J.-P. Eckmann, C. Poquet, {\it Non-equilibrium steady state and subgeometric ergodicity for a chain of three coupled rotors}, Nonlinearity, {\bf 28} (2015), 2397-2421.
\bibitem[DL]{DL} 
D.Dolgopyat, C. Liverani, 
{\it Energy Transfer in a Fast-Slow Hamiltonian System},
Commun. Math. Phys., {\bf 308} (2011), 201-225.
\bibitem[Dud]{Dud} R.M. Dudley, {\it Real Analysis and Probability}, Cambridge University Press, Cambridge (2002).
\bibitem[Dym12]{Dym12} A.V. Dymov, {\it Dissipative effects in a linear Lagrangian system with infinitely many degrees of freedom}, Izv. Math.  {\bf 76} (2012), 1116-1149.
\bibitem[Dym14]{Dym14} A. Dymov, {\it Nonequilibrium Statistical Mechanics of Hamiltonian Rotators with Alternated Spins}, 
Journal of Statistical Physics, {\bf 158} (2015), 968-1006.
\bibitem[EH]{EckH}  J.-P. Eckmann, M. Hairer, {\it Non-Equilibrium Statistical Mechanics of Strongly Anharmonic Chains of Oscillators}, Commun. Math. Phys., {\bf 212} (2000), 105-164.
\bibitem[EPRB]{Eck} J.-P. Eckmann, C.-A. Pillet, L. Rey-Bellet, {\it Non-equilibrium statistical mechanics of anharmonic chains coupled to two heat baths at different temperatures}, Commun. Math. Phys., {\bf 201} (1999), 657-697.
\bibitem[FW06]{FW06} M.I. Freidlin, A.D. Wentzell, 
{\it Long-Time Behavior of Weakly Coupled Oscillators},
Journal of Statistical Physics,
{\bf 123} (2006), 1311-1337.
\bibitem[FW12]{FW} M. Freidlin, A. Wentzell,
{\it Random Perturbations of Dynamical
Systems}, 3rd ed., Springer-Verlag, Berlin Heidelberg (2012).
\bibitem[HM]{HM} M. Hairer, J.C. Mattingly, {\it Slow energy dissipation in anharmonic oscillator chains}, Comm. Pure Appl. Math., {\bf 62} (2009), 999-1032.
\bibitem[KaSh]{KaSh} I. Karatzas, S. Shreve, {\it Brownian Motion and Stochastic Calculus}, 2nd ed., Springer Verlag, Berlin (1991).
\bibitem[Kha12]{Khb} R. Khasminskii, {\it Stochastic stability of differential equations}; 2nd ed., Springer Verlag, Berlin (2012).
\bibitem[Kuk10]{Kuk10} S.B. Kuksin, {\it Damped-driven KdV and effective equations for long-time behaviour of its solutions}, GAFA, {\bf 20} (2010), 1431-1463.
\bibitem[Kuk13]{Kuk13}  S.B. Kuksin, {\it Weakly nonlinear stochastic CGL equations}, Ann. IHP PR {\bf 49} (2013), 1033-1056.
\bibitem[KM]{KM} S. Kuksin, A. Maiocchi, {\it Resonant averaging for weakly nonlinear stochastic Schrodinger equations}, arXiv:1309.5022.
\bibitem[KP]{KuPi} S.B. Kuksin, A.L. Piatnitski, {\it Khasminskii-Witham averaging for randomly perturbed KdV equation}, J.Math. Pures Appl., {\bf 89} (2008), 400-428.
\bibitem[KS]{KuSh} S. Kuksin, A. Shirikyan, {\it Mathematics of two-dimensional turbulence}, Cambridge University Press, Cambridge (2012). 
\bibitem[LepLiPo]{LepLiPo} S. Lepri, R. Livi, A. Politi, {\it Thermal conduction in classical low-dimensional lattices}, Phys. Rep., {\bf 377} (2003).
\bibitem[LO]{LO} C. Liverani, S. Olla, {\it Toward the Fourier law for a weakly interacting anharmonic crystal}, AMS, {\bf 25} (2012), 555-583.
\bibitem[MSH]{MSH} J.C. Mattingly, A.M. Stuart, D.J. Higham, {\it Ergodicity for SDEs and approximations: locally
Lipschitz vector fields and degenerate noise},  Stochastic Processes and their Applications, {\bf 101} (2002), 185-232. 
\bibitem[{\O}ks]{Oks} B. {\O}ksendal, {\it Stochastic Differential Equations,} Springer-Verlag, Berlin (2003).
\bibitem[Pei]{Pei} R. Peierls,{\it On the kinetic theory of thermal conduction in crystals}, Selected
Scientific Papers of Sir Rudolf Peierls, with commentary, World
Scientific, Singapore (1997), 15-48.
\bibitem[PV]{PaVe} E. Pardoux, A.Yu. Veretennikov, {\it On the Poisson equation and diffusion approximation. I}, The Annals of Probability, {\bf 29} (2001), 1061-1085.
\bibitem[RBT]{RBT} L. Rey-Bellet,  L.E. Thomas, {\it Exponential convergence to non-equilibrium stationary states in classical statistical mechanics}, Commun. Math. Phys., {\bf 225} (2002),  305-329.
\bibitem[RSZ]{RSZ} M.Rockner, B.Schmuland, X.Zhang, {\it Yamada-Watanabe theorem for stochastic evolution equations in infinite 
dimensions}, Condensed Matter Physics {\bf 11} (2008), 247-259.
\bibitem[Ru]{R} D.A. Ruelle, {\it Mechanical Model for Fourier's Law of Heat Conduction}, 
Commun. Math. Phys., {\bf 311} (2012), 755-768.   
\bibitem[Sp]{Sp} H. Spohn, {\it Large scale dynamics of interacting particles},  Springer-Verlag, Heidelberg (1991).
\bibitem[Tay]{Tay} M.E. Taylor, {\it Partial Differential Equations I: Basic Theory}, Springer-Verlag, New York (1996).
\bibitem[Tr]{Tr} D. Treschev, {\it Oscillator and thermostat}, Discrete Contin. Dyn. Syst. {\bf 28}  (2010), 1693-1712.
\bibitem[Ver87]{Ver87} A. Veretennikov, {\it Bounds for the Mixing Rate in the Theory of Stochastic Equations}, Theory Probab. Appl., {\bf 32}  (1987), 273-281.
\bibitem[Ver97]{Ver97} A. Yu.  Veretennikov, {\it On polynomial mixing bounds for stochastic differential equations}, Stochastic Processes and their Applications,  {\bf 70} (1997), 115-127.
\bibitem[Yor]{Yor} M. Yor, {\it Existence et unicit\'e de diffusion \`a valeurs dans un espace de Hilbert}, Ann. IHP B, {\bf 10} (1974), 55-88. 
\end{thebibliography}
\end{document}